\documentclass[prd,preprintnumbers,nofootinbib,superscriptaddress,onecolumn,preprint]{revtex4-1}

\pdfoutput=1
\usepackage{graphicx}
\usepackage{amsmath}
\usepackage{amssymb}
\usepackage{slashed}
\usepackage{braket}
\usepackage{mathtools}
\usepackage[utf8]{inputenc}
\usepackage{color}
\usepackage[usenames,dvipsnames]{xcolor}
\usepackage[normalem]{ulem}
\usepackage{soul}
\usepackage{units}
\usepackage{rotating}
\usepackage{hhline}
\usepackage{multirow}
\usepackage{tabularx}
\usepackage{bm}
\usepackage{longtable}
\usepackage{makecell}
\usepackage{setspace}

\newcommand{\hxi}{\tilde{\xi}}
\newcommand{\hzeta}{\tilde{\zeta}}
\newcommand{\T}[2]{\boldsymbol{#1}_{#2\text{T}}}
\newcommand{\THD}[1]{\boldsymbol{#1}_{\text{H,T}}}
\newcommand{\THDsc}[1]{#1_{\text{H,T}}}

\newcommand{\Tsc}[2]{#1_{#2\text{T}}}
\newcommand{\Tscsq}[2]{#1^2_{#2\text{T}}}
\newcommand{\no}{\nonumber \\}
\newcommand{\parz}[1]{\ensuremath{\left(#1\right)}}
\newcommand{\order}[1]{\ensuremath{O\parz{#1}}}

\newcommand{\contractor}[1]{\ensuremath{{\rm P}}_{#1}}
\newcommand{\contractortot}[1]{\ensuremath{{\rm P}}_{#1}}

\newcommand{\ps}{\ensuremath{\Pi}}

\newcommand{\hard}{\ensuremath{\mathcal{H}}}

\newcommand{\proper}{\text{reduced}}
\newcommand{\improper}{\text{unreduced}}

\newcommand{\cosfun}{\ensuremath{\mathcal{ \, C}}}
\newcommand{\ffun}{\ensuremath{\mathcal{\, F}}}
\newcommand{\ggun}{\ensuremath{ \mathcal{\, G}}}
\newcommand{\llun}{\ensuremath{ \mathcal{\, L}}}
\newcommand{\kkun}{\ensuremath{ \mathcal{\, K}}}

\newcommand{\eref}[1]{Eq.~(\ref{e.#1})}
\newcommand{\erefs}[2]{Eqs.~(\ref{e.#1})--(\ref{e.#2})}
\newcommand{\fref}[1]{Fig.~\ref{f.#1}}

\newcommand{\aref}[1]{Appendix~\ref{a.#1}}
\newcommand{\sref}[1]{Sec.~\ref{s.#1}}

\allowdisplaybreaks
\begin{document}

\DeclareRobustCommand*\diff[2][]{%
   \mathop{
     \mathrm{d}^{#1}
     \mskip-0.2\thinmuskip
    #2}\nolimits
}
\newcommand{\di}[1]{\diff{#1}{}}

\newcommand{\intloop}{\int \frac{\diff{^{4 - 2 \epsilon} k}{}}{(2\pi)^{4 -2 \epsilon}}}
\newcommand{\intloopb}{\int \frac{\diff{^{4 - 2 \epsilon} l}{}}{(2\pi)^{4 -2 \epsilon}}}

\title{Large Transverse Momentum in Semi-Inclusive Deeply
Inelastic Scattering Beyond Lowest Order}

\preprint{JLAB-THY-19-2897}

\author{B.~Wang}
\email{0617626@zju.edu.cn}
\affiliation{Department of Physics, Old Dominion University, Norfolk, VA 23529, USA}
\affiliation{Jefferson Lab, 12000 Jefferson Avenue, Newport News, VA 23606, USA}
\affiliation{Zhejiang Institute of Modern Physics, Department of Physics, Zhejiang University,\\ Hangzhou 310027, China}
\author{J.~O.~Gonzalez-Hernandez}
\email{ joseosvaldo.gonzalez@to.infn.it}
\affiliation{Department of Physics, Old Dominion University, Norfolk, VA 23529, USA}
\affiliation{Jefferson Lab, 12000 Jefferson Avenue, Newport News, VA 23606, USA}
\affiliation{Dipartimento di Fisica, Universit\`{a} di Torino, Via P. Giuria 1, 1-10125, Torino, Italy}
\author{T.~C.~Rogers}
\email{tedconantrogers@gmail.com}
\affiliation{Department of Physics, Old Dominion University, Norfolk, VA 23529, USA}
\affiliation{Jefferson Lab, 12000 Jefferson Avenue, Newport News, VA 23606, USA}
\author{N.~Sato}
\email{nsato@jlab.org}
\affiliation{Jefferson Lab, 12000 Jefferson Avenue, Newport News, VA 23606, USA}

\date{4 March 2019}

\begin{abstract}
Motivated by recently observed tension between $\order{\alpha_s^2}$
calculations of very large transverse momentum dependence in both
semi-inclusive deep inelastic scattering and Drell-Yan scattering, we repeat the details of 
the calculation through $\order{\alpha_s^2}$ transversely differential cross section.  The 
results confirm earlier calculations, and provide further support to the observation that 
tension exists with current parton distribution and fragmentation functions.  
\end{abstract}

\maketitle

\section{Introduction}
In a previous article~\cite{Gonzalez-Hernandez:2018ipj}, we discussed
the semi-inclusive deep inelastic scattering (SIDIS) process: $$l(l) +
\text{Proton}(P) \to l^{\prime}(l^{\prime})+\text{Hadron}(P_H) + X,$$
and we highlighted the challenge of finding agreement between
$\order{\alpha_s^2}$ calculations and existing SIDIS data in the very large
transverse momentum transverse momentum limit where standard collinear
factorization is expected to be valid.  One motivation is that obtaining a  
description of the small transverse momentum behavior associated with 
nucleon structure requires a good understanding of the matching to large 
transverse momentum where a transverse momentum dependent (TMD) 
factorization description fails. 
Given the current focus on using deeply inelastic 
hadro-production to access nucleon structure sensitivity sensitivity,
it is imperative to 
examine theoretical framework for the full $\Tsc{q}{}$-range in greater detail.  In the 
language of Ref.~\cite{Gonzalez-Hernandez:2018ipj}, we are interested in this paper in 
what was there called ``region 3" behavior, corresponding to where $\Tsc{q}{}$ is so 
large that the small $\Tsc{q}{}/Q$ approximations associated with TMDs are not reliable, 
but where ordinary collinear factorization should be applicable and reliable.  
Calculations to $\order{\alpha_s^2}$ have existed for some 
time~\cite{Daleo:2004pn,Kniehl:2004hf}.  The main observation of 
Ref.~\cite{Gonzalez-Hernandez:2018ipj} was that, while the $\order{\alpha_s^2}$ 
correction gives an order of magnitude increase over the leading order, that still is not 
sufficient to achieve reasonable agreement with data for $Q$ in the region of one to 
several GeVs, transverse momentum of order $Q$, and for moderate Bjorken-$x$. 
Combined with similar observations concerning the Drell-Yan pointed out in 
\cite{Bacchetta:2019tcu}, this points to general tension between transverse momentum 
dependent cross sections at $\Tsc{q}{} \sim Q$ and collinear factorization. 

There are a number of potential explanations or solutions, including
a direct re-tuning of collinear parton distribution and/or
fragmentation functions to transversely differential SIDIS cross
sections. (A proposal to constrain gluon PDFs in transversely differential 
Drell-Yan cross sections was made already in~\cite{Berger:1998ev}.)
But before proceeding to consider these directions, 
it is important to validate the $O(\alpha_S^2)$ in \cite{Daleo:2004pn,Kniehl:2004hf}
that led to our conclusions in \cite{Gonzalez-Hernandez:2018ipj}.
Therefore, we have in this paper repeated the large transverse
momentum order $\order{\alpha_s^2}$ calculation following a slightly
different formal framework.  We are able to reproduce the results
in~\cite{Daleo:2004pn} very closely, thus bolstering the observations
we made earlier in~\cite{Gonzalez-Hernandez:2018ipj}. 

The specific purposes of this article are as follows:  i.) to lay out
the logical steps of the calculation with enough detail, we hope, to
lead to clues as to how to improve phenomenological agreement,
ii.) to make available a convenient numerical implementation for the large $\Tsc{q}{}$ 
region and, 
iii.) to present some
quantitative results relevant to current
experimental programs such as those at COMPASS and Jefferson Lab 12 GeV.
Overall, our results support the general observations made in
\cite{Gonzalez-Hernandez:2018ipj}.

In our calculations we use qgraf \cite{Nogueira:1991ex}
to generate Feynman graphs and FORM \cite{Kuipers:2012rf}
to carry out spinor and color traces for the amplitudes. The renormalization
counter terms are computed in MATHEMATICA with the packages 
FeynArts \cite{Kublbeck:1990xc,Hahn:2000kx} and FeynCalc \cite{Mertig:1990an,Shtabovenko:2016sxi}.
Further analytic manipulations were performed in MATHEMATICA. 
The code to compute the cross sections are publically
available at \cite{BigTMD}.

In \sref{overview}, we explain our setup and notation. In \sref{organization} we 
summarize the organization of our calculations. This includes a classification of
the partonic Feynman graphs needed at order $\alpha_s(Q)^2$ and a discussion of the 
phase space integrals for multiparton final states. We discuss the results of the 
calculation in \sref{results}, and give concluding remarks in \sref{conc}.

\section{Notation and Conventions}
\label{s.overview}

We will express quantities in terms of the conventional kinematical
variable $z \equiv P_H \cdot P/ (P \cdot q)$. $\T{P}{H,}$ is the Breit
frame transverse momentum of the produced hadron, and $P$ and $q$ are
the four-momenta of the incoming target hadron and the virtual photon
respectively.  We will focus on the unpolarized and
azimuthally-independent cross section since this is the most straight
forward observable to calculate.  (Although the results have potential
implications also for polarization dependent observables.) Many 
general treatments of SIDIS as a process are available~\cite{Meng:1991da,Levelt:1993ac,Meng:1995yn,Mulders:1996dh,Nadolsky:1999kb,Nadolsky:2000ky,Barone:2001sp,Ji:2004xq,Bacchetta:2004jz,Koike:2006fn,Bacchetta:2006tn,Bacchetta:2008xw}. Our notation is consistent with trends among these, with modifications as needed for our current purposes.

The unpolarized differential cross section is
\begin{equation}
4 P_{\rm H}^0 E^\prime 
\frac{\diff{\sigma_{H}}{}}
     {\diff{^3 {\bf l}{'}} \, \diff{^3 {\bf P}_{\rm H} }} 
= \frac{2 \, \alpha_{\rm em}^2 }{S_{lP} Q^4} \; 
L_{\mu \nu} W^{\mu \nu}\, ,
\label{e.colfac0}
\end{equation}
or
\begin{equation}
\frac{\diff{\sigma_{H}}}{\diff{x} \diff{y} \diff{z} \diff{\THDsc{P}^2}  } =
\frac{\pi^2 \alpha_{\rm em}^2 y}{2 Q^4 z }L_{\mu \nu} W^{\mu \nu} \, ,
\label{e.colfac}
\end{equation}
where $\sigma_{H}$ is the unpolarized hadronic cross section. $S_{lP}
= (l + P)^2$. The hadron transverse momentum $\T{P}{H,}$ is defined in
a frame where the photon and incoming hadron are back-to-back (a ``photon" frame).  
The Bjorken $x$ and the $y$ variable are the usual
definitions $x\equiv Q^2/2P\cdot q$ and $y\equiv P\cdot q/P\cdot l$.
$L_{\mu \nu}$ is the usual leptonic tensor
\begin{equation}
L_{\mu \nu} \equiv 2 (l_\mu l^\prime_\nu 
                + l^\prime_\mu l_\nu 
                - g_{\mu \nu} l \cdot l^\prime) \, ,
\label{e.lepttensor}
\end{equation}
and $W^{\mu \nu}$ is the hadronic tensor for SIDIS:
\begin{equation}
W^{\mu \nu}(P,q,P_H) \equiv 
  \frac{1}{(2 \pi)^4} \sum_{X} 
  \int \diff{^4 z} \, e^{i q \cdot z} 
  \langle P, S |j^{\mu}(z) | P_H,X \rangle \langle P_H,X | j^{\nu}(0) | P, S \rangle
  \, ,
\label{e.hadronictensor}
\end{equation}
where we have omitted spin and azimuthal angle dependent terms since
we do not consider these in this paper.  Also, we assume that $x$ and
$1/Q$ are small enough that both the proton and lepton mass can be
dropped in kinematical and phase space factors.  The normalization
convention in \eref{hadronictensor} is so that the prefactor on the
right hand side of \eref{colfac} is similar to the unpolarized case.

The unpolarized structure functions $F_1$ and $F_2$ are defined by the
usual gauge invariant decomposition
\begin{equation}
W^{\mu \nu} = 
  \left( -g^{\mu \nu} + \frac{q^{\mu} q^{\nu}}{q^2} \right) F_1 
  +\frac{(P^\mu - q^\mu P \cdot q / q^2)(P^\nu - q^\nu P \cdot q / q^2)}{P \cdot q} F_2
  \, .
\label{e.structdec}
\end{equation}
Then, the cross section is
\begin{equation}
\frac{\diff{\sigma_{H}}}{\diff{x} \diff{y} \diff{z} \diff{\THDsc{P}^2}}= 
  \frac{\pi^2 \alpha_{\rm em}^2}{z x y Q^2}
  \Biggl[ x y^2 F_1 + (1 - y) F_2 \Biggr]
  \, .
\end{equation}

In calculations, it is convenient to work with Lorentz invariant
structure function extraction tensors
$\contractortot{\Gamma}^{\mu\nu}$ with $\Gamma \in \{ g, PP \}$ 
where
\begin{align}
\contractortot{g}^{\mu\nu} = g^{\mu\nu} ,\;\quad
\contractortot{PP}^{\mu\nu} = P^\mu P^\nu \, .
\label{e.PgPPtot}
\end{align}
Then, 
\begin{equation}
F_1 = \contractortot{1}^{\mu \nu} W_{\mu \nu} \,, \qquad 
F_2 = \contractortot{2}^{\mu \nu} W_{\mu \nu} \,  \label{e.contractorstot}
\end{equation}
where in $4 - 2 \epsilon$ dimensions,
\begin{equation}
\contractortot{1}^{\mu\nu} = \frac{1}{1 - \epsilon}\left(-\frac{1}{2} \contractortot{g}^{\mu\nu} 
                          +\frac{2 x^2}{Q^2} \contractortot{PP}^{\mu\nu}  \right)
                          \, , \qquad
\contractortot{2}^{\mu\nu} = \left( \frac{3 - 2 \epsilon}{1 - \epsilon} \right) \frac{4  x^3}{Q^2} \contractortot{PP}^{\mu\nu} 
                          - \frac{x}{1 - \epsilon} \contractor{g}^{\mu\nu}
                          \, . 
\end{equation}
(We we always use the massless target approximation -- 
see discussion in \cite{Moffat:2019qll}.)
It is useful to express transverse momentum in terms of 
\begin{equation}
\T{q}{} = -\frac{\THD{P}}{z} \, . 
\end{equation}
In a frame where the incoming and outgoing hadrons are back-to-back,
$\T{q}{}$ is the transverse momentum of the virtual photon (assuming, as
always in this paper, that external hadron masses are negligible). 

Our treatment of factorization will follow the general style of \cite{Collins:2011qcdbook}.  
The factorization theorem that relates the hadronic and partonic differential cross
sections in SIDIS is
\begin{equation}
4 P_{\rm H}^0 E^\prime 
 \frac{\diff{\sigma_H}{}}
      {\diff{^{3- 2 \epsilon} {\bf l}{'}} \, \diff{^{3- 2 \epsilon} {\bf P}_{\rm H} }} 
=
  \int_{x}^{1} \frac{\diff{\xi}}{\xi} 
  \int_{z}^{1} \frac{\diff{\zeta}}{\zeta^2} 
  \left( 4 k_1^0 E^\prime 
        \frac{\diff{\hat{\sigma}_{ij}}{}}
             {\diff{^{3-2\epsilon} {\bf l}{'}} \, \diff{^{3-2\epsilon} {\bf k}_1 }} 
   \right) 
   f_{i/P}(\xi) d_{H/j}(\zeta) \, . 
\label{e.xsecfact}
\end{equation}
The $1/\xi$ is from the partonic flux factor, and the 1$/\zeta^2$ is
from the conversion between ${\bf k}_1$ and ${\bf P}_{\rm H}$.  The
indices $i$ and $j$ denote, respectively, the flavors of the parton in
the proton (with a momentum fraction $\xi$) and of the outgoing parton that
fragments into hadron $H$, whose momentum is a fraction $\zeta$ of
parton $j$ momentum. The incoming and outgoing parton momenta $p$ and
$k_1$ satisfy $p=\xi P$ and $k_1=P_H/\zeta$. (Indices $i$ and $j$ for
incoming and outgoing partons $p_i$ and $k_{1,j}$ are not shown
explicitly but are understood).  $f_{i/P}(\xi)$ and $d_{H/j}(\zeta)$
are the collinear parton distribution and fragmentation functions
respectively. It is also useful to define partonic variables 
\begin{equation}
\hat{x}\equiv \frac{Q^2}{(2p\cdot q)} = \frac{x}{\xi} \, , \qquad \hat{z} \equiv \frac{k_1\cdot p}{(p\cdot q)} = \frac{z}{\zeta} \, , \qquad  \Tsc{k}{1}\equiv \frac{\Tsc{P}{H,}}{\zeta} \, .
\end{equation}
The differential partonic hard part in \eref{xsecfact} is finite and
well-behaved, and at large transverse momentum it starts at
$\order{\alpha_s}$. 

The unpolarized partonic structure tensor for scattering off parton
$i$ into parton $j$ is defined in exact analogy with the hadronic
tensor:
\begin{equation}
4 k_1^0 E^\prime 
\frac{\diff{\hat{\sigma}_{ij}}{}}
     {\diff{^{3-2 \epsilon} {\bf l}{'}} \, \diff{^{3- 2 \epsilon} {\bf k}_1 }}
= \frac{2 \, \alpha_{\rm em}^2 }{\hat{s} Q^4} \; L_{\mu \nu} \hat{W}^{\mu \nu}\, ,
\label{e.colfac0p}
\end{equation}
with,
\begin{align}
\hat{W}_{\mu\nu, ij} \equiv 
  \frac{1}{2} \frac{1}{(2 \pi)^4} 
  \sum_{s,X} \int \diff{^4 w}{} \, e^{i q \cdot w} 
  \langle p_i,s| j_{\mu}(w) | k_{1j},X \rangle 
  \langle k_{1j},X | j_{\nu}(0) | p_i, s\rangle \, .
\label{e.prttsr}
\end{align}
Thus, from \eref{xsecfact}, 
\begin{equation}
W^{\mu \nu}(P,q,P_H) = 
  \int_{x}^{1} \frac{\diff{\xi}}{\xi} 
  \int_{z}^{1} \frac{\diff{\zeta}}{\zeta^2} 
  \hat{W}^{\mu \nu}_{ij}(q,x/\xi,z/\zeta) 
  f_{i/P}(\xi) d_{H/j}(\zeta) \, 
  \, . \label{e.factform}
\end{equation}
The partonic structure function decomposition is
\begin{equation}
\hat{W}^{\mu \nu}_{ij} =  
   \left(-g^{\mu \nu} + \frac{q^{\mu} q^{\nu}}{q^2} \right) \hat{F}_{1,ij} 
  +\frac{(p^\mu - q^\mu p \cdot q / q^2) (p^\nu - q^\nu p \cdot q / q^2)}
        {p \cdot q} \hat{F}_{2, ij} 
  \, .
\label{e.structdecpar}
\end{equation}
Then,
\begin{align}
F_1(x,z,Q^2,\T{q}{}) {}& = \int_{x}^{1} 
                         \frac{\diff{\xi}}{\xi} \int_{z}^{1} 
                         \frac{\diff{\zeta}}{\zeta^2} 
                         \hat{F}_{1, ij}(x/\xi,z/\zeta,Q^2,\T{q}{}) 
                         f_{i/P}(\xi) d_{H/j}(\zeta) 
                         \, , \\
F_2(x,z,Q^2,\T{q}{}) {}&=  \int_{x}^{1} \diff{\xi} 
                         \int_{z}^{1} \frac{\diff{\zeta}}{\zeta^2}
                         \hat{F}_{2, ij}(x/\xi,z/\zeta,Q^2,\T{q}{}) 
                         f_{i/P}(\xi) d_{H/j}(\zeta)
                         \,. 
\end{align}

\section{Organization}
\label{s.organization}

\subsection{Basic Setup}
For a process with $N$ final state partons,
$\contractortot{g}^{\mu\nu} W_{\mu\nu}$ and
$\contractortot{PP}^{\mu\nu} W_{\mu\nu}$ are squared amplitudes
integrated over the $N$-particle phase space for the outgoing partons,
\begin{align}
 \{ \contractor{g}^{\mu \nu}  \hat{W}^{(N)}_{\mu\nu}; 
    \contractor{PP}^{\mu \nu} \hat{W}^{(N)}_{\mu\nu}\}
&{} \equiv 
 \frac{1}{(2 \pi)^4} \int 
  \{\lvert M^{2\to N}_g\rvert^2;\lvert M^{2\to N}_{PP}\rvert^2\} \, 
  \diff{\ps}^{(N)} - \, {\rm Subtractions} \nonumber \\
&{} \equiv  \{ \contractor{g}^{\mu \nu}  \hat{W}^{(N)}_{\mu\nu}; 
    \contractor{PP}^{\mu \nu} \hat{W}^{(N)}_{\mu\nu}\}_{\rm unsub} - \, {\rm Subtractions} \, . 
\label{e.cntrns}
\end{align}
Here $\lvert M_g \rvert^2$ or $\lvert M_{PP} \rvert^2$ is the squared
amplitude for the process 
$$\gamma^\ast(q) + \text{parton}(p) \to \text{parton}(k_1) + \text{($N-1$) 
spectator partons},$$ with the polarization sum of the virtual photon
replaced by $g^{\mu\nu}$ or $P^{\mu}P^{\nu}$.  The subtraction terms
in \eref{cntrns} are needed to remove double counting with lower
orders of perturbation theory and cancel singularities in the first
term. The form of the subtraction term will be discussed
in~\sref{subtraction}. 
It is assumed in Eq.~\eqref{e.cntrns} that all integrals that allow
kinematical $\delta$-functions to be evaluated have been performed.
Also, the phase space factors associated with $k_1$ are excluded from
the partonic phase space since they give the $\hat{z}$ and
$\Tscsq{k}{1}$ dependence of the differential partonic cross section.
The $1/(2 \pi)^4$ on the right hand side of \eref{cntrns} is the same
factor in \eref{hadronictensor}. The $\diff{\ps}^{(N)}$ 
represents a generic phase
space factor for $2 \to N$ scattering (see \eref{2to2ps} and
\eref{2to3ps}).

Thus it is convenient to express the structure functions in the form
\begin{equation}
\contractortot{\Gamma}^{\mu \nu} W_{\mu \nu}(P,q,P_H)
= \sum_{N} \int_{x}^{1} \frac{\diff{\xi}}{\xi} 
  \int_{z}^{1} \frac{\diff{\zeta}}{\zeta^2}  \{ \contractor{g}^{\mu \nu}  \hat{W}^{(N)}_{\mu\nu}; 
    \contractor{PP}^{\mu \nu} \hat{W}^{(N)}_{\mu\nu}\} f_{i/P}(\xi) d_{H/j}(\zeta) \, ,
\label{e.had2par}
\end{equation}
where the $\sum_N$ is a sum over all possible partonic final states.

\begin{figure}
\centering
\begin{tabular}{c@{\hspace*{5mm}}c}
  \includegraphics[scale=0.5]{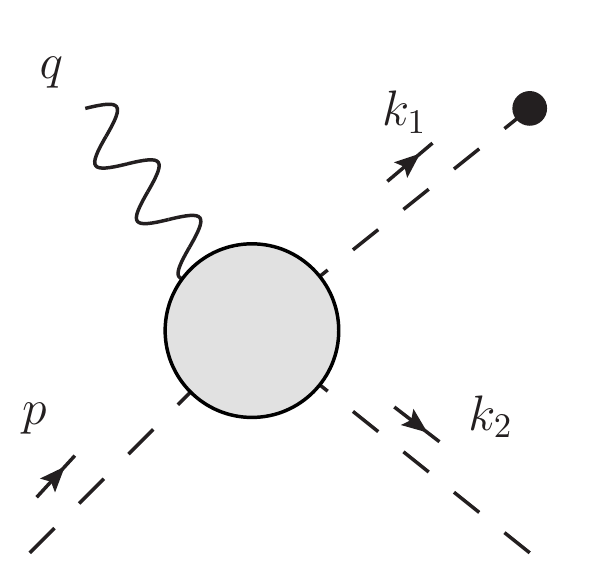} \hspace{0.5cm} &
  \hspace{0.5cm} \includegraphics[scale=0.5]{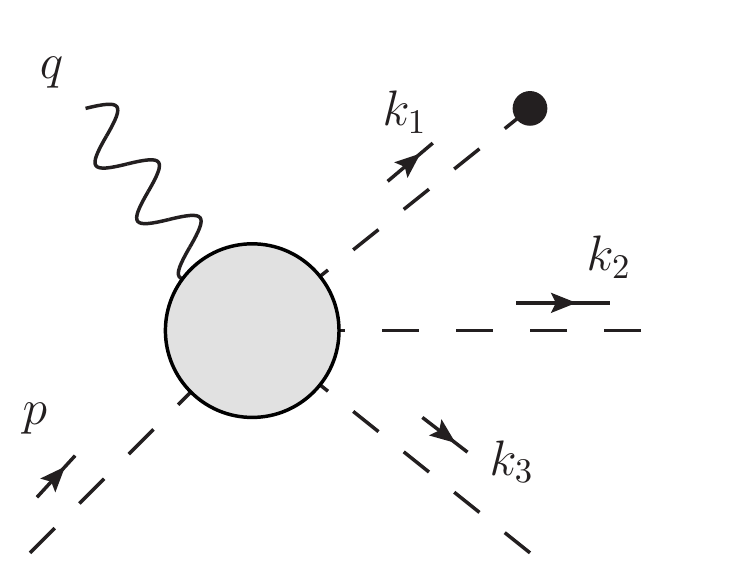} \\
  (a) & (b)
\end{tabular}
\caption{
  Momentum labels in amplitudes for (a) $2 \to 2$ and (b) $2 \to 3$
  partonic scattering. The dashed lines represent partons of unspecified
  flavor. The dot on the end of $k_1$ indicates this is the fragmenting
  parton. The other momenta are integrated in SIDIS.}
\label{f.basickinematics}
\end{figure}

We will express the phase space in terms of Mandelstam variables:
\begin{align}
s=&(p+q)^2 = 2 p \cdot q - Q^2, \\
t_i=&(q-k_i)^2 = -Q^2 - 2 q \cdot k_i, \\
u_i=&(p-k_i)^2 = -2 p \cdot k_i, \\
s_{ij}=&(k_i+k_j)^2 = 2 k_i \cdot k_j ,
\end{align}
where $i,j=1,2,3$, with the labeling in \fref{basickinematics}.   For
simplicity, $u_1$ and $t_1$ will be abbreviated as $u$ and $t$ from
here on. All Mandelstam variables will refer to the partonic cross
sections.  
Occasionally it will be useful to change kinematical variables, For example, 
\begin{align}
s ={}&(p+q)^2 
  = 2 p \cdot q - Q^2 
  = Q^2(1/(x/\xi) - 1), \\
t ={}&(q-k_1)^2 = -Q^2 - 2 q \cdot k_1 
  = -\frac{\Tscsq{k}{1}}{z/\zeta} - Q^2(1 - (z/\zeta)) 
  =-Q^2+\hat{z} \Biggl(Q^2-\frac{\Tscsq{k}{1}}{\hat{z}^2} \Biggr) \, , \\
u ={}&(p-k_1)^2 = -2 p \cdot k_1 
  = -\frac{(z/\zeta) Q^2}{x/\xi} 
  =-\frac{\hat{z}Q^2}{\hat{x}}\,, \\
\diff{t} \diff{u}={}&\frac{Q^2}{\hat{x}\hat{z}} \diff{\hat{z}} \diff{\Tscsq{k}{1}} \, .
\end{align}
The $O(\alpha_s^0)$ contribution to \eref{cntrns} is kinematically
constrained to $\Tsc{k}{1}=0$.  At $O(\alpha_s^1)$, only tree level
processes contribute, and no singularities appear. At $O(\alpha_s^2)$,
soft, collinear and UV singularities arise as $1/\epsilon$ and
$1/\epsilon^2$ poles in dimensional regularization with space-time
dimension $n\equiv4-2\epsilon$.

These singularities cancel in the sum of real, virtual and counterterm
graphs and after applying collinear factorization. More discussion of
the singularity structure at $O(\alpha_S^2)$ is available
in~\cite{Daleo:2004pn,Arnold:1988dp}.  The $O(\alpha_s)$ partonic
cross section is described in details in~\cite{Nadolsky:1999kb}, and
our results match that calculation. At $O(\alpha_s^2)$, details below
can also be found in, e.g.,
~\cite{Aversa:1988vb,Gordon:1993qc,Ellis:1979sj,Ellis:1981hk,Arnold:1988dp}.  
The change of variables in \eref{had2par}
from momentum fractions to Mandelstam variables is
\begin{align}
\int \diff{\xi}{} \diff{\zeta}{}  \cdots 
=\int_A^1 \diff{\xi}{} \int_0^B \diff{s_{23}}{} 
 \frac{\hat{x}^2\Tsc{P}{H,}^2+\hat{x}z^2Q^2(1-\hat{x})}
      {z(Q^2(1-\hat{x})-s_{23}\hat{x})^2} 
\cdots \, ,
\label{e.hdxsc}
\end{align}
where $\zeta$ is replaced by $s_{23}$, the virtuality of the spectator
parton system.  ($s_{23}$ is $k_X^2$ in the notation of 
\cite{Gonzalez-Hernandez:2018ipj}). Overall momentum conservation
$s+u+t=-Q^2+s_{23}$ ($s\equiv (p+q)^2=Q^2(1-\hat{x})/\hat{x}$) gives
\begin{align}
&\zeta=\frac{\hat{x}\Tsc{P}{H,}^2+z^2Q^2(1-\hat{x})}{z(Q^2(1-\hat{x})-s_{23}\hat{x})} 
\label{e.zeta2xz}\, , \\
&A=x+\frac{x\Tsc{P}{H,}^2}{z(1-z)Q^2} \, , \\
&B=Q^2\left (\frac{1}{\hat{x}}-1 \right) (1-z)-\frac{\Tsc{P}{H,}^2}{z} \, .
\label{e.limitsA}
\end{align}
%

\begin{figure}
\centering
\begin{tabular}{c@{\hspace*{.5cm}}c@{\hspace*{.5cm}}c@{\hspace*{.5cm}}c}
\hspace*{-0.3cm}
\includegraphics[width=0.21\textwidth]{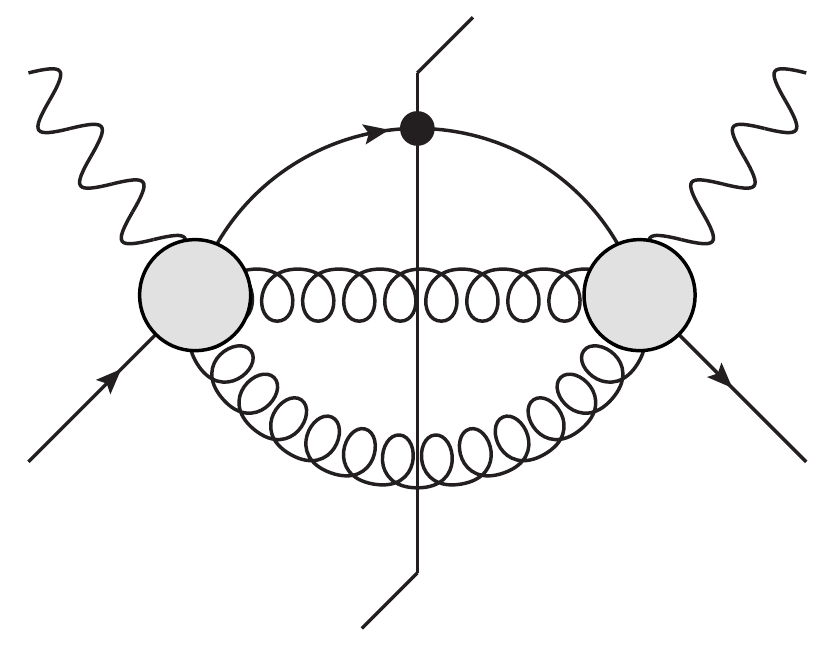}\ \
&
\includegraphics[width=0.21\textwidth]{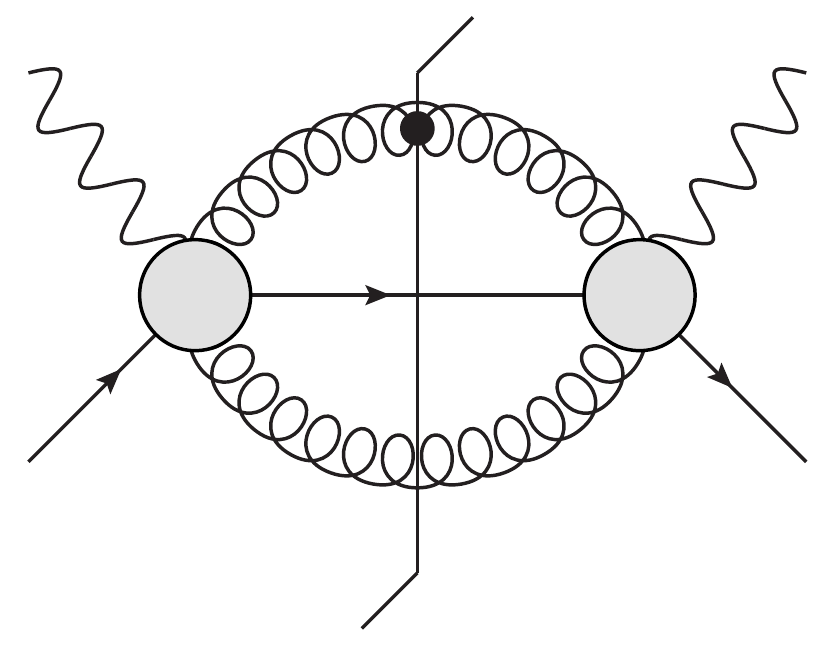}\ \
&
\includegraphics[width=0.21\textwidth]{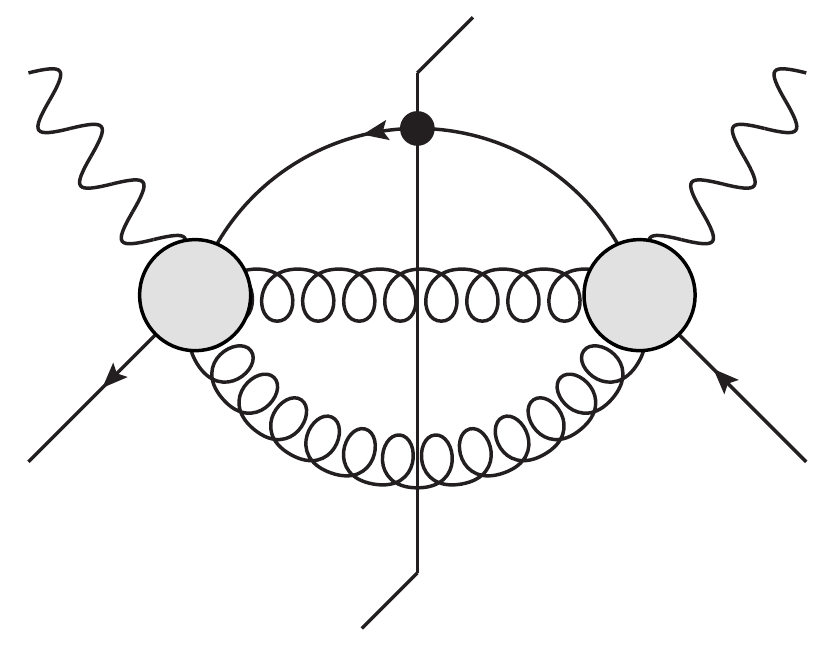}\ \
&
\includegraphics[width=0.21\textwidth]{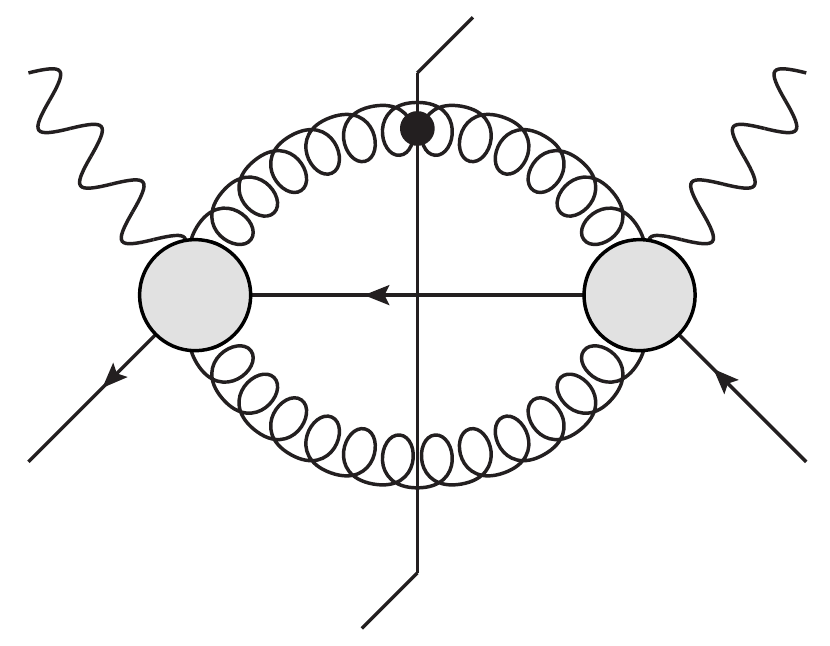}\\ 
(R-A) \ \ & (R-B) \ \ & (R-C) \ \ & (R-D) \vspace{10mm} \\ 
\hspace*{-0.3cm}
\includegraphics[width=0.21\textwidth]{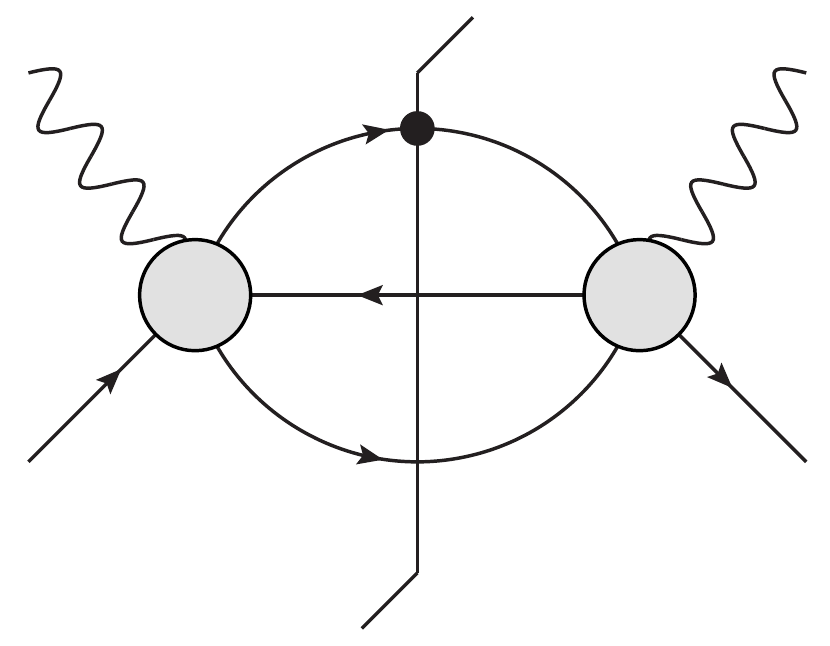}\ \
&
\includegraphics[width=0.21\textwidth]{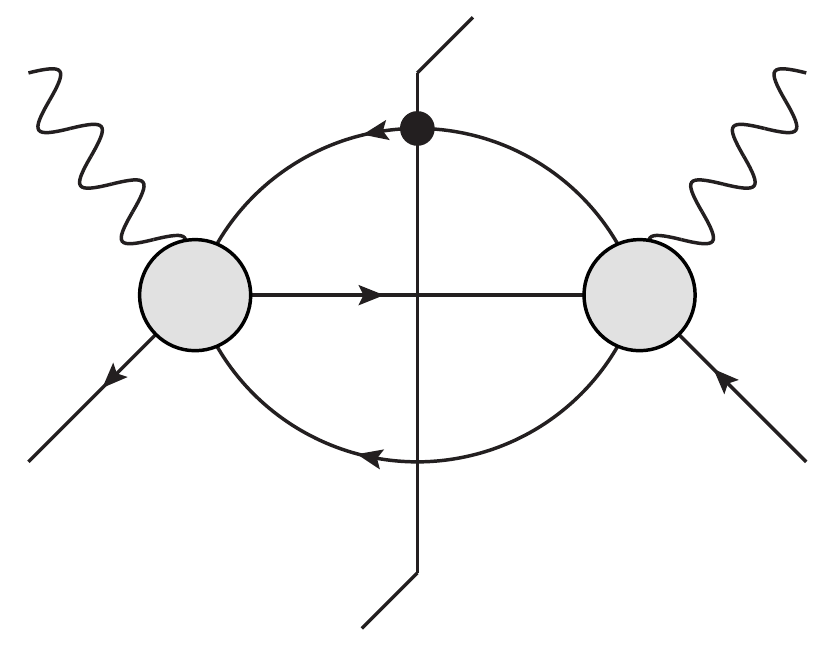}\ \
&
\includegraphics[width=0.21\textwidth]{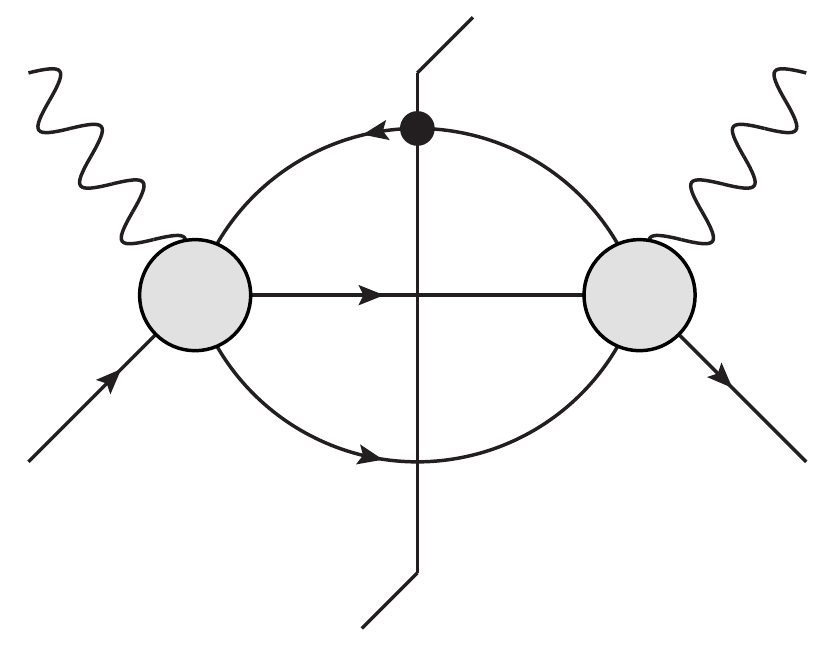}\ \
&
\includegraphics[width=0.21\textwidth]{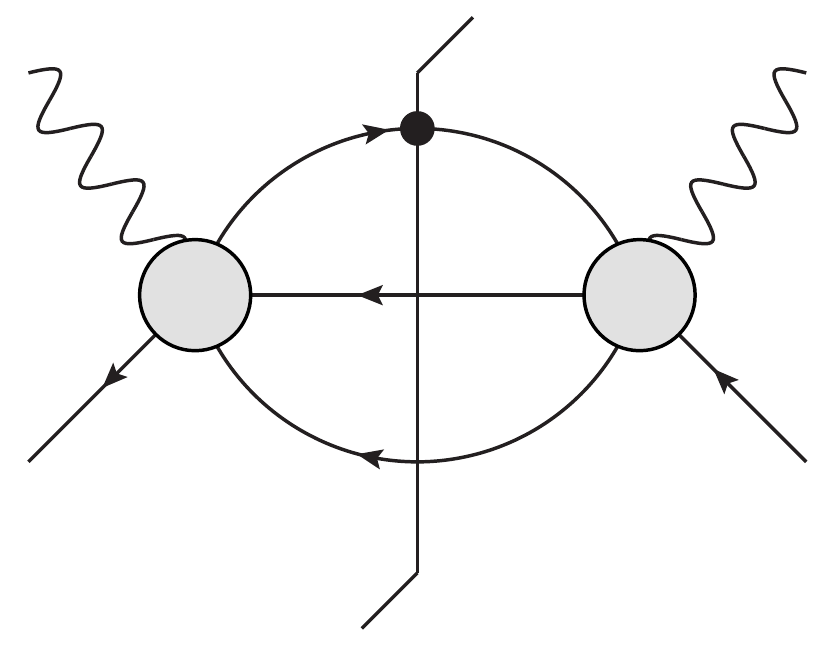}\\ 
(R-E) \ \ & (R-F) \ \ & (R-G) \ \ & (R-H) \vspace{10mm} \\ 
\multicolumn{4}{c}{
\begin{tabular}{c@{\hspace*{.5cm}}c@{\hspace*{.5cm}}c}
\includegraphics[width=0.21\textwidth]{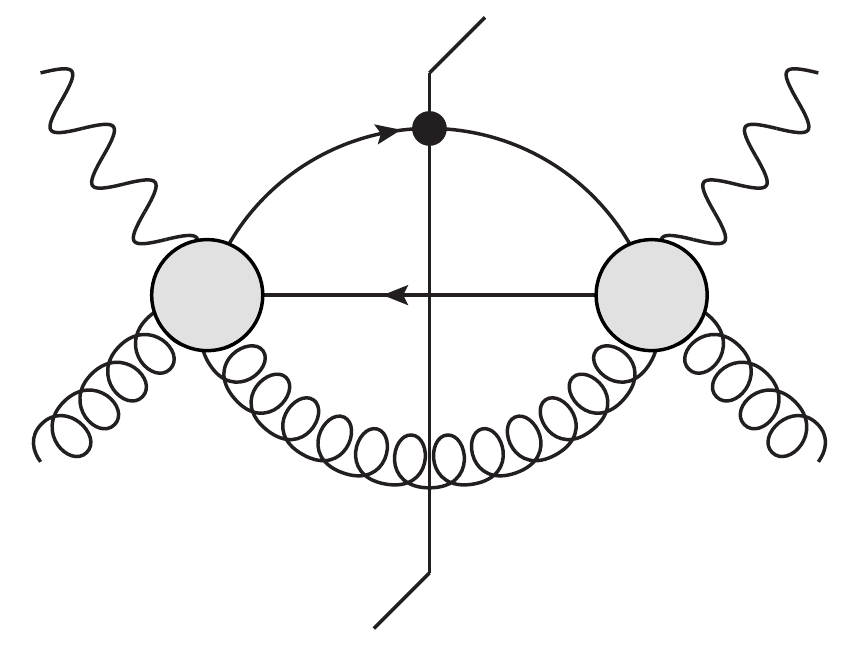}\ \
&
\includegraphics[width=0.21\textwidth]{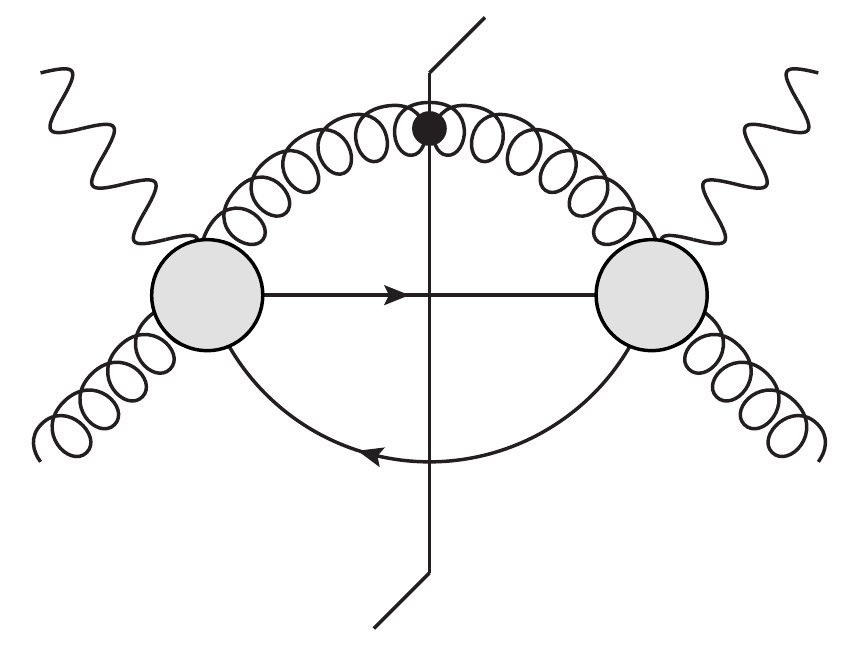}\ \
&
\includegraphics[width=0.21\textwidth]{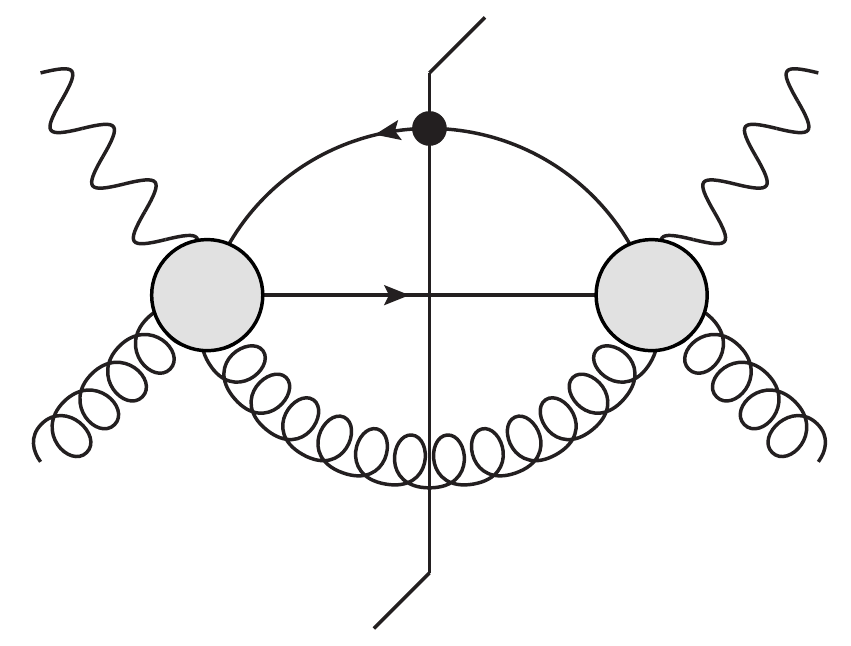}\\
(R-I) \ \ & (R-J) \ \ & (R-K) 
\end{tabular}
}
\vspace{10mm} \\ 
\hspace*{-0.3cm}
\includegraphics[width=0.21\textwidth]{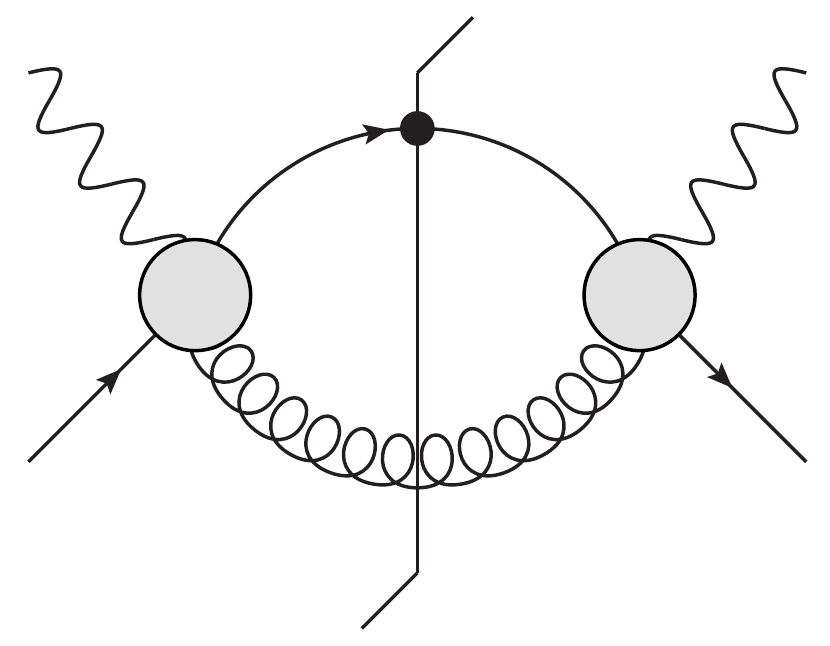}\ \
&
\includegraphics[width=0.21\textwidth]{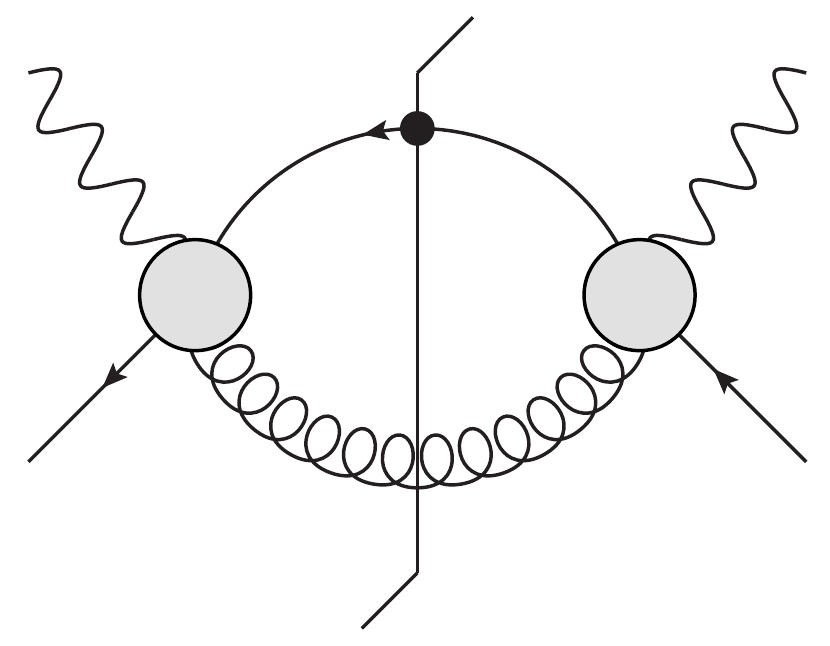}\ \
&
\includegraphics[width=0.21\textwidth]{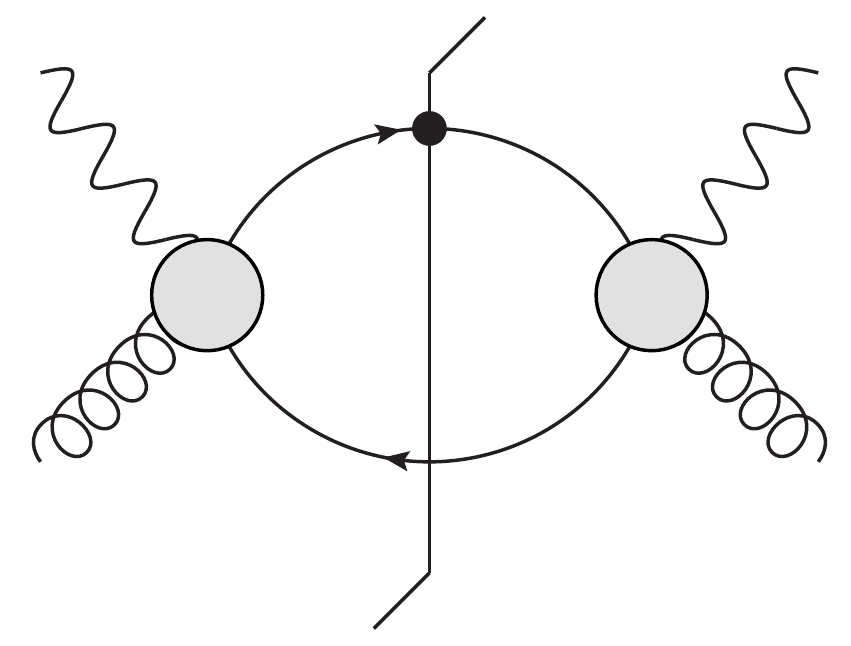}\ \
&
\includegraphics[width=0.21\textwidth]{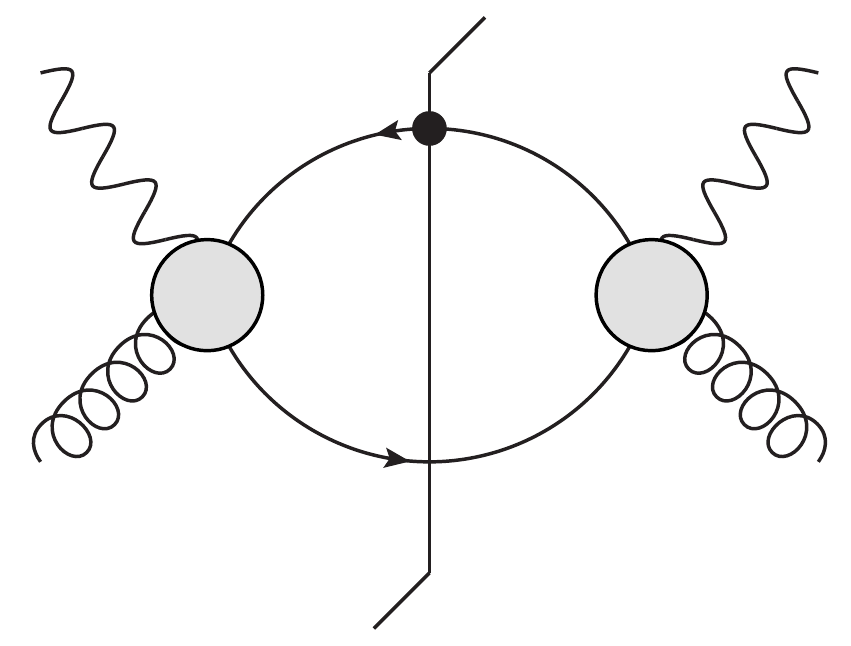}\\ 
(V-A) \ \ & (V-B) \ \ & (V-C) \ \ & (V-D) 
\vspace{10mm} \\ 
\multicolumn{4}{c}{
\begin{tabular}{c@{\hspace*{.5cm}}c}
\includegraphics[width=0.21\textwidth]{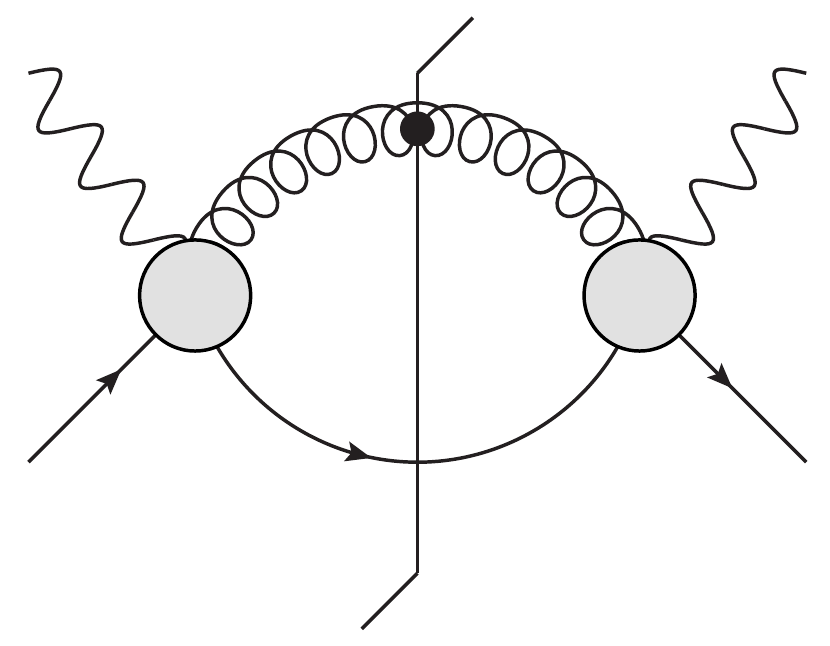}\ \
&
\includegraphics[width=0.21\textwidth]{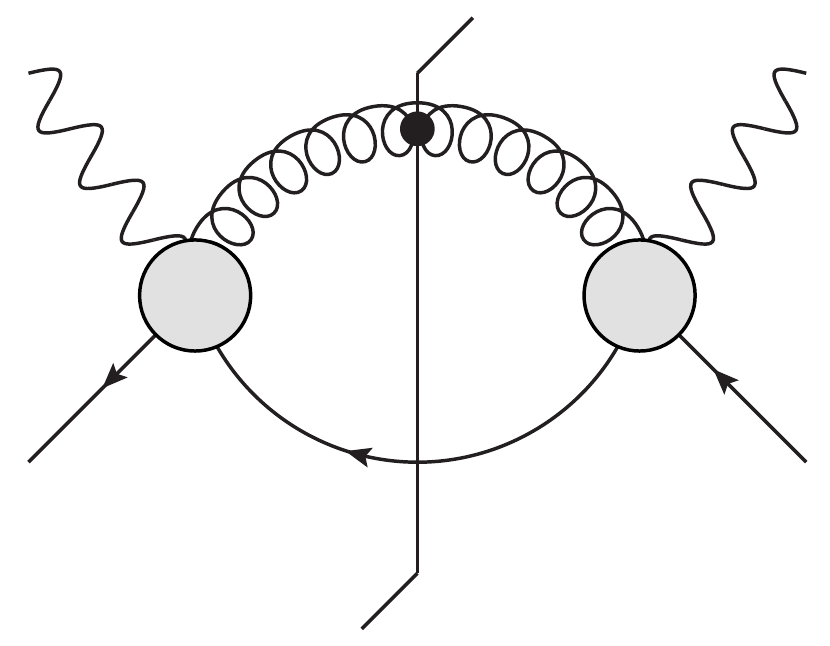}\\
(V-E) \ \ & (V-F) 
\end{tabular}
}
\end{tabular}
\caption{
  Structure of graphs needed at order $\alpha_s(Q)^2$.  The last six
  correspond to virtual corrections to $\alpha_s(Q)$ order graphs.}
\label{f.graphs}
\end{figure}

The graphical structures needed at order 
$\alpha_s(Q)^2$ can be
classified as in \fref{graphs}.  Each diagram corresponds to a
contribution to the squared amplitude, and the blobs include all
possible attachments. 
For the $2\to2$ subprocesses, the blobs include
up to one loop corrections in one side of the cut to give
$O(\alpha_S^2)$ accuracy to the hard parts.
The black dot marks the fragmenting parton and
the other lines are integrated over all phase space.

The following are assumed and not shown explicitly in the graphs:
\begin{itemize}
\item Subtractions, consistent with factorization, for all collinear divergences.
\item One QCD loop at all possible positions inside the blob in virtual processes.
\item Associated UV counter term graphs for virtual processes.
\end{itemize}
Define the hard parts $\hard$ for individual graphs as 
\begin{equation}
\hard^{\Gamma}_{ij;kl} \equiv
\contractor{\Gamma}^{\mu \nu} \hat{W}_{\mu \nu, i j} \, ,
\end{equation}
where $\Gamma \in \{g, pp\}$ and $i, j, k, l \in \{q, \bar{q}, g\}$.
As before, $i$ and $j$ label incoming and fragmenting parton flavors
respectively while $k$ and $l$ label the unobserved parton flavors.
The graph is virtual when only one flavor index appears after the
``$;$''. Figures \ref{f.graphs}(V-A) through \ref{f.graphs}(V-F)
correspond to the $\order{\alpha_s}$ graphs if all virtual loops in
the blobs are removed.

So, for example, $\hard^g_{qq;gg}$ is represented by
\fref{graphs}(R-A), when the partonic tensor is contracted with
$g^{\mu \nu}$. \fref{graphs}(R-E)  includes both
$\hard^g_{qq;q'\bar{q}'}$ and $\hard^g_{qq';q\bar{q}'}$ contributions
where the prime indicates a different flavor.  Note that graphs like
\fref{graphs}(R-G) give contributions like $\hard^g_{q\bar{q};q q}$ or
$\hard^g_{q\bar{q}';q' q}$. 

We work in the approximation that all particles are massless, except
for the photon which is highly virtual, $q^2=-Q^2$. In
\eref{cntrns}, for $2 \to 2$ scattering,
\begin{align}
&{}  
 \{ \contractor{g}^{\mu \nu}  \hat{W}^{(2)}_{\mu\nu}; 
    \contractor{PP}^{\mu \nu} \hat{W}^{(2)}_{\mu\nu}\}_{{\rm unsub}} \nonumber \\
&{}=  
  \frac{1}{(2 \pi)^4} \int 
  \{\lvert M^{2\to 2}_g\rvert^2;\lvert M^{2\to 2}_{PP}\rvert^2\} 
  \frac{\diff{^{n-1} k_2}{} }{(2 \pi)^{n-1} 2 k_2^0} 
  (2 \pi)^{n} \delta^{(n)}(q + p - k_1 - k_2) \nonumber \\
&{}=
  \frac{1}{(2 \pi)^4} \{\lvert M^{2\to 2}_g\rvert^2;\lvert M^{2\to 2}_{PP}\rvert^2\}  
  (2 \pi)\delta_+(k_2^2) \, .
\end{align}
Thus the $2 \to 2$ phase space factor is
\begin{align}
\diff{\ps}^{(2)}  &{}= (2 \pi) \delta_+(k_2^2)  \, \nonumber \\
                  &{}= \frac{2 \pi\hat{x}}{Q^2} 
\delta\left( \Bigl(1 - \hat{x} \Bigr)\Bigl( 1 - \hat{z} \Bigr) - \frac{\hat{x} \Tscsq{k}{1} }{\hat{z} Q^2} \right) \, . \label{e.2to2ps}
\end{align} 

For the $2 \to 3$ case,
we follow the strategy in \cite{Ellis:1979sj,Gordon:1993qc} where 
we work in the rest frame of the $k_2+k_3$
system, in which
\begin{align}
k_2=&\frac{\sqrt{s_{23}}}{2}
  (1,\bm{\hat{k}}_{n-3},\cos\beta_2\sin\beta_1,\cos\beta_1),
  \label{e.defk2} \\
k_3=&\frac{\sqrt{s_{23}}}{2}
  (1,-\bm{\hat{k}}_{n-3},-\cos\beta_2\sin\beta_1,-\cos\beta_1)
  \label{e.defk3},
\end{align}
where
$\bm{\hat{k}}_{n-3}$ denotes unit vectors for the first
$n-3$ components of the $n-1$ dimensional unit spatial vector in
spherical coordinates. See \fref{scatteringplane}. 

In the center-of-mass of $p$ and $q$, the spatial vectors $\bm{p}$,
$\bm{q}$, $\bm{k_1}$, and $\bm{k_2}+\bm{k_3}$ are in the same plane.
We then boost to the $k_2+k_3$ rest frame $\bm{p}$, where $\bm{q}$,
and $\bm{k_1}$ are still in the same plane -- see
\fref{scatteringplane}. We choose the spatial orientations of this
coordinate system so that $\bm{p}$, $\bm{q}$, and $\bm{k_1}$ are in
the plane created by the last two spatial axes. For these vectors, the
first $n-3$ spatial components are zero. Then the scattering
amplitudes do not depend on the first $n-3$ spatial components of
$k_2$ or $k_3$, and the 3-body phase space simplifies (see~\aref{PS3}
for useful identities relating to the 3-body final state phase space).
In this frame,
\begin{align}
&{}  
  \{\contractor{g}^{\mu \nu}  \hat{W}^{(3)}_{\mu\nu};
    \contractor{PP}^{\mu \nu} \hat{W}^{(3)}_{\mu\nu}\}_{{\rm unsub}}
  \nonumber \\
&{}=  
  \frac{1}{(2 \pi)^4} \int 
  \{\lvert M^{2\to 3}_g\rvert^2;\lvert M^{2\to 3}_{PP}\rvert^2\} 
  \frac{\diff{^{n-1} k_2}{} }{(2 \pi)^{n-1} 2 k_2^0} 
  \frac{\diff{^{n-1} k_3}{} }{(2 \pi)^{n-1} 2 k_3^0}
  (2 \pi)^{n} \delta^{(n)}(q + p - k_1 - k_2 - k_3)
  \nonumber\\ 
&{}=  
  \frac{1}{(2 \pi)^4} \int 
  \{\lvert M^{2\to 3}_g\rvert^2;\lvert M^{2\to 3}_{PP}\rvert^2\} 
  \frac{|{\bf k_2}|^{n-2}\diff{|{\bf k_2}|}{} }{(2 \pi)^{n-2} 2 k_2^0} 
  \delta_+(k_3^2)
  \diff{\Omega_{n-4}}
  \diff{\beta_1}
  \diff{\beta_2}
  \sin^{n-3}\beta_1 
  \sin^{n-4}\beta_2
  \nonumber\\ 
&{}=  
  \frac{s_{23}^{-\epsilon}2^{-2} \pi^{-\epsilon}}{(2\pi)^{6-2\epsilon}} 
   \frac{\Gamma(1 - \epsilon)}{\Gamma(1 - 2 \epsilon)}
  \int \{\lvert M^{2\to 3}_g\rvert^2;\lvert M^{2\to 3}_{PP}\rvert^2\} 
  \diff{\beta_1}
  \diff{\beta_2}
  \sin^{1-2\epsilon}\beta_1 
  \sin^{-2\epsilon}\beta_2
  \, , \label{e.23ps}
\end{align}
where in the third equality we use that the scattering amplitudes are
independent of the first $n-3$ spatial components of $k_2$. That is,
\begin{equation}
\diff{\ps}^{(3)}  =
\frac{s_{23}^{-\epsilon} 2^{-2} \pi^{-\epsilon}}{(2\pi)^{2-2\epsilon}} 
  \frac{\Gamma(1 - \epsilon)}{\Gamma(1 - 2 \epsilon)}
  \diff{\beta_1}
  \diff{\beta_2}
  \sin^{1-2\epsilon}\beta_1 
  \sin^{-2\epsilon}\beta_2
  \, . \label{e.2to3ps}
\end{equation} 
From overall momentum conservation
\begin{align}
s_{23}=
\frac{Q^2 (\hat{z} (1 - \hat{x}) - \hat{z}^2 (1 - \hat{x}) ) - \hat{x} \Tscsq{k}{1}}
     {\hat{x} \hat{z}} \, .
\end{align}
Since $s_{23}$ can become zero in certain regions of $\hat{x}$ and
$\hat{z}$, dimensional regularization is needed for the $s_{23}\to0$
behavior. Further details of the $2 \to 3$ phase space are provided in
the next subsection. 

\begin{figure}
\includegraphics[scale=0.5]{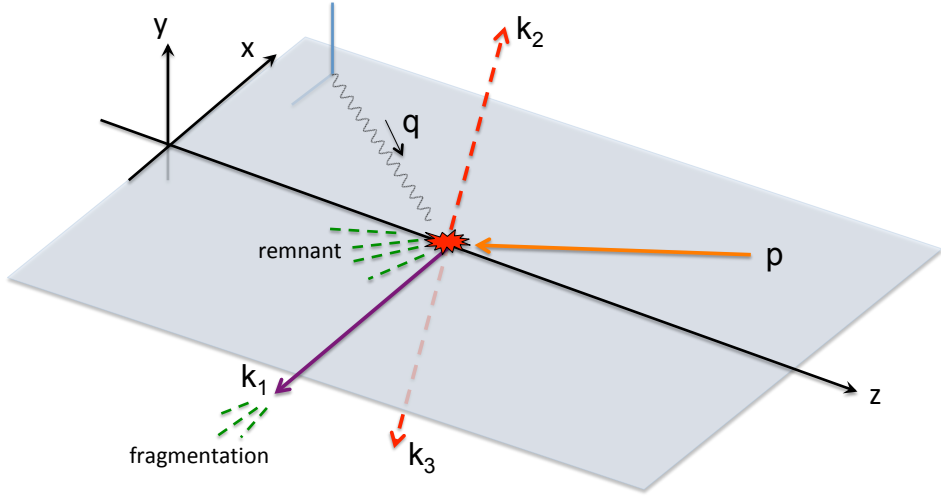}
\caption{Schematic of SIDIS with $2 \to 3$ partonic scattering, using the
  orientation of the vectors defined in \erefs{defk2}{defk3} 
  in 4-dimensions. 
  A $2 \to 2$ scattering configuration is recovered if $(k_2 +
  k_3)^2$ becomes much less than $Q^2$, or if $k_2$ or $k_3$ align with
  one of the other hadrons. The three frames defined in this section
  are obtain by rotating $p$, $q$ and $k_1$ inside the plane; frame 1
  corresponds to orienting the $z$-axis along $k_1$ (\eref{k1f1}), frame
  2 orients the $z$-axis along $p$ (\eref{pf2}), frame 3 orients the
  $z$-axis along $q$ (\eref{qf3}).}
\label{f.scatteringplane}
\end{figure}

To summarize the basic steps, integrals for the unobserved final state
phase space are first performed over angular variables.  Then,
integrals over momentum fractions are expressed via \eref{hdxsc} as
integrals over $\xi$ and $s_{23}$.  In the $s_{23} \to 0$ limit there
are soft and collinear singularities, proportional to
$\delta(s_{23})/\epsilon$ or $\delta(s_{23})/\epsilon^2$, that
cancel after combining $2 \to 3$ and
the corresponding virtual $2 \to 2$ processes. There are additional
collinear singularities in $2 \to 3$ and virtual $2 \to 2$ processes
which are ultimately canceled by subtraction terms in the
factorization algorithm. 
(These contain also $\delta(s_{23})$ and plus
distributions.)
All contributions are integrated over the physical range of
$s_{23}$ in~\erefs{hdxsc}{limitsA}.

\subsection{Factorization Order-By-Order}
\label{s.subtraction}

Once all real and virtual Feynman graphs have been computed in 
dimensional regularization, they must be combined in an algorithm consistent 
with a factorization theorem. 

According to the collinear factorization theorem, hard factors are
independent of the species of target and final state particles, so
hard scattering calculations can be performed for free and massless
partons as the target and final state without loss of generality.
(In other words, we will consider perturbatively calculable parton-in-parton PDFs and 
and parton-to-parton FFs.)
This simplifies calculations, so we start our calculations by
rewriting \eref{factform} as a factorization theorem for a partonic
initial and final state:
\begin{align}
W_{\mu \nu,i'j'}(p',q,k'_1) 
&{} = 
\int_{x-}^{1+} \frac{\diff{\xi}}{\xi} 
\int_{z-}^{1+} \frac{\diff{\zeta}}{\zeta^2} 
\hat{W}_{\mu \nu, ij}(q,x/\xi,z/\zeta) 
f_{i/i'}(\xi) d_{j'/j}(\zeta) 
\, \nonumber \\
&{}=  
\hat{W}_{\mu \nu, ij}(q,x/\xi,z/\zeta) \otimes
f_{i/i'}(\xi) \otimes d_{j'/j}(\zeta) 
\, \nonumber \\ 
&{}=  W^{(\text{LO})}_{\mu \nu,i'j'}(p',q,k'_1) 
+ W^{(\text{NLO})}_{\mu \nu,i'j'}(p',q,k'_1) 
+ O(\alpha_s^3) \, .
\label{e.factformpart}
\end{align}
All manipulations will be done in $n = 4 - 2 \epsilon$ dimensions
until the very end.  The left side is now an argument of $p'$, $q$,
and $k'_1$, and the PDFs and FFs have subscripts $i/i'$ and $j'/j$.
The momenta $p'$ and $k'_1$ play the roles that $P$ and $\THDsc{P}$
played earlier.  The initial $p'$ has a flavor $i'$ and the final
$k'_1$ has a flavor $j'$.  The second line in \eref{factformpart}
defines the ``$\otimes$" notation as the usual shorthand for the
convolution integrals in $\xi$ and $\zeta$. On the last line,
$W^{(\text{LO})}_{\mu \nu,i'j'}$ and $W^{(\text{NLO})}_{\mu \nu,i'j'}$
are hadronic tensors calculated to $\text{LO}$ and $\text{NLO}$
\emph{in the hard parts}, i.e. what one normally means when one speaks
of a ``leading order'' or ``next-to-leading order'' calculation in
collinear pQCD.

Expanding \eref{factformpart} to the relevant order in $\alpha_s$ gives
\begin{align}
W_{\mu \nu,i'j'}(p',q,k'_1)  
&{}= ( \hat{W}^{(\text{LO})}_{\mu \nu, ij}(q,x/\xi,z/\zeta) 
+ \hat{W}^{(\text{NLO})}_{\mu \nu, ij}(q,x/\xi,z/\zeta)) \nonumber \\    
&{} \otimes \left( f^{(1)}_{i/i'}(\xi) + f^{(\alpha_s)}_{i/i'}(\xi) + \cdots \right) 
\otimes \left( d^{(1)}_{j'/j}(\zeta) + d^{(\alpha_s)}_{j'/j}(\zeta) + \cdots \right) 
+ O(\alpha_s^3)\, . 
\label{e.expandW}
\end{align}
The aim is to get the hard parts, $\hat{W}^{(\text{LO})}_{\mu \nu,
	ij}(q,x/\xi,z/\zeta)$ and $\hat{W}^{(\text{NLO})}_{\mu \nu,
	ij}(q,x/\xi,z/\zeta)$. The leading order is just
\begin{align}
W^{(\text{LO})}_{\mu \nu,i'j'}(p',q,k'_1) 
&{}=  \hat{W}^{(\text{LO})}_{\mu \nu, ij}(q,x/\xi,z/\zeta) \nonumber \\
&{}   \otimes \left( f^{(1)}_{i/i'}(\xi) + f^{(\alpha_s)}_{i/i'}(\xi) + \cdots \right)
\otimes \left( d^{(1)}_{j'/j}(\zeta) + d^{(\alpha_s)}_{j'/j}(\zeta) + \cdots \right) \, ,
\label{e.expandWlead}
\end{align}
where the expression for 
$\hat{W}^{(\text{LO})}_{\mu \nu,ij}(q,x/\xi,z/\zeta)$ 
is simple and well-known -- it is just the sum
of tree level $2 \to 2$ graphs contributing to 
$\hat{W}_{\mu\nu,i'j'}(q,x/\xi,z/\zeta)$. 

To get $\hat{W}^{(\text{NLO})}_{\mu \nu, ij}(q,x/\xi,z/\zeta)$, first write
\begin{equation}
W_{\mu \nu,i'j'}(p',q,k'_1) =  W^{(\text{LO})}_{\mu \nu,i'j'}(p',q,k'_1) 
+ \left[ W_{\mu \nu,i'j'}(p',q,k'_1)  
- W^{(\text{LO})}_{\mu \nu,i'j'}(p',q,k'_1) \right] \, . 
\label{e.WpY}
\end{equation}
The term in braces is the correction to the leading order so it
vanishes by construction at order $\alpha_s$ in the hard part. Also,
it contains subtractions for the overlap with the $\text{LO}$, so it
is infrared safe through order $\alpha_s^2$ in hard scattering. Thus,
we may rewrite it as:
\begin{align}
&{} W_{\mu \nu,i'j'}(p',q,k'_1)  - W^{(\text{LO})}_{\mu \nu,i'j'}(p',q,k'_1)  = \nonumber \\
&{} \qquad \left[ W_{\mu \nu,i'j'}(q,x/\xi,z/\zeta)  
- W^{(\text{LO})}_{\mu \nu,i'j'}(q,x/\xi,z/\zeta) 
\right]^{\alpha_s^2} 
\otimes f_{i/i'}(\xi) \otimes d_{j'/j}(\zeta) 
+ O(\alpha_s^3) \, . 
\label{e.fdab}
\end{align}
The term in braces on the second line is equal to the first line
evaluated with hatted partonic variables and expanded only to order
$\alpha_s^2$ in the coupling. Comparing with \eref{expandW} and
\eref{WpY} shows that the factor in braces in \eref{fdab} is
$\hat{W}^{(\text{NLO})}_{\mu \nu, ij}(q,x/\xi,z/\zeta)$.  We use
\eref{expandWlead} to evaluate it explicitly in terms of low order
PDFs and FFs:
\begin{align}
\hat{W}^{(\text{NLO})}_{\mu \nu, ij}(q,\hat{x},\hat{z})  
&{}= W^{(\text{NLO})}_{\mu \nu,i'j'}(q,\hat{x},\hat{z})_{\rm unsub}  
-\hat{W}^{(\text{LO})}_{\mu \nu, ij}(q,\hat{x}/\hxi,\hat{z}/\hzeta) 
\otimes f_{i/i'}^{(\alpha_s)}(\hxi) \otimes d_{j'/j}^{(1)}(\hzeta) \nonumber \\
&{} \hspace{3.1cm} - \hat{W}^{(\text{LO})}_{\mu \nu, ij}(q,\hat{x}/\hxi,\hat{z}/\hzeta) 
\otimes f_{i/i'}^{(1)}(\hxi) \otimes d_{j'/j}^{(\alpha_s)}(\hzeta) \, \nonumber \\
&{}= W^{(\text{NLO})}_{\mu \nu,i'j'}(q,\hat{x},\hat{z})_{\rm unsub}  \nonumber \\
&{} \hspace{3.1cm} - \int_{x-}^{1+} \frac{\diff{\hxi}}{\hxi}  
\int_{z-}^{1+} \frac{\diff{\hzeta}}{\hzeta^2} 
\hat{W}^{(\text{LO})}_{\mu \nu, ij}(q,\hat{x}/\hxi,\hat{z}/\hzeta)  
f_{i/i'}^{(\alpha_s)}(\hxi)  d_{j'/j}^{(1)}(\hzeta) \nonumber \\
&{} \hspace{3.1cm}- \int_{x-}^{1+} \frac{\diff{\hxi}}{\hxi}  
\int_{z-}^{1+} \frac{\diff{\hzeta}}{\hzeta^2} 
\hat{W}^{(\text{LO})}_{\mu \nu, ij}(q,\hat{x}/\hxi,\hat{z}/\hzeta)  
f_{i/i'}^{(1)}(\hxi)  d_{j'/j}^{(\alpha_s)}(\hzeta)  \nonumber \\
&{}= W^{(\text{NLO})}_{\mu \nu,i'j'}(q,\hat{x},\hat{z})_{\rm unsub}  \nonumber \\
&{} \hspace{3.1cm}+\frac{\alpha_s}{4\pi} \frac{S_\epsilon}{\epsilon} 
\int_{x-}^{1+} \frac{\diff{\hxi}}{\hxi} 
\hat{W}^{(\text{LO})}_{\mu \nu, i''j'}(q,\hat{x}/\hxi,\hat{z}) 
P^{(0)}_{i''/i'}(\hxi) \nonumber \\
&{} \hspace{3.1cm}+ \frac{\alpha_s}{4\pi} \frac{S_\epsilon}{\epsilon} 
\int_{z-}^{1+} \frac{\diff{\hzeta}}{\hzeta^2} 
\hat{W}^{(\text{LO})}_{\mu \nu, i'j''}(q,\hat{x},\hat{z}/\hzeta) 
P^{(0)}_{j'/j''}(\hzeta) 	 \, . \,  
\label{e.Wnextorder}
\end{align}
Here, the $W^{(\text{NLO})}_{\mu \nu}(q,\hat{x},\hat{z})_{\rm unsub}$
is the sum of order $O(\alpha_s^2)$ graphs contributing to 
$W_{\mu\nu}(q,\hat{x},\hat{z})$ without any subtractions (i.e., graphs in
\fref{graphs}(R-A) through \fref{graphs}(V-F) 
integrated over phase space, 
but absent the
subtractions). The second two terms are obtained from
\eref{expandWlead}, expanded to order $\alpha_s^2$.  The last line in
\eref{Wnextorder} uses the known results for PDFs and FFs calculated
to $\order{\alpha_s}$ for massless quarks and gluons:
\begin{align}
f^{(1)}_{i/i'}(\xi)&{}=\delta_{ii'}\delta(1-\xi) \, ,  \\
d^{(1)}_{j'/j}(\zeta)&{}=\delta_{j'j}\delta(1-\zeta) \, , \\
f^{(\alpha_s)}_{i/i'}(\xi)&{}=-\frac{\alpha_s}{4\pi} 
\frac{S_\epsilon}{\epsilon} P^{(0)}_{i/i'}(\xi) \, , \\
d^{(\alpha_s)}_{j'/j}(\zeta)&{}=-\frac{\alpha_s}{4\pi} 
\frac{S_\epsilon}{\epsilon} P^{(0)}_{j'/j}(\zeta) \, .
\end{align}
Note that, consistent with the $\overline{\text{MS}}$ scheme, the
renormalized order $\alpha_s$ PDFs and FFs in dimensional
regularization for massless pQCD are just $S_\epsilon/\epsilon$ poles.
There are no $\mu^{\epsilon}$ factors or logarithms of $\mu$. Thus the
last line of \eref{Wnextorder} involves no factors of
$\mu^{\epsilon}$.  For completeness, the one-loop 
splitting functions~\cite{Altarelli:1977zs} are
\begin{align}
P^{(1)}_{qq'}(\xi)={}& 2C_F \delta_{qq'}
\left[ \frac{2}{(1-\xi)_{+}} 
-1 
-\xi +\frac{3}{2}\delta(1-\xi)
\right] \, , \\ 
P^{(1)}_{qg}(\xi)={}& 2 T_F \left[ (1 - \xi)^2 + \xi^2  \right] \, , \\
P^{(1)}_{gq}(\xi)={}& 2 C_F \left[ \frac{1 + (1 - \xi)^2}{\xi} \right] \, .
\end{align}

Our calculations need \eref{Wnextorder}, contracted with specific
$\contractortot{\Gamma}^{\mu\nu}$ tensors. The
substraction scheme has been carry out without contracting the Lorentz
indices of the partonic tensor. However the method is equally
applicable for any kind of contraction with external momenta. In our
calculations we carry out the substractions separately for the two 
extraction tensors in Eq.(\ref{e.PgPPtot}) and verified analytically
the cancellation of all infrared and collinear singularities.

\section{Results}
\label{s.results}

\subsection{Combining real and virtual contributions}
\label{s.combine}
After the hard real and virtual contributions are calculated, they
must be combined into infrared safe squared amplitudes.
Table~\ref{table:real-virt-corr} shows graphs from corresponding real
and virtual processes.  Note that the correspondence between real and
virtual processes is not one-to-one. In particular, infrared (soft and
collinear) singularities in $\hard_{qq;g}$ are canceled by three real
processes. A subtlety arises when a quark loop appears on the gluon
leg of $\hard_{qq;g}$. This creates a collinear pole term proportional
to $N_f$, the number of massless quark flavors in the loop. This pole
term is canceled after adding the real processes $\hard_{qq;q\bar{q}}$ and 
$\hard_{qq;q'\bar{q}'}$ with all massless flavors $q'$ 
(other than $q$)~\cite{Gordon:1993qc}. For processes
with a non-spectator gluon leg, the $N_f$ dependence of the collinear
pole is removed by factorization. Also notice that some real processes
have no corresponding virtual ones, and in these cases factorization
subtractions are sufficient to remove all infrared poles.  Since many
graphs need to be combined, the results are presented as the six
scattering hard parts listed in the last column of
Table~\ref{table:real-virt-corr}. 
We also need the hard parts for processes with reversed quark
number flow in one or two open fermion lines in the graph. That is, in
one or two open quark lines, quarks and their corresponding anti-quarks
are interchanged.  These are easily related to the
hard parts already listed in Tabel~\ref{table:real-virt-corr} by using
the fact that QED vertices acquire a minus sign under charge
conjugation. The results are summerized in
Tabel~\ref{table:real-virt-corr-idntcl-contrib}. Note that when a
quark line links an outgoing spectator quark anti-quark pair, there is
no need for the reversed quark flow. This is because the spectator
momenta are integrated over and interchanging the quark and anti-quark
will double count the contribution.\\
\begin{table}[!h]
\centering
\begin{tabular}{|c|c|c|c|c|}
  \hline
  \multicolumn{2}{|c|}{virtual} & \multicolumn{2}{|c|}{real} & combined\\
  \hline
  \multirow{2}{*}[1.4cm]{$\hard_{gq;\bar{q}}$} &
  \includegraphics[width=3cm, angle=0]{gallery/graph_topsN}  & 
  \multirow{2}{*}[1.4cm]{$\hard_{gq;\bar{q}g}$}  & 
  \includegraphics[width=3cm, angle=0]{gallery/graph_topsI} & 
  \multirow{2}{*}[1.4cm]{$\hard_{gq}$}\\
  \hline
  \multirow{2}{*}[1.4cm]{$\hard_{qq;g}$} &
  \includegraphics[width=3cm, angle=0]{gallery/graph_topsL} &
  \multirow{2}{*}[1.6cm]{\makecell{$\hard_{qq;gg}$\\$+$ 
                                   $\hard_{qq;q\bar{q}}$\\$+$ 
                                   $\hard_{qq;q'\bar{q}'}$}}& 
  \includegraphics[width=3cm, angle=0]{gallery/graph_topsA} 
  \includegraphics[width=3cm, angle=0]{gallery/graph_topsE} & 
  \multirow{2}{*}[1.4cm]{$\hard_{qq}$}\\
  \hline
  \multirow{2}{*}[1.4cm]{$\hard_{qg;q}$} &
  \includegraphics[width=3cm, angle=0]{gallery/graph_topsS}  & 
  \multirow{2}{*}[1.4cm]{$\hard_{qg;qg}$}  & 
  \includegraphics[width=3cm, angle=0]{gallery/graph_topsB} & 
  \multirow{2}{*}[1.4cm]{$\hard_{qg}$}\\
  \hline
  \multicolumn{2}{|c|}{\multirow{2}{*}[1.4cm]{N/A}} & 
  \multirow{2}{*}[1.4cm]{$\hard_{gg;q\bar{q}}$} & 
  \includegraphics[width=3cm, angle=0]{gallery/graph_topsJ} & 
  \multirow{2}{*}[1.4cm]{$\hard_{gg}$}\\
  \hline
  \multicolumn{2}{|c|}{\multirow{2}{*}[1.4cm]{N/A}} & 
  \multirow{2}{*}[1.4cm]{$\hard_{q\bar{q};qq}$} & 
  \includegraphics[width=3cm, angle=0]{gallery/graph_topsG} & 
  \multirow{2}{*}[1.4cm]{$\hard_{q\bar{q}}$}\\
  \hline
  \multicolumn{2}{|c|}{\multirow{2}{*}[1.4cm]{N/A}}  & 
  \multirow{2}{*}[1.4cm]{$\hard_{qq';q\bar{q}'}$} & 
  \includegraphics[width=3cm, angle=0]{gallery/graph_topsE} & 
  \multirow{2}{*}[1.4cm]{$\hard_{qq'}$}\\
  \hline
\end{tabular}
\caption{
  Correspondence between real and virtual graphs at order
  $\alpha_s(Q)^2$. The superscript of the hard part $\Gamma \in \{g,pp\}$ 
  is supressed. For virtual graphs, the additional gluon is
  inside one of the blobs.}
\label{table:real-virt-corr}
\end{table}

\subsection{Comparison with existing results}
\label{s.compare}
After combining the graphs in each of the six
scattering channels in Table~\ref{table:real-virt-corr}, we have
verified explicitly that all single and double poles cancel. The terms left are
finite in the limit $\epsilon \rightarrow 0$, and constitute the infrared
safe hard parts in the last column of
Table~\ref{table:real-virt-corr}. As a check, we have
compared with a (privately obtained) computer calculation that appears
to reproduce the results of Ref.~\cite{Daleo:2004pn} but is modified
to be consistent with the kinematics of the current experimental data.
For most kinematics, the calculations agree within experimental 
uncertainties, but we found several possible differences.
\begin{itemize}
	\item In the last row of Table~\ref{table:real-virt-corr}, our calculation of
	terms from the $\contractortot{g}^{\mu\nu}$ projection with the charge structure 
	$e_q^2$ and $e_qe_{q'}$ differs from
	the corresponding terms calculated in the previously existing code by a minus sign. 
	(By contrast, terms with
	$e_{q'}^2$ agree.) 
\item As is discussed in subsection~\ref{s.combine}, adding the real processes
     $\hard_{qq;q\bar{q}}$ and $\hard_{qq;q'\bar{q}'}$ with all massless flavors $q'$
	 gives terms proportional to $N_f$. The pole parts of these terms are canceled by 
	 the corresponding poles from $\hard_{qq;g}$. However, a finite part remains since
	 the real processes above with $N_f$ final state quark pairs give identical 
	 contribution even when the 
	 quark pairs are not collinear. These terms contribute to the third row 
		of 
	 Table~\ref{table:real-virt-corr}. In the previously existing code,
	 we did not find an explicit dependence on $N_f$ in this channel.
\end{itemize}
The numerical result of our calculation can in some cases differ from 
the previously existing code by as much as $\sim 100\%$ for 
certain individual subchannels. In the kinematics of the COMPASS 
experiment, 
this translates into a discrepancy of up to $\sim 20\%$ for the overall cross section. 

\subsection{Phenomenological Results}
\label{s.Pheno}

We examine the impact of the $O(\alpha_S^2)$
corrections in SIDIS by plotting the NLO to LO ratio $K$-factor for
the $F_1$ structure function.
We will consider values of $z$ between $0.2$ and $0.8$ since this is 
a region where factorization theorems based on current
fragmentation in SIDIS is conventionally expected to apply.

For $x$, $Q$, $q_{\rm T}$, we choose values of $Q=2,20 {\rm~GeV}$, 
$q_{\rm T}=Q, 2Q$ and $0.001\le x \le 1$, which correspond to
kinematics ranges accessible by existing experimental facilities such
as COMPASS and JLab 12.
\fref{kfactor} shows the $K$-factor across the aforementioned kinematical range, and it 
shows a clear concave up shape 
in its dependence on $x$ with values that decrease as $Q$ varies from 
$Q = 2$ to $20$~GeV. 

It is notable that even for $Q$ as large as  $2$~GeV, the $K$-factor
is greater than $2$, even at its minimum value.
Note also that this minimum is approximately in the region of $0.01\le x \le0.1$, close to 
the valence region relevant to many hadron structure studies. 
The $K$-factor increases at smaller values of $x$, 
indicating that the NLO corrections becomes increasingly important 
for describing regions with large $(P+q)^2$.  
It also increases at large $x$ 
until it reaches the kinematic boundary where the phase space  
for the hadron production vanishes. 

The enhancements at large $x$ can be traced to the logarithms of $B$ (see Eq. 32
and appendix D) which are associated with soft gluon effects near the
kinematical threshold. Thus it is likely that threshold resummation 
becomes important in this region. Point-by-point in $\Tsc{q}{}$, effects near the
kinematical boundaries are more apparent than in $\Tsc{q}{}$-inclusive cross sections. 
It is interesting to compare with observations found in \cite{Anderle:2012rq} where 
threshold resummation effects for $q_{\rm T}$ integrated SIDIS were found to 
give sizable corrections, but not the factor of $\gtrsim 5$ enhancement needed to 
explain the largest $\Tsc{q}{}$ present in \fref{kfactor}.

The minimum in the $K$-factor is helpful for identifying regions in $x$
where ordinary fixed order treatments are most likely to be sufficient.  
Given a set of values for $z$, $\Tsc{q}{}$, and $Q$,  
let $x_{0}$ denote the $x$ value corresponding to the minimum $K$-factor.  Then
bins in  $x$, $Q$, $z$ ,$\Tsc{q}{}$ can be classified according to $x<x_{0}$
and $x>x_{0}$.  
We show this in \fref{kin} for kinematical bins corresponding to recent COMPASS
$\Tsc{q}{}$-dependent SIDIS multiplicities $h^{\pm}$~\cite{Aghasyan:2017ctw} for 
$\Tsc{q}{}/Q>1$. 
The values of $x$ the for which the $K$-factor reaches its 
minimum for $\Tsc{q}{} \approx Q$ can be read from the plots. Away from these regions, 
additional resummation techniques may be needed.

It may also be necessary to update collinear PDFs outside these regions in 
order to fully describe the large $\Tsc{q}{}$ behavior. 
As mentioned in \cite{Gonzalez-Hernandez:2018ipj}, the large
$\Tsc{q}{}$ tails of SIDIS are sensitive to the large $\xi$ PDFs and
large $\zeta$ FFs. These could potentially open up new opportunities to constrain 
collinear PDFs at large momentum fractions.  

At the same time, factorization theorems for the full $\Tsc{q}{}$-dependent SIDIS 
spectrum, including $\Tsc{q}{} \approx 0$ with non-perturbative TMD PDFs and TMD 
FFs, need the large $\Tsc{q}{} \sim Q$ component in order to have a completely reliable 
tests of the factorization formalism. 
Ideally, such validation will take place when $\Tsc{q}{}$-dependent
SIDIS data are included in the simultaneous extraction of collinear PDFs and FFs
as well as non-perturbative TMD PDFs and FFs in QCD global analysis. 

\begin{table}[!h]
\centering
\begin{tabular}{|c|c|c|c|}
  \hline
  \multicolumn{2}{|c|}{virtual} & 
  \multicolumn{2}{|c|}{real} \\
  \hline
  $\hard_{gq;\bar{q}}$ & 
  $\hard_{g\bar{q};q}=\hard_{gq;\bar{q}}$  & 
  $\hard_{gq;\bar{q}g}$ & 
  $\hard_{g\bar{q};qg}=\hard_{gq;\bar{q}g}$ \\
  \hline
	\multirow{3}{*}{$\hard_{qq;g}$} & 
  \multirow{3}{*}{ $\hard_{\bar{q}\bar{q};g}=\hard_{qq;g}$} & 
  $\hard_{qq;gg}$ & 
  $\hard_{\bar{q}\bar{q};gg}=\hard_{qq;gg}$ \\ 
  & & 
  $\hard_{qq;q\bar{q}}$ & 
  $\hard_{\bar{q}\bar{q};q\bar{q}}=\hard_{qq;q\bar{q}}$  \\
	& & 
  $\hard_{qq;q'\bar{q}'}$ & 
  $\hard_{\bar{q}\bar{q};q'\bar{q}'}^{e_q^2}=\hard_{qq;q'\bar{q}'}^{e_q^2};
   \hard_{\bar{q}\bar{q};q'\bar{q}'}^{e_{q'}^2}=\hard_{qq;q'\bar{q}'}^{e_{q'}^2}$\\
  \hline
	$\hard_{qg;q}$ & 
  $\hard_{\bar{q}g;\bar{q}}=\hard_{qg;q}$ & 
  $\hard_{qg;qg}$ & 
  $\hard_{\bar{q}g;\bar{q}g}=\hard_{qg;qg}$ \\
  \hline
  \multicolumn{2}{|c|}{N/A} & 
  $\hard_{gg;q\bar{q}}$ & 
  N/A \\
  \hline
  \multicolumn{2}{|c|}{N/A} & 
  $\hard_{q\bar{q};qq}$  & 
  $\hard_{\bar{q}q;\bar{q}\bar{q}}=\hard_{q\bar{q};qq}$ \\
  \hline
  \multicolumn{2}{|c|}{\multirow{3}{*}{N/A}} &  
  \multirow{3}{*}{$\hard_{qq';q\bar{q}'}$} & 
  $\hard_{\bar{q}\bar{q}';\bar{q}q'}=\hard_{qq';q\bar{q}'}$ \\
	\multicolumn{2}{|c|}{\multirow{3}{*}{}} &  
  & 
  $\hard_{\bar{q}q';\bar{q}\bar{q}'}^{e_q^2}=\hard_{qq';q\bar{q}'}^{e_q^2};
   \hard_{\bar{q}q';\bar{q}\bar{q}'}^{e_{q'}^2}=\hard_{qq';q\bar{q}'}^{e_{q'}^2};
   \hard_{\bar{q}q';\bar{q}\bar{q}'}^{e_qe_{q'}}=-\hard_{qq';q\bar{q}'}^{e_qe_{q'}}$  \\
	\multicolumn{2}{|c|}{\multirow{3}{*}{}} &  
  & 
  $\hard_{q\bar{q}';qq'}^{e_q^2}=\hard_{qq';q\bar{q}'}^{e_q^2};
   \hard_{q\bar{q}';qq'}^{e_{q'}^2}=\hard_{qq';q\bar{q}'}^{e_{q'}^2};
   \hard_{q\bar{q}';qq'}^{e_qe_{q'}}=-\hard_{qq';q\bar{q}'}^{e_qe_{q'}}$  \\
  \hline
\end{tabular}
\caption{
  The first column of real and virtual are the hard parts as in
  Table~\ref{table:real-virt-corr}.  The second column gives the hard
  parts needed other than those in the first column.  The superscript of
  the hard part $\Gamma \in \{g, pp\}$ is supressed. $e_q^2$,
  $e_{q'}^2$, and $e_qe_{q'}$ are various combinations of quark electric
  charges. A charge combination as a superscript denotes terms
  proportional to that particular charge structure in a hard part.}
\label{table:real-virt-corr-idntcl-contrib}
\end{table}

\begin{figure}[!h]
\centering
\includegraphics[width=\textwidth, angle=0]{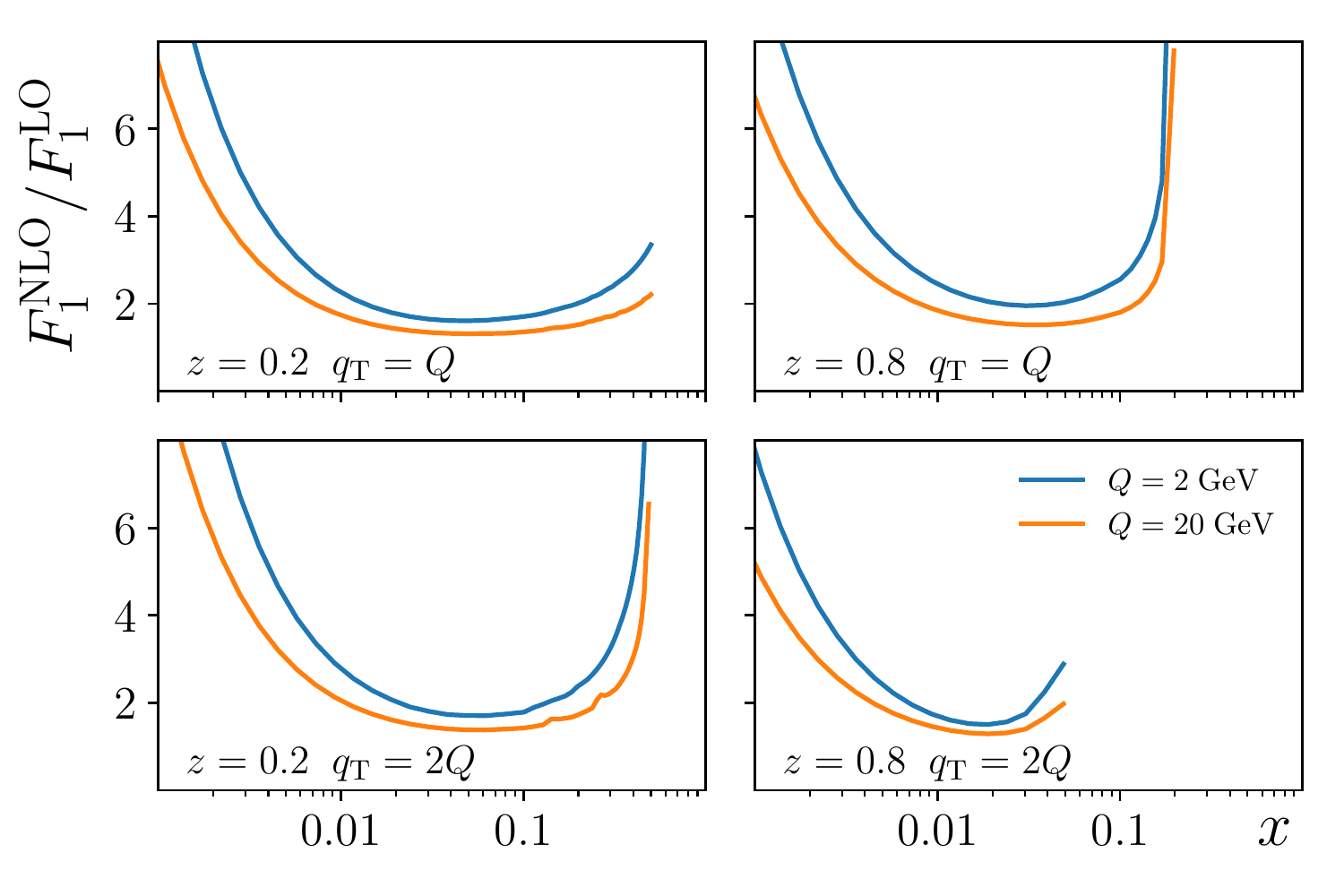}  
\caption{
  $K$-factor ratio. We use CJ15nlo PDFs \cite{Accardi:2016qay}
  and DSS09 FFs \cite{deFlorian:2007ekg}. 
  The ratio is bounded at large $x$ due to
  limiting phase space for hadron production. 
  The small noise in the lower left panel is due to the
  oscillatory behavior in the interpolation of FFs tables.}
\label{f.kfactor}
\end{figure}

\begin{figure}[!h]
\centering
\includegraphics[width=\textwidth, angle=0]{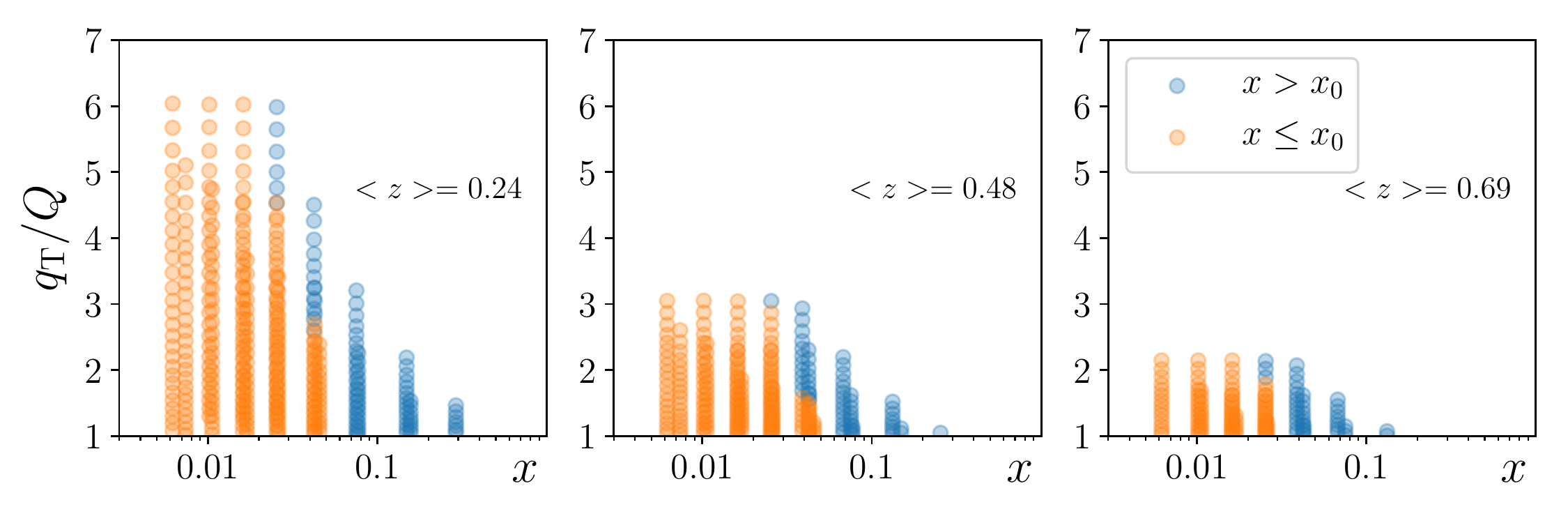}  
\caption{
  SIDIS kinematics of $h^{\pm}$ at COMPASS \cite{Aghasyan:2017ctw}. 
  $x_{0}$ is the location in $x$ where  the minumum of $K$-factor 
  is attained at a given $(Q,q_{\rm T},z)$. Samples with $x>x_{0}$ 
  indicates sensitivity to large threshold corrections.
}
\label{f.kin}
\end{figure}

\section{Conclusion}
\label{s.conc}
The computational tools necessary to reproduce \fref{kfactor}, \fref{kin} and other similar 
calculations are available at~\cite{BigTMD}, along with documentation. As indicated in the 
introduction, these generally confirm earlier calculations 
(e.g.,~\cite{Daleo:2004pn,Kniehl:2004hf}) of the overall cross section to within about 
$20\%$, well within present experimental uncertainties, and thus strengthens our 
earlier position~\cite{Gonzalez-Hernandez:2018ipj} 
that significant tension exists between existing SIDIS data and collinear factorization 
calculations at large $\Tsc{q}{} \gtrsim Q$. 

We have identified region 
where collinear factorization appears most reliable as the 
regions with minimal $K$-factors, with current sets of collinear PDFs and FFs. As 
discussed in \sref{Pheno}, minimal values for the $K$-factor
lie in the region $0.01<x<0.1$, that is, approximately in valence kinematics. The increase 
of the $K$-factor at both smaller and larger values of $x$, may point at the importance of 
resummation effects outside the valence region, or the need to refit collinear 
PDFs and FFs. 
This last interpretation is consistent with that of our previous 
study~\cite{Gonzalez-Hernandez:2018ipj}. It is also important to note that the $2 \to 2$ 
kinematics of the $\order{\alpha_s}$ contribution places severe kinematical constraints 
on the relationship between the initial and final state partons. Therefore, it is likely that 
the generally large $K$-factors are at least partly due simply to a kinematical suppression 
of the $\order{\alpha_s}$ contribution, and are not a fundamental problem with the 
convergence of the perturbation series. We note that 
Reference~\cite{Balitsky:2017gis} has addressed somewhat similar issues, but in the 
region where $\Tsc{q}{}$ is still small enough that a $\Tsc{q}{}/Q$ power expansion 
is still meaningful, and also in the limit of $Q^2 \gg s$.

As a next step, we plan to refit collinear functions in the large $\Tsc{q}{}$ region, using 
SIDIS and $e^+ e^-$-annihilation to back-to-back, and Drell-Yan scattering to explore 
the possibility that this allows the region of minimum $K$-factor in \fref{kfactor} to be 
fully accommodated with minimal modification to existing fits. 

	Since the pioneering work in presented in~\cite{Meng:1995yn,Graudenz:1996an,Nadolsky:1999kb}, there has been a large number of
	studies on unpolarized SIDIS cross sections~\cite{Koike:2006fn,Anselmino:2006rv,Anselmino:2013lza,Signori:2013mda,Sun:2013hua,Su:2014wpa,Bacchetta:2017gcc}. Unpolarized SIDIS is however, only one component in a broad program of phenomenolical studies where the universality of parton correlation functions plays a central role in testing pictures of nucleon structure~\cite{Anselmino:2016uie,Anselmino:2015sxa,Bacchetta:2015ora,Scimemi:2017etj,Kang:2017btw,Kang:2014zza,Landry:2002ix,Sun:2016kkh,Sun:2011iw,Echevarria:2015uaa,Guzzi:2013aja,Echevarria:2016scs,Echevarria:2014xaa,Boglione:2018dqd,Boglione:2017jlh,Kang:2015msa,Lin:2017stx,Ye:2016prn,Bacchetta:2019tcu,Bastami:2018xqd,Bertone:2019nxa}. In order for this program to be successful, the precision in the determination of parton correlation functions needs to meet the demands of the increasingly precision of current and future experimental programs
	~\cite{Airapetian:2012ki,Adolph:2013stb,Adolph:2016bga,Seidl:2019jei,Aschenauer:2019kzf,Bradamante:2018ick}. This demands a satisfactory resolution to problems in the $\Tsc{q}{} \sim Q$ region for which the pQCD calculation at large $\Tsc{q}{}$ is crucial.

\appendix
\begin{acknowledgments}	
We thank J. Owens, J.-W. Qiu, and W. Vogelsang for very useful 
discussions.  We thank R. Sassot for explanations regarding the code in \cite{Daleo:2004pn}. 
T.~Rogers's work was supported by the U.S. Department of Energy, Office of 
Science, Office of Nuclear Physics, under Award Number DE-SC0018106. This work was 
also supported by the DOE Contract No. DE- AC05-06OR23177, under which Jefferson 
Science Associates, LLC operates Jefferson Lab. 
N.~Sato was partially supported by DE-FG-04ER41309, DE-SC0018106 and 
the DOE Contract No. DE- AC05-06OR23177, under which Jefferson 
Science Associates, LLC operates Jefferson Lab. 
B.~Wang was supported in part by the National Science Foundation of China 
(11135006, 11275168, 11422544, 11375151, 11535002) and the Zhejiang University Fundamental
Research Funds for the Central Universities (2017QNA3007). 
J.~O.~Gonzalez-Hernandez work was partially supported
by Jefferson Science Associates, LLC under  U.S. DOE Contract \#DE-AC05-06OR23177 and by the 
U.S. DOE Grant \#DE-FG02-97ER41028.
\end{acknowledgments}

\section{Identities for three-body partonic phase space}
\label{a.PS3}

The phase space of three massless particles has the form
\begin{align}
\int PS_3=&\int\frac{\diff{^{n-1}k_1}}{(2\pi)^{n-1}2k_1^0}\frac{\diff{^{n-1}k_2}}{(2\pi)^{n-1}2k_2^0}\frac{\diff{^{n-1}k_3}}{(2\pi)^{n-1}2k_3^0}(2\pi)^n\delta^{(n)}(p+q-k_1-k_2-k_3) \cdots \, .
\end{align}
The $k_1$ integral can be performed in the center-of-mass frame of $p$
and $q$
\begin{align}
p=&(P,\bm{0}_{n-2},-P),\label{eq:pmom} \\
q=&(E_1,\bm{0}_{n-2},P),\label{eq:qmom}
\\
k_1=&E(1,\bm{k}_{n-2},\cos\theta),
\end{align}
where $\bm{k}_{n-2}$ denotes the first $n-2$ components of the $n-1$
dimensional unit spatial vector in spherical coordinates, and
\begin{align}
P=\frac{s+Q^2}{2\sqrt{s}}, \\
E_1=\frac{s-Q^2}{2\sqrt{s}}.
\end{align}
In $k_1$ phase space, $E$ and $\theta$ are related with
observables. The ``azimuthal'' angles
in $\bm{k}_{n-2}$ can be integrated to give
\begin{align}
\int \frac{\diff{^{n-1} k_1}}{2 k_1^0} \cdots 
=& - \int \frac{1}{2} \Omega_{n-3} E^{1-2\epsilon} \sin^{-2\epsilon} \theta
\diff{E} \, \diff{\cos\theta} \cdots \, ,
\label{eq:ps1}
\end{align}
where $\Omega_m$ is the $m$ dimensional angular volume,
\begin{align}
\Omega_m=&2^m\pi^{m/2}\frac{\Gamma (m/2)}{\Gamma (m)} \, .
\end{align}
$E$ and $\theta$ are related to Lorentz invariants $u$ and $t$ via
\begin{align}
t=&-2(E_1 E-PE\cos\theta)-Q^2, \\
u=&-2(PE+PE\cos\theta).
\end{align}
In terms of $t$ and $u$, the $k_1$ phase space becomes
\begin{align}
\int\frac{\diff{^{n-1}k_1}}{2k_1^0} \cdots =&- \int \frac{\Omega_{n-3}}{4(s+Q^2)}
\left[\frac{(s+Q^2)}{u(st+Q^2s_{23})}\right ]^{\epsilon} \diff{u}{} \diff{t}{} \cdots \, .
\label{e.psk1}
\end{align}
To simplify the $k_2$ and $k_3$ phase space, we work in the $k_2$ and
$k_3$ center-of-mass frame where $k_2$ and $k_3$ take the form of
Eqs.~\eqref{e.defk2} and~\eqref{e.defk3}. The choice of the spatial orientation of axes
also follows the discussion below Eq.~\eqref{e.defk2}. In this frame
\begin{align}
&{} \int PS_{k_2k_3} \cdots \no 
&{}\equiv\int\frac{\diff{^{n-1}k_2}}{2k_2^0}\frac{\diff{^{n-1}k_3}}{2k_3^0}\delta^{(n)}(p+q-k_1-k_2-k_3) \cdots =
\int\frac{\diff{^{n-1}} k_2}{2k_2^0} \delta^+((k_3^0)^2-\lvert
\bm{k_3}\rvert^2) \cdots \, . 
\end{align}
Since the scattering amplitudes are independent of 
the first $n-3$ spatial components of $k_2$, the ``azimuthal'' part of
$\diff{^{n-1}k_2}$ can be integrated to give
\begin{align}
\int PS_{k_2k_3} \cdots =\int
2^{2\epsilon-3} \Omega_{n-4}s_{23}^{-\epsilon}\sin^{1-2\epsilon}\beta_1\sin^{-2\epsilon}\beta_2 \diff{\beta_1}{} \diff{\beta_2} \cdots \, .
\label{eq:ps23}
\end{align}


\section{Phase space integration for $2 \to 3$ processes in SIDIS}
\label{s.integral}

The main effort of the calculation is in the 2 $\rightarrow$ 3 process
$p+q \rightarrow k_1+k_2+k_3$.  After partial fractioning (to be
discussed~\aref{parfrac}), each term from the squared amplitudes
$\lvert M^{2\to 3}_g \rvert^2$,$\lvert M^{2\to 3}_{PP}\rvert^2$
in~\eref{23ps} contains at most two Mandelstam variables depending on
the angles $\beta_1$ and $\beta_2$.  So, up to overall factors,
\eref{23ps} takes the form
\begin{align}
\int_0^{\pi} \diff{\beta_1}{} \int_0^{\pi} \diff{\beta_2}{} 
\frac{\sin^{1-2\epsilon}\beta_1\sin^{-2\epsilon}\beta_2}
{(a+b\cos\beta_1)^j(A+B\cos\beta_1+C\sin\beta_1\cos\beta_2)^l} \, ,
\label{e.int_gen}
\end{align}
where $j$ and $l$ are integers. The coefficients $a$, $b$, $A$, $B$,
and $C$ are specific to the squared amplitude and do not depend on
$\beta_1$ and $\beta_2$. They are determined by the spatial
orientations of axes in the $k_2+k_3$ rest frame. Note that
\erefs{defk2}{defk3} do not determine uniquely the components of $p$,
$q$, and $k_1$. There can still be a rotation in the plane determined
by the last two spatial axes. Taking advantage of this freedom, we
specify three frames.
\begin{itemize}
	\item[] \textbf{Frame 1:}
	\begin{align}
	&p=p_0(1,\bm{0}_{n-3},\sin\alpha_1,\cos\alpha_1),\label{e.pf1} \\
	&q=(q_0,\bm{0}_{n-3},q^{\prime}\sin\theta_1,q^{\prime}\cos\theta_1)\label{e.qf1},\\
	&k_1=k_{10}(1,\bm{0}_{n-3},0,1)\label{e.k1f1},
	\end{align}
	\item[] \textbf{Frame 2:}
	\begin{align}
	&p=p_0(1,\bm{0}_{n-3},0,1)\label{e.pf2},\\
	&q=(q_0,\bm{0}_{n-3},q^{\prime}\sin\alpha_2,q^{\prime}\cos\alpha_2)\label{e.qf2},\\
	&k_1=k_{10}(1,\bm{0}_{n-3},\sin\theta_2,\cos\theta_2),\label{e.k1f2} 
	\end{align}
	\item[] \textbf{Frame 3:}
	\begin{align}
	&p=p_0(1,\bm{0}_{n-3},\sin\alpha_3,\cos\alpha_3),\label{e.pf3} \\
	&q=(q_0,\bm{0}_{n-3},0,q^{\prime})\label{e.qf3},\\
	&k_1=k_{10}(1,\bm{0}_{n-3},\sin\theta_3,\cos\theta_3),\label{e.k1f3} 
	\end{align}
	where
	\begin{align}
	&p_0=\frac{s_{23}-t}{2\sqrt{s_{23}}}, \\
	&q_0=\frac{s+t}{2\sqrt{s_{23}}}, \\
	&q^{\prime}=\sqrt{q_0^2+Q^2}, \\
	&k_{10}=-\frac{s_{23}-s}{2\sqrt{s_{23}}}, \\
	&\sin\alpha_1=\sin\theta_2
	=\frac{2\sqrt{s_{23}u(Q^2s_{23}+st)}}{(s-s_{23})(s_{23}-t)},\\
	&\sin\alpha_2=\sin\alpha_3
	=\frac{2\sqrt{s_{23}u(Q^2s_{23}+st)}}{(s_{23}-t)\sqrt{4Q^2s_{23}+(s+t)^2}},\\
	&\sin\theta_1=\sin\theta_3
	=\frac{2\sqrt{s_{23}u(Q^2s_{23}+st)}}{(s-s_{23})\sqrt{4Q^2s_{23}+(s+t)^2}}.
	\end{align}
\end{itemize}

There are three cases (see~\aref{abABC} for a proof) for the
coefficients in the integral in \eref{int_gen}: 
(1) $a^2=b^2$ and $A^2=B^2+C^2$; 
(2) $a^2 > b^2$ and $A^2=B^2+C^2$; 
(3) $a^2= b^2$ and $A^2 > B^2+C^2$. 
Integrals in case (3) can be transformed to case (2) by switching to
another frame. With multiple frames, the number of integrals to
compute can be reduced. In the following we show how the integrals in
case (1) and (2) are obtained. 
\begin{itemize}
	\item[]
	\textbf{Case (1):}
	In this case the integral in~\eref{int_gen} has a closed form~\cite{vanNeerven:1985xr}
	\begin{align}
	&\int_0^{\pi} \di{\beta_1} \int_0^{\pi} \di{\beta_2} \frac{\sin^{1-2\epsilon}
		\beta_1\sin^{-2\epsilon}\beta_2}
	{(1-\cos\beta_1)^j(1-\cos\chi\cos\beta_1-\sin\chi\sin\beta_1\cos\beta_2)^l}
	\nonumber \\
	&=2\pi\frac{\Gamma(1-2\epsilon)}{\Gamma(1-\epsilon)^2}2^{-j-l}B(1-\epsilon-j,1-
	\epsilon-l)_2F_1\Bigl(j,l,1-\epsilon;\cos^2\frac{\chi}{2} \Bigr),
	\label{e.case1}
	\end{align}
	where for simplicity an overall factor  is
	not shown and we choose $a=A=-b=1$, $B=\cos\chi$, and $C=\sin\chi$.
	In the frames we choose above, $\chi$ can be $\alpha_1$ or $\theta_2$.
	Here the signs of the trigonometric function terms in the denominator
	are chosen to be negative. All other choices can be transformed to
	this one by applying substitutions $\beta_{1,2}\rightarrow \pi
	-\beta_{1,2}$, and/or $\chi \rightarrow \pi +\chi$. The derivation of
	this integral can be found in appendix A of~\cite{vanNeerven:1985xr}. 
	\item[] \textbf{Case (2):}
	In this case the integral no longer has a closed form. Following 
	steps similar to appendix A of~\cite{vanNeerven:1985xr}, we arrive at 
	\begin{align}
	&I_{j,l} = 
	\int_0^{\pi} \di{\beta_1} \int_0^{\pi} \di{\beta_2} 
	\frac{\sin^{1-2\epsilon}\beta_1\sin^{-2\epsilon}\beta_2}
	{(D-\cos\beta_1)^j(1-\cos\chi\cos\beta_1-\sin\chi\sin\beta_1\cos\beta_2)^l}
	\nonumber\\[10pt]
	&{}=(-1)^{l+1}2^{1-l-j}\pi
	\frac{\Gamma(n-3)\Gamma(2+l-n/2)}{\Gamma(n/2-1)\Gamma(l)\Gamma(n/2-2)\Gamma(3-n/2)}\nonumber
	\\[10pt]
	&\hspace{100pt} \times
	\int_0^1dx\int_0^1dz
	\frac{x^{l-1}z^{n/2-2}(1-x)^{n/2-l-2}(1-z)^{n/2-l-2}}
	{[z+(D-1)/2-(1+\cos\chi)xz/2]^j} \nonumber \\[10pt]
	&{}=(-1)^{l+1}2^{1-l-j}\pi
	\frac{\Gamma(n-3)\Gamma(2+l-n/2)\Gamma(n/2-l-1)}
	{\Gamma(n/2-1)^2\Gamma(n/2-2)\Gamma(3-n/2)}\nonumber\\[10pt]
	&\hspace{100pt} \times \int_0^1 \di{z} \frac{z^{n/2-2}(1-z)^{n/2-l-2}}{(z+(D-1)/2)^j}
	\,_2F_1\Bigl(j,l,n/2-1,w\Bigr), \label{e.case2}\\[10pt]
	w&{}\equiv\frac{(1+\cos\chi)z}{D-1+2z},
	\end{align}
	where $D>1$. The integral representation of hypergeometric function is
	used. Note the $I_{j,l}$ notation for integrals with different
	denominator structures. The result of evaluating~\eref{case2}
	for specific $j$ and $l$ is given in~\aref{drv_ecase2}.  Here $\chi$
	can be $\alpha_3$ or $\theta_3$. The remaining $z$ integral has to be
	performed by expanding the integrand into series of $\epsilon$, and
	computing the integral order by order. The subtlety comes from the
	treatment of the factor $(1-z)^{n/2-l-2}$, which can produce poles for
	$l\geq 1$. The integral has the form
	\begin{align}
	\int_0^1\di{z} (1-z)^{-\epsilon-l}f(z),
	\label{e.zint}
	\end{align}
	where $f(z)$ is a regular function at $z=1$. When $l=1$, the pole term
	can be made explicit using the identity
	\begin{align}
	(1-z)^{-\epsilon-1}=-\frac{\delta(1-z)}{\epsilon}+\frac{1}{(1-z)_+}-\epsilon\left (\frac{\log
		(1-z)}{1-z}\right )_+ + O(\epsilon^2),
	\label{e.pole_exp}
	\end{align}
	where the ``plus" functions are defined, for a function $p(\xi)$, through
	\begin{align}
	\int_x^1 \di{\xi} (p(\xi))_+ q(\xi) = \int_0^1 \di{\xi}
	p(\xi)(q(\xi)-q(1))-\int_0^x \di{\xi} p(\xi)q(\xi).
	\end{align}
	When $l>1$, the integral in~\eref{zint} is divergent.
	Nonetheless, it can be analytically continued. To do this, we write 
	\begin{align}
	&\int_0^1\di{z} (1-z)^{-\epsilon-l}f(z)\nonumber \\
	&=\int_0^1\di{z} (1-z)^{-\epsilon-l+1}f_1(z)+f(1)\int_0^1\di{z} (1-z)^{-\epsilon-l},
	\label{e.intz2} \\
	&{} \text{with} \; f_1(z)\equiv \frac{(f(z)-f(1))}{1-z} \, .
	\end{align}
	Note that, in the first term on the right-hand side of~\eref{intz2},
	the power of $1-z$ is increased by 1, with a regular function
	$f_1(z)$. The second term is divergent, but can be analytically
	continued using $\Gamma$ functions:
	\begin{align}
	\int_0^1\di{z} (1-z)^{-\epsilon-l}=B(1,-\epsilon-l+1)
	=\frac{\Gamma(-\epsilon-l+1)}
	{\Gamma(-\epsilon-l+2)}=\frac{1}{-\epsilon-l+1} \, .
	\end{align}
	The manipulation in \eref{intz2} can be repeated until the power of
	$1-z$ in the integral becomes $-\epsilon-1$, in which case
	\eref{pole_exp} can be used. To expand the hypergeometric function in
	the integrand in \eref{case2}, we note that if either $j$ or $l$ is
	less than 1, the hypergeometric series terminates and the function
	reduces to a polynomial. For $j=1$ and $l=1$ we use the expansion
	\begin{align}
	_2F_1(1,1,1-\epsilon,w)=(1-w)^{-1-\epsilon}(1+\epsilon^2 \text{Li}_2(w)+O(\epsilon^3)).
	\end{align}
	For the case that one of $j$ and $l$ is larger than 1 and the other
	is at least 1, we use Gauss's contiguous relations to reduce $j$ or
	$l$.  For example, for $F(1,2,1-\epsilon)$ and $F(2,2,1-\epsilon)$
	(with the shorthand notation $F(j,l,k)\equiv \,_2F_1(j,l,k,w)$), we use
	\begin{align}
	&(k-2j+(j-l)w)F(j,l,k)+j(1-w)F(j+1,l,k)-(k-j)F(j-1,l,k)=0, \\
	&(k-j-l)F(j,l,k)+l(1-w)F(j,l+1,l)-(l-j)F(j-1,l,k)=0 
	\end{align}
	to get
	\begin{align}
	F(1,2,1-\epsilon)&=\frac{1}{1-w}\left
	(-\epsilon+\left ( \epsilon+1\right
	)F(1,1,1-\epsilon)  \right ), \\
	F(2,2,1-\epsilon)&= -\frac{1}{1-w}\biggl (\epsilon+1 \biggr )
	F(1,1,1-\epsilon) \nonumber \\
	&\qquad \qquad +\frac{1}{(1-w)^2}\biggl (\epsilon+2\biggr )
	\biggl [-\epsilon+\biggl (\epsilon+1 \biggr)F(1,1,1-\epsilon) \biggr ] \, .
	\end{align}
	
\end{itemize}

Other well-known non-trivial aspects of the computation are the 
conversion of the expressions in the squared amplitude
into forms that allow \eref{case1} or \eref{case2} to be used, which are reviewed in 
\aref{parfrac}, and the algorithm for calculating virtual corrections, which are reviewed 
in \aref{virt}. Finally the procedure for combining all Feynman graph calculations 
consistently into a a factorized cross section is reviewed in \aref{subtraction}.


\section{Proof that $a^2 \geq b^2$ and $A^2 \geq B^2+C^2$ in~\eref{int_gen}}
\label{a.abABC}

First, we show that $t_i<0$. In the center-of-mass frame of $p$ and
$q$,
\begin{align}
p=&\parz{\frac{W^2+Q^2}{2W},\bm{0}_{n-2},-\frac{W^2+Q^2}{2W}} \, ,\label{e.pmom} \\
q=&\parz{\frac{W^2-Q^2}{2W},\bm{0}_{n-2},\frac{W^2+Q^2}{2W}},\label{e.qmom} \\
k=&E(1,\bm{k}_{n-2},\cos\theta) \, ,
\end{align}
where $W=\sqrt{s}$. $k$ can be any one of $k_i$, $i=1,2,3$, with the subscript
suppressed above. In this frame, $E<\frac{1}{2}\sqrt{s}=\frac{1}{2}W$ and
\begin{align}
t=&(q-k)^2=-2\left
[\frac{(W^2-Q^2)E}{2W}-\frac{(W^2+Q^2)E\cos\theta}{2W}   \right
]-Q^2 \,  \\
=&-\left [WE(1-\cos\theta)-\frac{EQ^2}{W}(1+\cos\theta)   \right ]-Q^2
\,  \\
<&-\left [ 0-\frac{\frac{1}{2}WQ^2}{W}(1+\cos\theta)\right ]-Q^2 \\
=&\frac{1}{2}Q^2(\cos\theta-1)\\
\leq &0.
\end{align}
The $t_i$ are Lorentz invariants and thus $t_i<0$ holds in any
frame.
Since the signs of $t_i$ do not change and
\begin{align}
-\lvert b\rvert \leq b\cos\beta_1 \leq \lvert b\rvert, \quad -\sqrt{B^2+C^2}\leq
B\cos\beta_1+C\sin\beta_1\cos\beta_2 \leq \sqrt{B^2+C^2},
\end{align}
we conclude that $a^2 \geq b^2$ and $A^2 \geq B^2+C^2$, where the ``$=$'' 
holds when the corresponding variable is $u_i$ or $s_{ij}$, and the
``$>$'' holds for $t_i$.

\section{Partial fraction algorithm}
\label{a.parfrac}

To perform the three-body phase space integrations as described \sref{integral},
it is necessary to first convert  expressions in the squared amplitude
into forms that allow \eref{case1} or \eref{case2} to be used. In this
appendix, we review the way this can be automized. 

To simplify the discussion, we will refer to angle dependent
Mandelstam variables (ADMV): 
\begin{equation} 
\textbf{\text{ADMV}} = \text{``Angle Dependent Mandelstam Variables"} 
            = \left\{ t_2,t_3,u_2,u_3,s_{12},s_{13} \right\} \, .
\end{equation}
The angle independent Mandelstam variables (AIMVs) are: 
\begin{equation} 
\textbf{\text{AIMV}} = \text{``Angle Independent Mandelstam Variables"} 
            = \left\{ t_1,u_1,s,s_{23} \right\} \, .
\end{equation}
A general term in the squared amplitude starts as simple
products of AIMVs and ADMVs. A conversion to a useful form is
accomplished with a set partial fractions such that each term in the
squared amplitude can be expressed as 
\begin{equation} 
\frac{[{\rm AIMVs}]}{X^{\alpha_1}Y^{\alpha_2}} \, ,
\label{eq:ready}
\end{equation}
with $\alpha_1$ and $\alpha_2$ being real numbers that could be
positive, negative or zero.  The quantity in the brackets is some
combination of the AIMVs while  $X$ and $Y$ are two different types of
ADMV that belong to to either $t_i$, $u_i$ or $s_{ij}$. A term in such
a form is called a  ``\proper'' term.  Otherwise, we will call it
``\improper''. Example are:
\begin{itemize}
\item \textbf{$\proper$ terms}: (constant), ($t_2^{-2}u_3^{-1}$), ($s_{12}^{-2}$),($u_2$)
\item \textbf{$\improper$ terms}: ($t_2^{-2}t_3^{-1}$), ($u_2u_3$), 
                                     ($t_2^{-2}u_3^{-1}s_{12}^{-1}$)
\end{itemize}
We need the following further sets of definitions:
\begin{itemize}
\item \textbf{\emph{t-type, u-type, s-type Mandelstam variables}}: An ADMV is
      referred to as t-type if it is $t_2$ or $t_3$, and similarly for
      u-type and s-type.
\item The ADMVs are not independent variables; it is possible to construct
      linear relations relating them. This is achieved by squaring different
      rearrangements of the momentum conservation equation:
      \begin{equation} 
      p+q-k_1=k_2+k_3.
      \label{eq:mom_consv}
      \end{equation}
      \begin{itemize}
      \item[] \textbf{\emph{Two ADMV relations (2ARs)}}: One such type of  relation is
            between two ADMVs of the same type. These are obtained by moving one
            of $p$, $q$, and $k_1$ in Eq.~\eqref{eq:mom_consv} to the right-hand
            side and squaring the equation. For instance, moving $p$ to the right
            and squaring gives
            \begin{equation}
            t_1=u_2+u_3+s_{23}, \label{e.mrel1}
            \end{equation}
            relating two u-type ADMVs. Similarly, moving $q$ or $k_1$ to the right
            gives the relations for $t$-type or $s$-type ADMVs respectively. There
            are three 2ARs. The other two are:
            \begin{align}
            s &{}= s_{12} + s_{23} + s_{13} \\
            u_1 &{}= t_2 + t_3 + s_{23} + Q^2 \, . \label{e.u1id}
            \end{align}
      \item[] \textbf{\emph{Three ADMV relations (3ARs)}}: Another type of relation is between
            three ADMVs, each from a different ADMV type. We can get such
            relations for any three ADMVs of different types. To do this notice
            that exchanging the positions of $k_2$ or $k_3$ with $k_1$ in
            Eq.~\eqref{eq:mom_consv} and squaring will give one of such relation.
            For instance, squaring $p+q-k_2=k_1+k_3$:
            \begin{equation} 
            s_{13}=s+t_2+u_2+Q^2 \, , \label{e.3arex}
            \end{equation} 
            a relations between $s_{13}$, $t_2$, and $u_2$. After one
            such relation is obtained, the others are obtained by replacing
            the ADMVs in that relation using three 2ARs.
      \end{itemize}
\end{itemize}
Now the partial fraction is implemented in two steps:
\begin{itemize}
\item[] \textbf{Step 1}: Consider a term $T_1$ in the squared amplitude.  The
      first step is to reduce the number of ADMVs in the denominator to two
      or fewer, and to ensure that no two ADMVs of the same type appear in
      the denominator. This is done in two substeps:
      \begin{itemize}
      \item[] \textbf{Substep 1}: First, separate any ADMVs of the same type in the
            denominator. If two ADMVs of the same type appear in the denominator
            of $T_1$, use a 2AR for the two ADMVs to write a factor of unity in
            the form
            \begin{equation} 
            1=\frac{\sum_i c_i {\rm ADMV}_i}{[{\rm AIMVs}]}
            \label{e.unity}
            \end{equation}
            and write
            \begin{equation} 
            T_1 = T_1 \parz{ \frac{\sum_i c_i {\rm ADMV}_i}{[{\rm AIMVs}]} } \, , 
            \label{e.unity2}
            \end{equation}
            with $c_j = \pm 1$. Expanding the right side of~\eref{unity2}
            produces a sum of new terms.  Each will have one power lower of these
            two ADMVs in its denominator.  Further multiplication by unity factors
            like~\eref{unity} can be repeated until in all terms only one of the
            two ADMVs appears in any denominator.  As an example, consider 
            \begin{equation}
            T_1 = \frac{s}{t_2 t_3 u_2} \, . 
            \end{equation}
            $t_2$ and $t_3$ both appear in the denominator,
            so we will use~\eref{u1id} to write~\eref{unity}:
            \begin{equation}
            1 = \frac{u_1 - s_{23} - Q^2}{u_1- s_{23} - Q^2} 
            = \frac{t_2 + t_3}{u_1- s_{23} - Q^2} \, .
            \end{equation}
            Then
            \begin{align}
            T_1 {}&= \parz{ \frac{s}{t_2 t_3 u _2} } 
                     \parz{\frac{t_2 + t_3}{u_1- s_{23} - Q^2}} \\
                  &= \frac{s}{t_3 u_2 \parz{u_1- s_{23} - Q^2} } 
                   + \frac{s}{t_2 u_2 \parz{u_1- s_{23} - Q^2} }\, .
            \end{align}
       \item[] \textbf{Substep 2}: After Substep 1 is done for all three types, the
       leftover terms have denominators with at most three ADMVs, all from
       different types. For terms with two or fewer ADMVs in the denominator,
       nothing more needs to be done in this step.  For terms with three
       ADMVs in the denominator, one may write a 3AR in the form
       of~\eref{unity} and reduce the number of ADMVs in the denominator to
       two or fewer in a way similar to Substep 1. For example, say that 
       \begin{equation}
       T_1 = \frac{1}{s_{13} u_2 t_2} \, .
       \end{equation}
       Then it is possible to eliminate $s_{13}$, $u_2$ and $t_2$ by
       using~\eref{3arex} to construct~\eref{unity}. Then,
       \begin{align}
       T_1 &{}= \parz{ \frac{1}{s_{13} u_2 t_2} } \parz{\frac{s + Q^2}{s + Q^2}} \no
       &{}= \parz{ \frac{1}{s_{13} u_2 t_2} } \parz{\frac{s_{13} - t_2 - u_2}{s + Q^2}}  \no
       &{}= \frac{1}{u_2 t_2} \frac{1}{\parz{s + Q^2}} 
           - \frac{1}{s_{13} u_2} \frac{1}{\parz{s + Q^2}}
           - \frac{1}{s_{13} t_2} \frac{1}{\parz{s + Q^2}} \, .
       \end{align}
    \end{itemize}
\item[] \textbf{Step 2}: After Step 1 is completed, all terms have two or fewer
      ADMVs in the denominator. To get to the final form in
      Eq.~\eqref{eq:ready}, we need to make sure no third ADMV
      appears in the numerator. This is
      done by noticing that the ADMVs in the numerator can be written in
      terms of the ADMVs of the denominator using 2ARs and 3ARs.  For
      instance, if one term has $t_2u_3$ in the denominator, then any t-type
      ADMV in the numerator can be replaced by $t_2$, and u-type by $u_3$,
      using 2ARs. Any s-type ADMV can be replaced by a linear combination of
      $t_2$ and $u_3$ using the corresponding 3AR involving $t_2$ and $u_3$.
      This puts $T_1$ in the form of Eq.~\eqref{eq:ready}. (Recall
      that $\alpha_1$ and $\alpha_2$ can be negative or zero.) This example
      had two ADMVs in the denominator. If the number of ADMVs in the
      denominator is 1, one can express the ADMVs in the numerator in terms
      of two ADMVs of different types, where one is the ADMV in the
      denominator and the other is chosen randomly. If the number of ADMVs
      in the denominator is 0, then the ADMVs in the numerator can be
      expressed by any two randomly chosen ADMVs of different types.
      There are six ADMVs, but the four relations from
      \erefs{mrel1}{3arex} eliminate four in terms of the other two.
\end{itemize}

\section{Virtual Contributions}
\label{a.virt}

At order $O(\alpha_s^2)$, virtual corrections to the partonic cross
section involve only one-loop integrals.  Comprehensive reviews of the
methods for this type of calculation can be found in
\cite{Berger:2009zb,Britto:2010xq,Ellis:2011cr}.  In this article, we
use the traditional Passarino-Veltman (PV)
approach~\cite{Passarino:1978jh}, for which we closely follow the
notation of \cite{Ellis:2011cr}. 

Our $O(\alpha_s^2)$ virtual contributions are the interference terms
between tree-level $2 \to 2$ amplitudes and $2 \to 2$ amplitudes with
an additional virtual loop, averaged(summed) over initial(final)
states.  The structure of each such interference term is
\begin{align}
\label{e.virtualform}
{\cal M}_{\rm V}\,{\cal M}^*_{tree} 
  =\mu^{2 \epsilon}\,\intloopb 
   \frac{{\rm Tr}\{...\}_{\mu\nu} \contractortot{\Gamma}^{\mu \nu}}
        {d_1 ... d_n} + {\rm Hermitian \; Conjugate} \, .
\end{align}
The denominators contain massless propagators of the form
$d_i=(l+\Delta_i)^2+i0$, where $\Delta_i$ depends only on external
momenta, which we will call $v_i$. That is, $v_i \in \{p,q,k_1\}$ or
combinations thereof.  The trace in \eqref{e.virtualform} contains
Dirac gamma matrices contracted with either the external momenta
$v_i$, or with the loop momentum $l$. Both IR and UV divergences are
handled by standard dimensional regularization techniques.  After
these steps, the numerator in Eq~\eqref{e.virtualform} is a collection
of terms with Lorentz invariant products of loop and external momenta.
All integrals needed to calculate the virtual corrections in
Figure~\ref{f.graphs} can be written in one of the following ways:
\begin{align}
B_{\{0,\alpha\}}
  \equiv& \mu^{2 \epsilon}\,\intloopb
          \frac{\{1,l_\alpha\}}
          {l^2(k+v_1)^2}\,,\nonumber\\
C_{\{0,\alpha,\alpha\beta\}}
  \equiv&\mu^{2 \epsilon}\,\intloopb
         \frac{\{1,l_\alpha,l_\alpha l_\beta\}\hspace{40pt}}
              {l^2(l+v_1)^2(l+v_1+v_2)^2}\,,\nonumber\\
D_{\{0,\alpha,\alpha\beta,\alpha\beta\delta\}}
  \equiv&\mu^{2 \epsilon}\,\intloop
  \frac{\{1,l_\alpha,l_\alpha l_\beta,l_\alpha l_\beta l_\delta\}\hspace{98pt}}
       {l^2(l+v_1)^2(l+v_1+v_2)^2(l+v_1+v_2+v_3)^2}\,,
\label{e.virtualtypical}
\end{align} 
where higher rank integrals cannot appear since, in every case,
virtual correction diagrams at one-loop involve at least one gluon
propagator.  We have left the Feynman prescription implicit in the
denominators in  Eq.~\eqref{e.virtualtypical}, and partonic cross
section calculations are done in massless QCD.  These tensor integrals
can be written in terms the scalar box ($D_0$), triangle ($C_0$) and
bubble ($B_0$) integrals using the Passarino-Veltman reduction
procedure\cite{Passarino:1978jh}. \\

As an example, consider the steps to reduce $C^{\alpha\beta}$.
Symmetry properties allow $C^{\alpha\beta}$, $C^{\beta}$ and
$B^{\beta}$ to be written in terms of form factors ${\rm B}_1(v_1)$,
${\rm C}_1(v_1,v_2)$,  ${\rm C}_2(v_1,v_2)$, and ${\rm C}_{ij}(v_1,v_2)$:
\begin{subequations}
\begin{align}
B^{\beta}(v_1)     & \equiv v_1^\beta {\rm B}_1(v_1) \,,\\
\label{e.formfactorsa}
C^{\beta}(v_1,v_2) & \equiv v_1^\beta {\rm C}_1(v_1,v_2) + v_2^\beta {\rm C}_2(v_2,v_2) \, ,\\
C^{\alpha\beta}(v_1,v_2) &\equiv g^{\alpha\beta} {\rm C}_{00}(v_1,v_2) 
                          +\sum_{ij=1}^2v_i^\alpha v_j^\beta {\rm C}_{ij}(v_1,v_2) \,.
\label{e.formfactorsb}
\end{align}
\end{subequations}
Defining the Graham matrix $\mathbf{G}_{i,j}=v_i\cdot v_j$ and the row
vectors $\vec{{\rm C}}^{(1)}_{i}={\rm C}_{i1}$, $\vec{{\rm
C}}^{(2)}_{i}={\rm C}_{i2}$, the contraction of both sides of
Eq.~\eqref{e.formfactorsb} with $v_i{}_\alpha$ and $g_{\alpha \beta}$
respectively gives
\begin{align}\label{e.PVlineareqs}
\begin{pmatrix}
\mathbf{G}\,\vec{{\rm C}}^{(1)}_{i}&\mathbf{G}\,\vec{{\rm C}}^{(2)}_{i} 
\end{pmatrix}
=\mathbf{R}\,,\qquad (4-2\epsilon) \,{\rm C}_{00}=B_0(v_2)-{\rm Tr}\{\mathbf{R}\}\,,
\end{align}
where $\mathbf{R}$ is defined by the relations
\begin{align}\label{e.rhs}
R_{11}=&\frac{1}{2}\left({\rm B}_1(u)
                         +B_0(v_2)
                         -v_1^2{\rm C}_1(v_1,u)
                         -2{\rm C}_{00}(v_1,u)
                    \right) \, , \nonumber \\
R_{12}=&\frac{1}{2}\left( {\rm B}_1(u)
                         -{\rm B}_1(v_2)
                         -v_1^2{\rm C}_2(v_1,u)
                    \right) \, , \nonumber \\
R_{21}=&\frac{1}{2}\left( {\rm B}_1(v_1)
                         -{\rm B}_1(u)-(v_1^2+2v_1\cdot v_2){\rm C}_1(v_1,u)
                   \right) \, , \nonumber\\
R_{22}=&\frac{1}{2}\left(-{\rm B}_1(u)
                         -(v_1^2+2v_1\cdot v_2){\rm C}_2(v_1,u)
                         -2{\rm C}_{00}\right)\, .
\end{align}
Here we have used the notation $u = v_1 + v_2$.  The results for
$R_{ij}$ above are obtained by contracting $v_i{}_\alpha$ with the
right-hand side of Eq.~\eqref{e.formfactorsb} and using the property
that the products $v_i\cdot l$ can always be written in terms of the
denominators $d_i=(l+v_1+...+v_{i+1})^2$ as
\begin{align}
v_1\cdot l=\frac{1}{2}\left(d_2-d_1-v_1^2\right)\, , \qquad
v_2\cdot l=\frac{1}{2}\left(d_3-d_2-v_2^2-2v_1\cdot v_2\right).
\label{e.kdotpi}
\end{align}
This allows $v_i{}_\alpha \, C^{\alpha\beta}$ to be reduced to rank-1
integrals, whereupon expansion according to
Eqs.~\eqref{e.formfactorsa}, lead to Eqs.~\eqref{e.rhs}. Solving
Eq.~\eqref{e.PVlineareqs} gives ${\rm C}_{ij}$ and ${\rm C}_{00}$ in
terms of form factors corresponding to lower rank tensor integrals. In
a similar way, one may reduce residual ${\rm C}_i$ and ${\rm B}_1$
form factors so that only the scalar integrals $C_0$ and $B_0$ appear at the end of the
reduction procedure. Equations analogous to~\eqref{e.PVlineareqs} and
\eqref{e.rhs} for all cases in Eq.~\eqref{e.virtualtypical} are
provided in Appendix A of \cite{Ellis:2011cr}.

After the reduction procedure, all virtual contributions from
Figure~\ref{f.graphs} are  in terms of scalar integrals, whose values
depend only on Lorentz invariants constructed from momenta that appear
in the denominators. The complete set of scalar integrals are in
~\cite{Ellis:2007qk}.  In our calculation, all of these contain single
or double pole singularities. 

The singular behavior of $D_0$ and $C_0$ corresponds to soft
and collinear divergences that exactly cancel the soft and
collinear singularities in the $2 \to 3$ calculation of the
corresponding channel. \\

For our computations, we only need expressions for a reduced number of
cases for $C_0$ and $D_0$, following the notation of
~\cite{Ellis:2007qk}: $C_0$ divergent type 1 and type 2, $D_0$ divergent type 2 

The usual UV singularities introduced by virtual corrections ultimately are 
all produced by $B_0$-type integrals. UV divergent tadpole integrals $A_0$ also appear 
but are zero in a massless theory. Thus, keeping track of $B_0$ integrals accounts for all 
UV singularities, which cancel in the sum of virtual graphs. Self-energy diagrams on 
external legs enter for each channel via the corresponding factors of field strength 
normalization $Z_i^{1/2}$, as prescribed by the LSZ theorem. \\

At order $O(\alpha_s^2)$, $Z_i$ deviates from unity only for
$2\to2$ partonic scattering. Denoting the sum of amputated
diagrams for leading order, virtual, and real emission by ${\cal
M}_{LO}^{(a)}$, ${\cal M}_V^{(a)}$ and ${\cal M}_R^{(a)}$, the
square-modulus amplitude for \emph{partonic} scattering is
\begin{align}
 |{\cal M}|^{\,2}
 =&\,\,  |{\cal M}_{LO}^{(a)}|^{\,2}+|{\cal M}_{R}^{(a)}|^{\,2}
 +2 \Re \{{\cal M}_{LO}^{(a)}{}^{\dagger}{\cal M}_{V}^{(a)}\} \,.
\label{e.amplitudeUVfinite}
\end{align}

Explicit solutions for scalar integrals given in \cite{Ellis:2007qk}
hold when all of the relevant Lorentz invariants are space-like.
When needed, we perform analytic continuation on the
logarithms, by restoring the Feynman prescription
$\log \left(\frac{a}{b}\right) 
  \longrightarrow \log \left(a+i 0\right)+\log \left(\frac{1}{b+i 0}\right)$.
For the dilogarithms in our calculation, we find the following relation useful
\begin{align}
 {\rm Li}_2\left(1+\frac{a-i0}{b+i0}\right) = &
\, \frac{\pi^2}{3}
 -\frac{1}{2}\log^2\left(1+\frac{a}{b}\right)
 -i\pi \log\left(1+\frac{a}{b}\right)
 -{\rm Li}_2\left(\frac{1}{1+\frac{a}{b}}\right)\, 
\end{align}
where both $a$ and $b$ are positive.

\section{Angular integrals with a virtual photon}
\label{a.drv_ecase2}

For completeness we provide
a complete list of the relevant angular integrals. 
The method to compute these integrals is described in detail in
Sec~\sref{integral}. The alternatives for some of the integrals here can be
found from appendix C of Ref.~\cite{Beenakker:1988bq}. But there the list of integrals
is computed for heavy quark production process and is not adequate
for SIDIS. To compactify notation, we define
\begin{align}
\cosfun &{} = \cos \chi \\
\ffun &{} = \cosfun^2 \left(1-3 D^2\right)+4 \cosfun D+D^2-3 \\
\ggun &{} = \cosfun^2 \left(3 D^2-1\right)-4 \cosfun D-D^2+3 \\ 
\llun &{} = \log \left(\frac{D+1}{D-1}\right) \\ 
\kkun &{} = \left(\text{Li}_2\left(\frac{2}{D+1}\right)-\text{Li}_2\left(-\frac{2}{D-1}\right)\right) 
\,.
\end{align}
The integrals are computed up to the needed powers in $\epsilon$. The expressions 
for $I_{j,l}$ from Eq.~\ref{e.case2} needed for order $\alpha_s^2$ SIDIS are then
\begin{align}
	        I{}_{1,-4}  &{} = \frac{\pi}{72} \Bigl( -630 \cosfun^4 D^3+330 \cosfun^4 D+1440 \cosfun^3 D^2-384 \cosfun^3+540 \cosfun^2 D^3-1764 \cosfun^2 D \nonumber \\ 
                &{} \qquad +9 \Bigl( \cosfun^4 \left(35 D^4-30 D^2+3\right)+16 \cosfun^3 D \left(3-5 D^2\right)-6 \cosfun^2 \left(5 D^4-18 D^2+5\right) \nonumber \\ 
                &{} \qquad \qquad +16 \cosfun D \left(3 D^2-5\right)+3 D^4-30 D^2+35 \Bigr) \llun \nonumber \\
                &{} \qquad -864 \cosfun D^2+1152 \cosfun-54 D^3+522 D \Bigr) \nonumber \\
			    &{} +\frac{\pi \epsilon}{72} \Bigl(-1773 \cosfun^4 D^3+943 \cosfun^4 D+3744 \cosfun^3 D^2-1024 \cosfun^3+1674 \cosfun^2 D^3 \nonumber \\
                &{} \qquad +\frac{27}{2} \left(\cosfun^2-1\right) \left(D^2-1\right) \left(\cosfun^2 \left(19 D^2-3\right)-32 \cosfun D-3 D^2+19\right) \llun \nonumber \\
                &{} \qquad -4398 \cosfun^2 D+9 \Bigl( \cosfun^4 \left(35 D^4-30 D^2+3\right)+16 \cosfun^3 D \left(3-5 D^2\right) \nonumber \\
                &{} \qquad \qquad -6 \cosfun^2 \left(5 D^4-18 D^2+5\right)+16 \cosfun D \left(3 D^2-5\right) \nonumber \\ 
                &{} \qquad \qquad +3 D^4-30 D^2+35 \Bigr) \kkun \nonumber \\ 
									 &{} \qquad -2592 \cosfun D^2+2688 \cosfun-189 D^3+1599 D \Bigr) \\
	        I{}_{2,-4}  &{} = \frac{\pi}{3 \left(D^2-1\right)}  \Bigl( 36 \cosfun \left(3-5 \cosfun^2\right) D^3+12 \cosfun \left(13 \cosfun^2-11\right) D \nonumber \\ 
                &{} \qquad +3 \left(35 \cosfun^4-30 \cosfun^2+3\right) D^4 +\left(-115 \cosfun^4+222 \cosfun^2-51\right) D^2+16 \left(\cosfun^4-6 \cosfun^2+3\right) \Bigr) \nonumber \\
                &{} \qquad +\frac{1}{2} \pi  \Bigl( 5 \cosfun^4 D \left(3-7 D^2\right)+12 \cosfun^3 \left(5 D^2-1\right)+6 \cosfun^2 D \left(5 D^2-9\right) \nonumber \\ 
                &{} \qquad \qquad +4 \cosfun \left(5-9 D^2\right)-3 D \left(D^2-5\right) \Bigr) \llun \nonumber \\
                &{} + \frac{\pi \epsilon}{18} \Bigl( 72 \left(21-29 \cosfun^2\right) \cosfun D+32 \left(39-8 \cosfun^2\right) \cosfun^2+54 \left(27 \cosfun^4-26 \cosfun^2+3\right) D^2 \nonumber \\ 
                &{} \qquad  - 9 \left(5 \cosfun^4 D \left(7 D^2-3\right)+\cosfun^3 \left(12-60 D^2\right) + 6 \cosfun^2 D \left(9-5 D^2\right)+4 \cosfun \left(9 D^2-5\right)  \right. \nonumber \\  
                &{} \qquad \left. +3 D \left(D^2-5\right)\right) \kkun-528 \Bigr) \nonumber \\
                &{} +\frac{\pi  \epsilon}{2 \left(D^2-1\right)}  \Bigl( \cosfun^4 \left(-11 D^4+30 D^2-15\right) D-4 \cosfun^3 \left(D^4+6 D^2-3\right) \nonumber \\
                &{} \qquad +6 \cosfun^2 \left(3 D^4-2 D^2+3\right) D-4 \cosfun \left(3 D^4-2 D^2+3\right) \nonumber \\
                &{} \qquad -3 D^5+6 D^3+D \Bigr) \llun  \\ 
I{}_{1,-3}  &{} = \pi  \Bigl( 5 \cosfun^3 D^2-\frac{4}{3} \cosfun^3-\frac{1}{2} (\cosfun D-1) \left(5 \cosfun^2 D^2-3 \cosfun^2-4 \cosfun D-3 D^2+5\right) \llun \nonumber \\
                &{} \qquad -9 \cosfun^2 D-3 \cosfun D^2+8 \cosfun+3 D \Bigr) \nonumber \\ 
               &{} \qquad + \pi  \epsilon  \left(-\frac{1}{2} (\cosfun D-1) \left(\cosfun^2 \left(5 D^2-3\right)-4 \cosfun D-3 D^2+5\right) \kkun \right. \nonumber \\ 
               &{} \left. \qquad + \frac{1}{9} \Bigl( 117 \cosfun^3 D^2-32 \cosfun^3-\frac{27}{2} (\cosfun D-1) \left(\cosfun^2 D^2-\cosfun^2-D^2+1\right) \llun \right. \nonumber \\
               &{} \left. \qquad -189 \cosfun^2 D-81 \cosfun D^2+156 \cosfun+81 D\right) \Bigr)  \\
				  I{}_{2,-3}  &{} = \frac{3}{2} \pi  \left(\cosfun \left(\cosfun^2 \left(5 D^2-1\right)-6 \cosfun D-3 D^2+3\right)+2 D\right) \llun \nonumber \\
                &{} \qquad +\frac{\pi  \left(\cosfun^3 D \left(13-15 D^2\right)+6 \cosfun^2 \left(3 D^2-2\right)+3 \cosfun D \left(3 D^2-5\right)-6 D^2+8\right)}{D^2-1} \nonumber \\
                &{} \qquad + \frac{\pi \epsilon}{2} \Bigl( -58 \cosfun^3 D \nonumber \\ 
                &{} \qquad \qquad +3 \left(\cosfun \left(\cosfun^2 \left(5 D^2-1\right)-6 \cosfun D-3 D^2+3\right)+2 D\right) \kkun \nonumber \\
                &{} \qquad \qquad +48 \cosfun^2+42 \cosfun D-24\Bigr) \\
I{}_{-2,-2}  &{}= \frac{4 \pi}{15} \left( \cosfun^2+10 \cosfun D+10 D^2+3 \right) + \frac{4 \pi}{225} \epsilon  \left(46  \cosfun^2+400  \cosfun D+325 D^2+123 \right)  \\
I{}_{-1,-2}  &{}= \frac{4 \pi}{3} ( \cosfun+2 D) + \frac{\epsilon \pi}{9} \left(32 \cosfun+52 D\right)  \\
I{}_{1,-2}   &{} =  \pi \Bigl( D - 3   \cosfun^2 D+4  \cosfun-\frac{1}{2}  \ffun \llun \Bigr) \nonumber \\
                 &{} \qquad+ \epsilon \pi \left\{-\frac{1}{4}  \ggun \llun \log \left[\frac{1}{16} (D-1) (D+1)^3\right] 
                 \right. \nonumber \\  &{} \qquad \left. - \frac{1}{2} \llun \left(-\cosfun^2 D^2+ \cosfun^2+D^2-1 \right) \right. \nonumber \\
                 &{} \qquad\left. + \frac{1}{6} \pi ^2 \left(\cosfun^2 \left(3 D^2-1\right)-4 \cosfun D-D^2+3\right)
                 \right. \nonumber \\  &{} \qquad \left.  + (\cosfun (8-7 \cosfun D)+3 D) - \ggun \text{Li}_2\left[\frac{D-1}{D+1}\right]  \right\} \\
				  I{}_{2,-2}   &{} = \frac{\pi  \left(\cosfun^2 \left(6 D^2-4\right)-\left(D^2-1\right) (\cosfun (3 \cosfun D-2)-D) \llun-4 \cosfun D-2 D^2+4\right)}{D^2-1} \nonumber \\ 
                 &{} + \pi  \epsilon  \left(4 \left(2 \cosfun^2+\frac{D (\cosfun D-1)^2 \llun}{2 \left(D^2-1\right)}-1\right) \right. \nonumber \\ 
                 &{} \left. \qquad -(\cosfun (3 \cosfun D-2)-D) \kkun\right) \\
I{}_{-2,-1}  &{}= \frac{2 \pi}{3} \left(2  \cosfun D+3 D^2+ 1 \right) + \epsilon \pi \left(\frac{32 \cosfun D}{9}+4 D^2+\frac{16}{9}\right)  \\
I{}_{-1,-1}  &{}=  2 \pi \Bigl( \frac{\cosfun }{3}+D \Bigr) + \epsilon \pi \left(\frac{16  \cosfun}{9}+4  D\right)  \\
I{}_{-2,0}  &{}= 2 \pi  \left( D^2+\frac{1}{3} \right) +  4 \pi \epsilon \left( D^2+\frac{4}{9}\right) \\
I{}_{-1,0}  &{}= 2 \pi  D + 4 \pi  D \epsilon \\
I{}_{1,0}  &{}=  \pi  \llun  +  \epsilon \pi \kkun \\
I{}_{2,0}  &{}=  \frac{2 \pi }{D^2-1}  + \epsilon \frac{2 \pi  D \llun}{D^2-1} \\
I{}_{1,-1}   &{} = \pi \left(-(\cosfun D-1) \llun+2 \cosfun\right) \nonumber \\
                 &{} \qquad + \pi \epsilon \left\{ 4 \cosfun-\frac{\pi^2}{3} (\cosfun D-1)+\frac{1}{2} (\cosfun D-1) \llun \log \left[\frac{1}{16} (D-1) (D+1)^3\right] \right. \nonumber \\
                 &{} \qquad \left. + 2 (\cosfun D-1) \text{Li}_2\left[\frac{D-1}{D+1}\right]\right\}  \\
I{}_{2,-1}   &{} = -\frac{\pi  \left(-\cosfun \left(D^2-1\right) \llun+2 \cosfun D-2\right)}{D^2-1} \nonumber \\
                 &{} \qquad + \epsilon \pi \left\{ \frac{-\llun \left(\cosfun \left(D^2-1\right) \log \left[\frac{1}{16} (D-1) (D+1)^3\right]
                 +4 D (\cosfun D-1)\right)}{2 (D^2-1)} \right. \nonumber \\ 
                 &{} \left. \qquad + \frac{\pi ^2}{3} \cosfun  -2 \cosfun \text{Li}_2\left[\frac{D-1}{D+1}\right] \right\}  \\
I{}_{-2,1}   &{}= -\pi \frac{(\cosfun-D)^2}{\epsilon } + \pi \left(-3 \cosfun^2+4  \cosfun D+1 \right) +  \epsilon \pi \left(-7 \cosfun^2+8 \cosfun D+3 \right) \\
I{}_{-1,1}   &{}= \pi \frac{(\cosfun - D)}{\epsilon } + 2 \pi \cosfun +  4 \epsilon  \pi \cosfun  \\
I{}_{1,1}   &{} = \frac{\pi }{\epsilon  (\cosfun-D)} + \frac{\pi  \log \left[\frac{D^2-1}{(D-\cosfun)^2}\right]}{\cosfun-D} 
                      + \frac{2 \pi \epsilon}{(\cosfun-D)} \left\{\text{Li}_2\left(\frac{\cosfun-1}{D-1}\right)-\text{Li}_2\left(\frac{D-\cosfun}{D+1}\right) \right. \nonumber \\ 
                &{} \qquad \left. + \log (\cosfun+1) \log \left(\frac{D+1}{D-\cosfun}\right)+\log (D-\cosfun) \log \left(\frac{D-\cosfun}{D-1}\right) \right. \nonumber \\
                &{} \qquad \left. - \frac{1}{4} \llun \log \left((D-1) (D+1)^3\right)+ \frac{\pi^2}{6} \right\} \\
	       I{}_{2,1}
&{} = -\frac{\pi }{\epsilon }\frac{1}{(D-\cosfun)^2} 
+\frac{\pi  \left(\left(D^2-1\right) \log \left[\frac{(D-\cosfun)^2}{-1+D^2}\right]-2 \cosfun D+2\right)}{\left(D^2-1\right) (\cosfun-D)^2} \nonumber \\
&{} \qquad + \frac{2 \pi  \epsilon  \left(\text{Li}_2\left[\frac{D-\cosfun}{D+1}\right]-\text{Li}_2\left[\frac{\cosfun-1}{D-1}\right]\right)}{(\cosfun-D)^2} \nonumber \\
&{} \qquad-\frac{\pi  \epsilon}{6 \left(D^2-1\right) (\cosfun-D)^2}  \Bigl\{12 \log [D+1] \left(\left(D^2-1\right) \log [\cosfun+1]+\cosfun+D^2-D-1\right) \nonumber \\ 
&{} \qquad\left. -12 \log [D-1] \left(\left(D^2-1\right) \log [D-\cosfun]+\cosfun-D^2-D+1\right) \right. \nonumber \\
&{} \qquad \left.+\left(D^2-1\right) \left(2 \left(6 \log [D-\cosfun] \left(\log \left[\frac{D-\cosfun}{\cosfun+1}\right]-2\right)+\pi ^2\right) \right. \right. \nonumber \\
&{} \qquad \left. -3 \llun \log \left[(D-1) (D+1)^3\right]\right) \Bigr\} \\
I{}_{-2,2} &{}= \frac{\pi}{\epsilon } \left( 3 \cosfun^2 - 2 \cosfun D - 1\right) + \pi \left( 3 \cosfun^2 + 2 \cosfun D -D^2 -2  \right) \nonumber \\ 
               &{} \qquad +   \epsilon \pi \left( 9 \cosfun^2 - 2 \cosfun D+ D^2-4 \right)  \\
I{}_{-1,2} &{}=-\frac{\pi  \cosfun}{\epsilon } + \pi ( \cosfun- D) + \epsilon \pi (D- \cosfun) \\
I{}_{1,2}  &{}= \frac{\pi}{\epsilon} \frac{(1 - \cosfun D )}{ (\cosfun-D)^3} \nonumber \\ 
               &{} \qquad + \frac{\pi}{(\cosfun-D)^3}  \left(-\cosfun^2-2 \cosfun D +D^2+2 + (\cosfun D-1) \log \left[\frac{(D-\cosfun)^2}{-1+D^2}\right] \right)  \\
I{}_{2,2}  &{} = \frac{\pi}{\epsilon} \frac{\cosfun^2+2 \cosfun D-3}{ (\cosfun-D)^4} + \frac{\pi}{\left(D^2-1\right) (\cosfun-D)^4}  \Bigr( \cosfun^2 \left(7 D^2-5\right) \nonumber \\ 
               &{} \qquad +  \left(D^2-1\right) \left(\cosfun^2  + 2 \cosfun D-3\right) \log \left[\frac{D^2-1}{(D-\cosfun)^2}\right]  \nonumber \\
               &{} \qquad +  2 \left(D^2-3\right) \cosfun D-D^2 \left(D^2+5\right)+8 \Bigl) \, .
\end{align}
Certain order $\epsilon$ terms are not needed and may be dropped in calculations at $\order{\alpha_s^2}$. 

\bibliography{bibliography}

\providecommand{\noopsort}[1]{}
\begin{thebibliography}{77}%
\makeatletter
\providecommand \@ifxundefined [1]{%
 \@ifx{#1\undefined}
}%
\providecommand \@ifnum [1]{%
 \ifnum #1\expandafter \@firstoftwo
 \else \expandafter \@secondoftwo
 \fi
}%
\providecommand \@ifx [1]{%
 \ifx #1\expandafter \@firstoftwo
 \else \expandafter \@secondoftwo
 \fi
}%
\providecommand \natexlab [1]{#1}%
\providecommand \enquote  [1]{``#1''}%
\providecommand \bibnamefont  [1]{#1}%
\providecommand \bibfnamefont [1]{#1}%
\providecommand \citenamefont [1]{#1}%
\providecommand \href@noop [0]{\@secondoftwo}%
\providecommand \href [0]{\begingroup \@sanitize@url \@href}%
\providecommand \@href[1]{\@@startlink{#1}\@@href}%
\providecommand \@@href[1]{\endgroup#1\@@endlink}%
\providecommand \@sanitize@url [0]{\catcode `\\12\catcode `\$12\catcode
  `\&12\catcode `\#12\catcode `\^12\catcode `\_12\catcode `\%12\relax}%
\providecommand \@@startlink[1]{}%
\providecommand \@@endlink[0]{}%
\providecommand \url  [0]{\begingroup\@sanitize@url \@url }%
\providecommand \@url [1]{\endgroup\@href {#1}{\urlprefix }}%
\providecommand \urlprefix  [0]{URL }%
\providecommand \Eprint [0]{\href }%
\providecommand \doibase [0]{http://dx.doi.org/}%
\providecommand \selectlanguage [0]{\@gobble}%
\providecommand \bibinfo  [0]{\@secondoftwo}%
\providecommand \bibfield  [0]{\@secondoftwo}%
\providecommand \translation [1]{[#1]}%
\providecommand \BibitemOpen [0]{}%
\providecommand \bibitemStop [0]{}%
\providecommand \bibitemNoStop [0]{.\EOS\space}%
\providecommand \EOS [0]{\spacefactor3000\relax}%
\providecommand \BibitemShut  [1]{\csname bibitem#1\endcsname}%
\let\auto@bib@innerbib\@empty
\bibitem [{\citenamefont {Gonzalez-Hernandez}\ \emph
  {et~al.}(2018{\natexlab{a}})\citenamefont {Gonzalez-Hernandez}, \citenamefont
  {Rogers}, \citenamefont {Sato},\ and\ \citenamefont
  {Wang}}]{Gonzalez-Hernandez:2018ipj}%
  \BibitemOpen
  \bibfield  {author} {\bibinfo {author} {\bibfnamefont {J.~O.}\ \bibnamefont
  {Gonzalez-Hernandez}}, \bibinfo {author} {\bibfnamefont {T.~C.}\ \bibnamefont
  {Rogers}}, \bibinfo {author} {\bibfnamefont {N.}~\bibnamefont {Sato}}, \ and\
  \bibinfo {author} {\bibfnamefont {B.}~\bibnamefont {Wang}},\ }\href {\doibase
  10.1103/PhysRevD.98.114005} {\bibfield  {journal} {\bibinfo  {journal} {Phys.
  Rev.}\ }\textbf {\bibinfo {volume} {D98}},\ \bibinfo {pages} {114005}
  (\bibinfo {year} {2018}{\natexlab{a}})},\ \Eprint
  {http://arxiv.org/abs/1808.04396} {arXiv:1808.04396 [hep-ph]} \BibitemShut
  {NoStop}%
\bibitem [{\citenamefont {Daleo}\ \emph {et~al.}(2005)\citenamefont {Daleo},
  \citenamefont {de~Florian},\ and\ \citenamefont {Sassot}}]{Daleo:2004pn}%
  \BibitemOpen
  \bibfield  {author} {\bibinfo {author} {\bibfnamefont {A.}~\bibnamefont
  {Daleo}}, \bibinfo {author} {\bibfnamefont {D.}~\bibnamefont {de~Florian}}, \
  and\ \bibinfo {author} {\bibfnamefont {R.}~\bibnamefont {Sassot}},\ }\href
  {\doibase 10.1103/PhysRevD.71.034013} {\bibfield  {journal} {\bibinfo
  {journal} {Phys. Rev.}\ }\textbf {\bibinfo {volume} {D71}},\ \bibinfo {pages}
  {034013} (\bibinfo {year} {2005})},\ \Eprint
  {http://arxiv.org/abs/hep-ph/0411212} {arXiv:hep-ph/0411212 [hep-ph]}
  \BibitemShut {NoStop}%
\bibitem [{\citenamefont {Kniehl}\ \emph {et~al.}(2005)\citenamefont {Kniehl},
  \citenamefont {Kramer},\ and\ \citenamefont {Maniatis}}]{Kniehl:2004hf}%
  \BibitemOpen
  \bibfield  {author} {\bibinfo {author} {\bibfnamefont {B.~A.}\ \bibnamefont
  {Kniehl}}, \bibinfo {author} {\bibfnamefont {G.}~\bibnamefont {Kramer}}, \
  and\ \bibinfo {author} {\bibfnamefont {M.}~\bibnamefont {Maniatis}},\ }\href
  {\doibase 10.1016/j.nuclphysb.2005.01.031, 10.1016/j.nuclphysb.2005.05.017}
  {\bibfield  {journal} {\bibinfo  {journal} {Nucl. Phys.}\ }\textbf {\bibinfo
  {volume} {B711}},\ \bibinfo {pages} {345} (\bibinfo {year} {2005})},\
  \bibinfo {note} {[Erratum: Nucl. Phys.B720,231(2005)]},\ \Eprint
  {http://arxiv.org/abs/hep-ph/0411300} {arXiv:hep-ph/0411300 [hep-ph]}
  \BibitemShut {NoStop}%
\bibitem [{\citenamefont {Bacchetta}\ \emph {et~al.}(2019)\citenamefont
  {Bacchetta}, \citenamefont {Bozzi}, \citenamefont {Lambertsen}, \citenamefont
  {Piacenza}, \citenamefont {Steiglechner},\ and\ \citenamefont
  {Vogelsang}}]{Bacchetta:2019tcu}%
  \BibitemOpen
  \bibfield  {author} {\bibinfo {author} {\bibfnamefont {A.}~\bibnamefont
  {Bacchetta}}, \bibinfo {author} {\bibfnamefont {G.}~\bibnamefont {Bozzi}},
  \bibinfo {author} {\bibfnamefont {M.}~\bibnamefont {Lambertsen}}, \bibinfo
  {author} {\bibfnamefont {F.}~\bibnamefont {Piacenza}}, \bibinfo {author}
  {\bibfnamefont {J.}~\bibnamefont {Steiglechner}}, \ and\ \bibinfo {author}
  {\bibfnamefont {W.}~\bibnamefont {Vogelsang}},\ }\href@noop {} {\  (\bibinfo
  {year} {2019})},\ \Eprint {http://arxiv.org/abs/1901.06916} {arXiv:1901.06916
  [hep-ph]} \BibitemShut {NoStop}%
\bibitem [{\citenamefont {Berger}\ \emph {et~al.}(1998)\citenamefont {Berger},
  \citenamefont {Gordon},\ and\ \citenamefont {Klasen}}]{Berger:1998ev}%
  \BibitemOpen
  \bibfield  {author} {\bibinfo {author} {\bibfnamefont {E.~L.}\ \bibnamefont
  {Berger}}, \bibinfo {author} {\bibfnamefont {L.~E.}\ \bibnamefont {Gordon}},
  \ and\ \bibinfo {author} {\bibfnamefont {M.}~\bibnamefont {Klasen}},\ }\href
  {\doibase 10.1103/PhysRevD.58.074012} {\bibfield  {journal} {\bibinfo
  {journal} {Phys. Rev.}\ }\textbf {\bibinfo {volume} {D58}},\ \bibinfo {pages}
  {074012} (\bibinfo {year} {1998})},\ \Eprint
  {http://arxiv.org/abs/hep-ph/9803387} {arXiv:hep-ph/9803387 [hep-ph]}
  \BibitemShut {NoStop}%
\bibitem [{\citenamefont {Nogueira}(1993)}]{Nogueira:1991ex}%
  \BibitemOpen
  \bibfield  {author} {\bibinfo {author} {\bibfnamefont {P.}~\bibnamefont
  {Nogueira}},\ }\href {\doibase 10.1006/jcph.1993.1074} {\bibfield  {journal}
  {\bibinfo  {journal} {J. Comput. Phys.}\ }\textbf {\bibinfo {volume} {105}},\
  \bibinfo {pages} {279} (\bibinfo {year} {1993})}\BibitemShut {NoStop}%
\bibitem [{\citenamefont {Kuipers}\ \emph {et~al.}(2013)\citenamefont
  {Kuipers}, \citenamefont {Ueda}, \citenamefont {Vermaseren},\ and\
  \citenamefont {Vollinga}}]{Kuipers:2012rf}%
  \BibitemOpen
  \bibfield  {author} {\bibinfo {author} {\bibfnamefont {J.}~\bibnamefont
  {Kuipers}}, \bibinfo {author} {\bibfnamefont {T.}~\bibnamefont {Ueda}},
  \bibinfo {author} {\bibfnamefont {J.~A.~M.}\ \bibnamefont {Vermaseren}}, \
  and\ \bibinfo {author} {\bibfnamefont {J.}~\bibnamefont {Vollinga}},\ }\href
  {\doibase 10.1016/j.cpc.2012.12.028} {\bibfield  {journal} {\bibinfo
  {journal} {Comput. Phys. Commun.}\ }\textbf {\bibinfo {volume} {184}},\
  \bibinfo {pages} {1453} (\bibinfo {year} {2013})},\ \Eprint
  {http://arxiv.org/abs/1203.6543} {arXiv:1203.6543 [cs.SC]} \BibitemShut
  {NoStop}%
\bibitem [{\citenamefont {Kublbeck}\ \emph {et~al.}(1990)\citenamefont
  {Kublbeck}, \citenamefont {Bohm},\ and\ \citenamefont
  {Denner}}]{Kublbeck:1990xc}%
  \BibitemOpen
  \bibfield  {author} {\bibinfo {author} {\bibfnamefont {J.}~\bibnamefont
  {Kublbeck}}, \bibinfo {author} {\bibfnamefont {M.}~\bibnamefont {Bohm}}, \
  and\ \bibinfo {author} {\bibfnamefont {A.}~\bibnamefont {Denner}},\ }\href
  {\doibase 10.1016/0010-4655(90)90001-H} {\bibfield  {journal} {\bibinfo
  {journal} {Comput. Phys. Commun.}\ }\textbf {\bibinfo {volume} {60}},\
  \bibinfo {pages} {165} (\bibinfo {year} {1990})}\BibitemShut {NoStop}%
\bibitem [{\citenamefont {Hahn}(2001)}]{Hahn:2000kx}%
  \BibitemOpen
  \bibfield  {author} {\bibinfo {author} {\bibfnamefont {T.}~\bibnamefont
  {Hahn}},\ }\href {\doibase 10.1016/S0010-4655(01)00290-9} {\bibfield
  {journal} {\bibinfo  {journal} {Comput. Phys. Commun.}\ }\textbf {\bibinfo
  {volume} {140}},\ \bibinfo {pages} {418} (\bibinfo {year} {2001})},\ \Eprint
  {http://arxiv.org/abs/hep-ph/0012260} {arXiv:hep-ph/0012260 [hep-ph]}
  \BibitemShut {NoStop}%
\bibitem [{\citenamefont {Mertig}\ \emph {et~al.}(1991)\citenamefont {Mertig},
  \citenamefont {Bohm},\ and\ \citenamefont {Denner}}]{Mertig:1990an}%
  \BibitemOpen
  \bibfield  {author} {\bibinfo {author} {\bibfnamefont {R.}~\bibnamefont
  {Mertig}}, \bibinfo {author} {\bibfnamefont {M.}~\bibnamefont {Bohm}}, \ and\
  \bibinfo {author} {\bibfnamefont {A.}~\bibnamefont {Denner}},\ }\href
  {\doibase 10.1016/0010-4655(91)90130-D} {\bibfield  {journal} {\bibinfo
  {journal} {Comput. Phys. Commun.}\ }\textbf {\bibinfo {volume} {64}},\
  \bibinfo {pages} {345} (\bibinfo {year} {1991})}\BibitemShut {NoStop}%
\bibitem [{\citenamefont {Shtabovenko}\ \emph {et~al.}(2016)\citenamefont
  {Shtabovenko}, \citenamefont {Mertig},\ and\ \citenamefont
  {Orellana}}]{Shtabovenko:2016sxi}%
  \BibitemOpen
  \bibfield  {author} {\bibinfo {author} {\bibfnamefont {V.}~\bibnamefont
  {Shtabovenko}}, \bibinfo {author} {\bibfnamefont {R.}~\bibnamefont {Mertig}},
  \ and\ \bibinfo {author} {\bibfnamefont {F.}~\bibnamefont {Orellana}},\
  }\href {\doibase 10.1016/j.cpc.2016.06.008} {\bibfield  {journal} {\bibinfo
  {journal} {Comput. Phys. Commun.}\ }\textbf {\bibinfo {volume} {207}},\
  \bibinfo {pages} {432} (\bibinfo {year} {2016})},\ \Eprint
  {http://arxiv.org/abs/1601.01167} {arXiv:1601.01167 [hep-ph]} \BibitemShut
  {NoStop}%
\bibitem [{\citenamefont {Gonzalez-Hernandez}\ \emph
  {et~al.}(2018{\natexlab{b}})\citenamefont {Gonzalez-Hernandez}, \citenamefont
  {Rogers}, \citenamefont {Sato},\ and\ \citenamefont {Wang}}]{BigTMD}%
  \BibitemOpen
  \bibfield  {author} {\bibinfo {author} {\bibfnamefont {J.~O.}\ \bibnamefont
  {Gonzalez-Hernandez}}, \bibinfo {author} {\bibfnamefont {T.~C.}\ \bibnamefont
  {Rogers}}, \bibinfo {author} {\bibfnamefont {N.}~\bibnamefont {Sato}}, \ and\
  \bibinfo {author} {\bibfnamefont {B.}~\bibnamefont {Wang}},\ }\href
  {https://github.com/JeffersonLab/BigTMD} {\enquote {\bibinfo {title}
  {Jeffersonlab/bigtmd},}\ } (\bibinfo {year} {2018}{\natexlab{b}})\BibitemShut
  {NoStop}%
\bibitem [{\citenamefont {Meng}\ \emph {et~al.}(1992)\citenamefont {Meng},
  \citenamefont {Olness},\ and\ \citenamefont {Soper}}]{Meng:1991da}%
  \BibitemOpen
  \bibfield  {author} {\bibinfo {author} {\bibfnamefont {R.}~\bibnamefont
  {Meng}}, \bibinfo {author} {\bibfnamefont {F.~I.}\ \bibnamefont {Olness}}, \
  and\ \bibinfo {author} {\bibfnamefont {D.~E.}\ \bibnamefont {Soper}},\ }\href
  {\doibase 10.1016/0550-3213(92)90230-9} {\bibfield  {journal} {\bibinfo
  {journal} {Nucl. Phys.}\ }\textbf {\bibinfo {volume} {B371}},\ \bibinfo
  {pages} {79} (\bibinfo {year} {1992})}\BibitemShut {NoStop}%
\bibitem [{\citenamefont {Levelt}\ and\ \citenamefont
  {Mulders}(1994)}]{Levelt:1993ac}%
  \BibitemOpen
  \bibfield  {author} {\bibinfo {author} {\bibfnamefont {J.}~\bibnamefont
  {Levelt}}\ and\ \bibinfo {author} {\bibfnamefont {P.~J.}\ \bibnamefont
  {Mulders}},\ }\href {\doibase 10.1103/PhysRevD.49.96} {\bibfield  {journal}
  {\bibinfo  {journal} {Phys. Rev.}\ }\textbf {\bibinfo {volume} {D49}},\
  \bibinfo {pages} {96} (\bibinfo {year} {1994})},\ \Eprint
  {http://arxiv.org/abs/hep-ph/9304232} {arXiv:hep-ph/9304232 [hep-ph]}
  \BibitemShut {NoStop}%
\bibitem [{\citenamefont {Meng}\ \emph {et~al.}(1996)\citenamefont {Meng},
  \citenamefont {Olness},\ and\ \citenamefont {Soper}}]{Meng:1995yn}%
  \BibitemOpen
  \bibfield  {author} {\bibinfo {author} {\bibfnamefont {R.}~\bibnamefont
  {Meng}}, \bibinfo {author} {\bibfnamefont {F.~I.}\ \bibnamefont {Olness}}, \
  and\ \bibinfo {author} {\bibfnamefont {D.~E.}\ \bibnamefont {Soper}},\ }\href
  {\doibase 10.1103/PhysRevD.54.1919} {\bibfield  {journal} {\bibinfo
  {journal} {Phys. Rev.}\ }\textbf {\bibinfo {volume} {D54}},\ \bibinfo {pages}
  {1919} (\bibinfo {year} {1996})},\ \Eprint
  {http://arxiv.org/abs/hep-ph/9511311} {arXiv:hep-ph/9511311} \BibitemShut
  {NoStop}%
\bibitem [{\citenamefont {Mulders}\ and\ \citenamefont
  {Tangerman}(1996)}]{Mulders:1996dh}%
  \BibitemOpen
  \bibfield  {author} {\bibinfo {author} {\bibfnamefont {P.~J.}\ \bibnamefont
  {Mulders}}\ and\ \bibinfo {author} {\bibfnamefont {R.~D.}\ \bibnamefont
  {Tangerman}},\ }\href {\doibase 10.1016/0550-3213(95)00632-X} {\bibfield
  {journal} {\bibinfo  {journal} {Nucl. Phys.}\ }\textbf {\bibinfo {volume}
  {B461}},\ \bibinfo {pages} {197} (\bibinfo {year} {1996})},\ \Eprint
  {http://arxiv.org/abs/hep-ph/9510301} {arXiv:hep-ph/9510301 [hep-ph]}
  \BibitemShut {NoStop}%
\bibitem [{\citenamefont {Nadolsky}\ \emph {et~al.}(1999)\citenamefont
  {Nadolsky}, \citenamefont {Stump},\ and\ \citenamefont
  {Yuan}}]{Nadolsky:1999kb}%
  \BibitemOpen
  \bibfield  {author} {\bibinfo {author} {\bibfnamefont {P.}~\bibnamefont
  {Nadolsky}}, \bibinfo {author} {\bibfnamefont {D.~R.}\ \bibnamefont {Stump}},
  \ and\ \bibinfo {author} {\bibfnamefont {C.~P.}\ \bibnamefont {Yuan}},\
  }\href {\doibase 10.1103/PhysRevD.64.059903, 10.1103/PhysRevD.61.014003
  10.1103/PhysRevD.64.059903, 10.1103/PhysRevD.61.014003} {\bibfield  {journal}
  {\bibinfo  {journal} {Phys. Rev.}\ }\textbf {\bibinfo {volume} {D61}},\
  \bibinfo {pages} {014003} (\bibinfo {year} {1999})},\ \Eprint
  {http://arxiv.org/abs/hep-ph/9906280} {arXiv:hep-ph/9906280 [hep-ph]}
  \BibitemShut {NoStop}%
\bibitem [{\citenamefont {Nadolsky}\ \emph {et~al.}(2001)\citenamefont
  {Nadolsky}, \citenamefont {Stump},\ and\ \citenamefont
  {Yuan}}]{Nadolsky:2000ky}%
  \BibitemOpen
  \bibfield  {author} {\bibinfo {author} {\bibfnamefont {P.~M.}\ \bibnamefont
  {Nadolsky}}, \bibinfo {author} {\bibfnamefont {D.~R.}\ \bibnamefont {Stump}},
  \ and\ \bibinfo {author} {\bibfnamefont {C.~P.}\ \bibnamefont {Yuan}},\
  }\href@noop {} {\bibfield  {journal} {\bibinfo  {journal} {Phys. Rev.}\
  }\textbf {\bibinfo {volume} {D64}},\ \bibinfo {pages} {114011} (\bibinfo
  {year} {2001})},\ \Eprint {http://arxiv.org/abs/hep-ph/0012261}
  {hep-ph/0012261} \BibitemShut {NoStop}%
\bibitem [{\citenamefont {Barone}\ \emph {et~al.}(2002)\citenamefont {Barone},
  \citenamefont {Drago},\ and\ \citenamefont {Ratcliffe}}]{Barone:2001sp}%
  \BibitemOpen
  \bibfield  {author} {\bibinfo {author} {\bibfnamefont {V.}~\bibnamefont
  {Barone}}, \bibinfo {author} {\bibfnamefont {A.}~\bibnamefont {Drago}}, \
  and\ \bibinfo {author} {\bibfnamefont {P.~G.}\ \bibnamefont {Ratcliffe}},\
  }\href@noop {} {\bibfield  {journal} {\bibinfo  {journal} {Phys. Rept.}\
  }\textbf {\bibinfo {volume} {359}},\ \bibinfo {pages} {1} (\bibinfo {year}
  {2002})},\ \Eprint {http://arxiv.org/abs/hep-ph/0104283} {hep-ph/0104283}
  \BibitemShut {NoStop}%
\bibitem [{\citenamefont {Ji}\ \emph {et~al.}(2004)\citenamefont {Ji},
  \citenamefont {Ma},\ and\ \citenamefont {Yuan}}]{Ji:2004xq}%
  \BibitemOpen
  \bibfield  {author} {\bibinfo {author} {\bibfnamefont {X.-D.}\ \bibnamefont
  {Ji}}, \bibinfo {author} {\bibfnamefont {J.-P.}\ \bibnamefont {Ma}}, \ and\
  \bibinfo {author} {\bibfnamefont {F.}~\bibnamefont {Yuan}},\ }\href {\doibase
  10.1016/j.physletb.2004.07.026} {\bibfield  {journal} {\bibinfo  {journal}
  {Phys. Lett.}\ }\textbf {\bibinfo {volume} {B597}},\ \bibinfo {pages} {299}
  (\bibinfo {year} {2004})},\ \Eprint {http://arxiv.org/abs/hep-ph/0405085}
  {arXiv:hep-ph/0405085 [hep-ph]} \BibitemShut {NoStop}%
\bibitem [{\citenamefont {Bacchetta}\ \emph {et~al.}(2004)\citenamefont
  {Bacchetta}, \citenamefont {D'Alesio}, \citenamefont {Diehl},\ and\
  \citenamefont {Miller}}]{Bacchetta:2004jz}%
  \BibitemOpen
  \bibfield  {author} {\bibinfo {author} {\bibfnamefont {A.}~\bibnamefont
  {Bacchetta}}, \bibinfo {author} {\bibfnamefont {U.}~\bibnamefont {D'Alesio}},
  \bibinfo {author} {\bibfnamefont {M.}~\bibnamefont {Diehl}}, \ and\ \bibinfo
  {author} {\bibfnamefont {C.~A.}\ \bibnamefont {Miller}},\ }\href@noop {}
  {\bibfield  {journal} {\bibinfo  {journal} {Phys. Rev.}\ }\textbf {\bibinfo
  {volume} {D70}},\ \bibinfo {pages} {117504} (\bibinfo {year} {2004})},\
  \Eprint {http://arxiv.org/abs/hep-ph/0410050} {hep-ph/0410050} \BibitemShut
  {NoStop}%
\bibitem [{\citenamefont {Koike}\ \emph {et~al.}(2006)\citenamefont {Koike},
  \citenamefont {Nagashima},\ and\ \citenamefont {Vogelsang}}]{Koike:2006fn}%
  \BibitemOpen
  \bibfield  {author} {\bibinfo {author} {\bibfnamefont {Y.}~\bibnamefont
  {Koike}}, \bibinfo {author} {\bibfnamefont {J.}~\bibnamefont {Nagashima}}, \
  and\ \bibinfo {author} {\bibfnamefont {W.}~\bibnamefont {Vogelsang}},\ }\href
  {\doibase 10.1016/j.nuclphysb.2006.03.009} {\bibfield  {journal} {\bibinfo
  {journal} {Nucl. Phys.}\ }\textbf {\bibinfo {volume} {B744}},\ \bibinfo
  {pages} {59} (\bibinfo {year} {2006})},\ \Eprint
  {http://arxiv.org/abs/hep-ph/0602188} {arXiv:hep-ph/0602188 [hep-ph]}
  \BibitemShut {NoStop}%
\bibitem [{\citenamefont {Bacchetta}\ \emph {et~al.}(2007)\citenamefont
  {Bacchetta} \emph {et~al.}}]{Bacchetta:2006tn}%
  \BibitemOpen
  \bibfield  {author} {\bibinfo {author} {\bibfnamefont {A.}~\bibnamefont
  {Bacchetta}} \emph {et~al.},\ }\href {\doibase 10.1088/1126-6708/2007/02/093}
  {\bibfield  {journal} {\bibinfo  {journal} {JHEP}\ }\textbf {\bibinfo
  {volume} {02}},\ \bibinfo {pages} {093} (\bibinfo {year} {2007})},\ \Eprint
  {http://arxiv.org/abs/hep-ph/0611265} {arXiv:hep-ph/0611265} \BibitemShut
  {NoStop}%
\bibitem [{\citenamefont {Bacchetta}\ \emph {et~al.}(2008)\citenamefont
  {Bacchetta}, \citenamefont {Boer}, \citenamefont {Diehl},\ and\ \citenamefont
  {Mulders}}]{Bacchetta:2008xw}%
  \BibitemOpen
  \bibfield  {author} {\bibinfo {author} {\bibfnamefont {A.}~\bibnamefont
  {Bacchetta}}, \bibinfo {author} {\bibfnamefont {D.}~\bibnamefont {Boer}},
  \bibinfo {author} {\bibfnamefont {M.}~\bibnamefont {Diehl}}, \ and\ \bibinfo
  {author} {\bibfnamefont {P.~J.}\ \bibnamefont {Mulders}},\ }\href {\doibase
  10.1088/1126-6708/2008/08/023} {\bibfield  {journal} {\bibinfo  {journal}
  {JHEP}\ }\textbf {\bibinfo {volume} {08}},\ \bibinfo {pages} {023} (\bibinfo
  {year} {2008})},\ \Eprint {http://arxiv.org/abs/0803.0227} {arXiv:0803.0227
  [hep-ph]} \BibitemShut {NoStop}%
\bibitem [{\citenamefont {Moffat}\ \emph {et~al.}(2019)\citenamefont {Moffat},
  \citenamefont {Rogers}, \citenamefont {Melnitchouk}, \citenamefont {Sato},\
  and\ \citenamefont {Steffens}}]{Moffat:2019qll}%
  \BibitemOpen
  \bibfield  {author} {\bibinfo {author} {\bibfnamefont {E.}~\bibnamefont
  {Moffat}}, \bibinfo {author} {\bibfnamefont {T.~C.}\ \bibnamefont {Rogers}},
  \bibinfo {author} {\bibfnamefont {W.}~\bibnamefont {Melnitchouk}}, \bibinfo
  {author} {\bibfnamefont {N.}~\bibnamefont {Sato}}, \ and\ \bibinfo {author}
  {\bibfnamefont {F.}~\bibnamefont {Steffens}},\ }\href@noop {} {\  (\bibinfo
  {year} {2019})},\ \Eprint {http://arxiv.org/abs/1901.09016} {arXiv:1901.09016
  [hep-ph]} \BibitemShut {NoStop}%
\bibitem [{\citenamefont {Collins}(2011)}]{Collins:2011qcdbook}%
  \BibitemOpen
  \bibfield  {author} {\bibinfo {author} {\bibfnamefont {J.~C.}\ \bibnamefont
  {Collins}},\ }\href {\doibase 10.1017/CBO9780511975592} {\emph {\bibinfo
  {title} {Foundations of Perturbative QCD}}}\ (\bibinfo  {publisher}
  {Cambridge University Press},\ \bibinfo {address} {Cambridge},\ \bibinfo
  {year} {2011})\BibitemShut {NoStop}%
\bibitem [{\citenamefont {Arnold}\ and\ \citenamefont
  {Reno}(1989)}]{Arnold:1988dp}%
  \BibitemOpen
  \bibfield  {author} {\bibinfo {author} {\bibfnamefont {P.~B.}\ \bibnamefont
  {Arnold}}\ and\ \bibinfo {author} {\bibfnamefont {M.~H.}\ \bibnamefont
  {Reno}},\ }\href {\doibase 10.1016/0550-3213(90)90311-Z,
  10.1016/0550-3213(89)90600-7} {\bibfield  {journal} {\bibinfo  {journal}
  {Nucl. Phys.}\ }\textbf {\bibinfo {volume} {B319}},\ \bibinfo {pages} {37}
  (\bibinfo {year} {1989})},\ \bibinfo {note} {[Erratum: Nucl.
  Phys.B330,284(1990)]}\BibitemShut {NoStop}%
\bibitem [{\citenamefont {Aversa}\ \emph {et~al.}(1989)\citenamefont {Aversa},
  \citenamefont {Chiappetta}, \citenamefont {Greco},\ and\ \citenamefont
  {Guillet}}]{Aversa:1988vb}%
  \BibitemOpen
  \bibfield  {author} {\bibinfo {author} {\bibfnamefont {F.}~\bibnamefont
  {Aversa}}, \bibinfo {author} {\bibfnamefont {P.}~\bibnamefont {Chiappetta}},
  \bibinfo {author} {\bibfnamefont {M.}~\bibnamefont {Greco}}, \ and\ \bibinfo
  {author} {\bibfnamefont {J.~P.}\ \bibnamefont {Guillet}},\ }\href {\doibase
  10.1016/0550-3213(89)90288-5} {\bibfield  {journal} {\bibinfo  {journal}
  {Nucl. Phys.}\ }\textbf {\bibinfo {volume} {B327}},\ \bibinfo {pages} {105}
  (\bibinfo {year} {1989})}\BibitemShut {NoStop}%
\bibitem [{\citenamefont {Gordon}\ and\ \citenamefont
  {Vogelsang}(1993)}]{Gordon:1993qc}%
  \BibitemOpen
  \bibfield  {author} {\bibinfo {author} {\bibfnamefont {L.~E.}\ \bibnamefont
  {Gordon}}\ and\ \bibinfo {author} {\bibfnamefont {W.}~\bibnamefont
  {Vogelsang}},\ }\href {\doibase 10.1103/PhysRevD.48.3136} {\bibfield
  {journal} {\bibinfo  {journal} {Phys. Rev.}\ }\textbf {\bibinfo {volume}
  {D48}},\ \bibinfo {pages} {3136} (\bibinfo {year} {1993})}\BibitemShut
  {NoStop}%
\bibitem [{\citenamefont {Ellis}\ \emph {et~al.}(1980)\citenamefont {Ellis},
  \citenamefont {Furman}, \citenamefont {Haber},\ and\ \citenamefont
  {Hinchliffe}}]{Ellis:1979sj}%
  \BibitemOpen
  \bibfield  {author} {\bibinfo {author} {\bibfnamefont {R.~K.}\ \bibnamefont
  {Ellis}}, \bibinfo {author} {\bibfnamefont {M.~A.}\ \bibnamefont {Furman}},
  \bibinfo {author} {\bibfnamefont {H.~E.}\ \bibnamefont {Haber}}, \ and\
  \bibinfo {author} {\bibfnamefont {I.}~\bibnamefont {Hinchliffe}},\ }\href
  {\doibase 10.1016/0550-3213(80)90010-3} {\bibfield  {journal} {\bibinfo
  {journal} {Nucl. Phys.}\ }\textbf {\bibinfo {volume} {B173}},\ \bibinfo
  {pages} {397} (\bibinfo {year} {1980})}\BibitemShut {NoStop}%
\bibitem [{\citenamefont {Ellis}\ \emph {et~al.}(1983)\citenamefont {Ellis},
  \citenamefont {Martinelli},\ and\ \citenamefont {Petronzio}}]{Ellis:1981hk}%
  \BibitemOpen
  \bibfield  {author} {\bibinfo {author} {\bibfnamefont {R.~K.}\ \bibnamefont
  {Ellis}}, \bibinfo {author} {\bibfnamefont {G.}~\bibnamefont {Martinelli}}, \
  and\ \bibinfo {author} {\bibfnamefont {R.}~\bibnamefont {Petronzio}},\ }\href
  {\doibase 10.1016/0550-3213(83)90188-8} {\bibfield  {journal} {\bibinfo
  {journal} {Nucl. Phys.}\ }\textbf {\bibinfo {volume} {B211}},\ \bibinfo
  {pages} {106} (\bibinfo {year} {1983})}\BibitemShut {NoStop}%
\bibitem [{\citenamefont {Altarelli}\ and\ \citenamefont
  {Parisi}(1977)}]{Altarelli:1977zs}%
  \BibitemOpen
  \bibfield  {author} {\bibinfo {author} {\bibfnamefont {G.}~\bibnamefont
  {Altarelli}}\ and\ \bibinfo {author} {\bibfnamefont {G.}~\bibnamefont
  {Parisi}},\ }\href@noop {} {\bibfield  {journal} {\bibinfo  {journal} {Nucl.
  Phys.}\ }\textbf {\bibinfo {volume} {B126}},\ \bibinfo {pages} {298}
  (\bibinfo {year} {1977})}\BibitemShut {NoStop}%
\bibitem [{\citenamefont {Anderle}\ \emph {et~al.}(2013)\citenamefont
  {Anderle}, \citenamefont {Ringer},\ and\ \citenamefont
  {Vogelsang}}]{Anderle:2012rq}%
  \BibitemOpen
  \bibfield  {author} {\bibinfo {author} {\bibfnamefont {D.~P.}\ \bibnamefont
  {Anderle}}, \bibinfo {author} {\bibfnamefont {F.}~\bibnamefont {Ringer}}, \
  and\ \bibinfo {author} {\bibfnamefont {W.}~\bibnamefont {Vogelsang}},\ }\href
  {\doibase 10.1103/PhysRevD.87.034014} {\bibfield  {journal} {\bibinfo
  {journal} {Phys. Rev.}\ }\textbf {\bibinfo {volume} {D87}},\ \bibinfo {pages}
  {034014} (\bibinfo {year} {2013})},\ \Eprint {http://arxiv.org/abs/1212.2099}
  {arXiv:1212.2099 [hep-ph]} \BibitemShut {NoStop}%
\bibitem [{\citenamefont {Aghasyan}\ \emph {et~al.}(2018)\citenamefont
  {Aghasyan} \emph {et~al.}}]{Aghasyan:2017ctw}%
  \BibitemOpen
  \bibfield  {author} {\bibinfo {author} {\bibfnamefont {M.}~\bibnamefont
  {Aghasyan}} \emph {et~al.} (\bibinfo {collaboration} {COMPASS}),\ }\href
  {\doibase 10.1103/PhysRevD.97.032006} {\bibfield  {journal} {\bibinfo
  {journal} {Phys. Rev.}\ }\textbf {\bibinfo {volume} {D97}},\ \bibinfo {pages}
  {032006} (\bibinfo {year} {2018})},\ \Eprint
  {http://arxiv.org/abs/1709.07374} {arXiv:1709.07374 [hep-ex]} \BibitemShut
  {NoStop}%
\bibitem [{\citenamefont {Accardi}\ \emph {et~al.}(2016)\citenamefont
  {Accardi}, \citenamefont {Brady}, \citenamefont {Melnitchouk}, \citenamefont
  {Owens},\ and\ \citenamefont {Sato}}]{Accardi:2016qay}%
  \BibitemOpen
  \bibfield  {author} {\bibinfo {author} {\bibfnamefont {A.}~\bibnamefont
  {Accardi}}, \bibinfo {author} {\bibfnamefont {L.~T.}\ \bibnamefont {Brady}},
  \bibinfo {author} {\bibfnamefont {W.}~\bibnamefont {Melnitchouk}}, \bibinfo
  {author} {\bibfnamefont {J.~F.}\ \bibnamefont {Owens}}, \ and\ \bibinfo
  {author} {\bibfnamefont {N.}~\bibnamefont {Sato}},\ }\href {\doibase
  10.1103/PhysRevD.93.114017} {\bibfield  {journal} {\bibinfo  {journal} {Phys.
  Rev.}\ }\textbf {\bibinfo {volume} {D93}},\ \bibinfo {pages} {114017}
  (\bibinfo {year} {2016})},\ \Eprint {http://arxiv.org/abs/1602.03154}
  {arXiv:1602.03154 [hep-ph]} \BibitemShut {NoStop}%
\bibitem [{\citenamefont {de~Florian}\ \emph {et~al.}(2007)\citenamefont
  {de~Florian}, \citenamefont {Sassot},\ and\ \citenamefont
  {Stratmann}}]{deFlorian:2007ekg}%
  \BibitemOpen
  \bibfield  {author} {\bibinfo {author} {\bibfnamefont {D.}~\bibnamefont
  {de~Florian}}, \bibinfo {author} {\bibfnamefont {R.}~\bibnamefont {Sassot}},
  \ and\ \bibinfo {author} {\bibfnamefont {M.}~\bibnamefont {Stratmann}},\
  }\href {\doibase 10.1103/PhysRevD.76.074033} {\bibfield  {journal} {\bibinfo
  {journal} {Phys. Rev.}\ }\textbf {\bibinfo {volume} {D76}},\ \bibinfo {pages}
  {074033} (\bibinfo {year} {2007})},\ \Eprint {http://arxiv.org/abs/0707.1506}
  {arXiv:0707.1506 [hep-ph]} \BibitemShut {NoStop}%
\bibitem [{\citenamefont {Balitsky}\ and\ \citenamefont
  {Tarasov}(2018)}]{Balitsky:2017gis}%
  \BibitemOpen
  \bibfield  {author} {\bibinfo {author} {\bibfnamefont {I.}~\bibnamefont
  {Balitsky}}\ and\ \bibinfo {author} {\bibfnamefont {A.}~\bibnamefont
  {Tarasov}},\ }\href {\doibase 10.1007/JHEP05(2018)150} {\bibfield  {journal}
  {\bibinfo  {journal} {JHEP}\ }\textbf {\bibinfo {volume} {05}},\ \bibinfo
  {pages} {150} (\bibinfo {year} {2018})},\ \Eprint
  {http://arxiv.org/abs/1712.09389} {arXiv:1712.09389 [hep-ph]} \BibitemShut
  {NoStop}%
\bibitem [{\citenamefont {Graudenz}(1997)}]{Graudenz:1996an}%
  \BibitemOpen
  \bibfield  {author} {\bibinfo {author} {\bibfnamefont {D.}~\bibnamefont
  {Graudenz}},\ }\href {\doibase 10.1016/S0370-2693(97)00667-9} {\bibfield
  {journal} {\bibinfo  {journal} {Phys. Lett.}\ }\textbf {\bibinfo {volume}
  {B406}},\ \bibinfo {pages} {178} (\bibinfo {year} {1997})},\ \Eprint
  {http://arxiv.org/abs/hep-ph/9606470} {arXiv:hep-ph/9606470 [hep-ph]}
  \BibitemShut {NoStop}%
\bibitem [{\citenamefont {Anselmino}\ \emph {et~al.}(2007)\citenamefont
  {Anselmino}, \citenamefont {Boglione}, \citenamefont {Prokudin},\ and\
  \citenamefont {Turk}}]{Anselmino:2006rv}%
  \BibitemOpen
  \bibfield  {author} {\bibinfo {author} {\bibfnamefont {M.}~\bibnamefont
  {Anselmino}}, \bibinfo {author} {\bibfnamefont {M.}~\bibnamefont {Boglione}},
  \bibinfo {author} {\bibfnamefont {A.}~\bibnamefont {Prokudin}}, \ and\
  \bibinfo {author} {\bibfnamefont {C.}~\bibnamefont {Turk}},\ }\href {\doibase
  10.1140/epja/i2007-10003-9} {\bibfield  {journal} {\bibinfo  {journal} {Eur.
  Phys. J.}\ }\textbf {\bibinfo {volume} {A31}},\ \bibinfo {pages} {373}
  (\bibinfo {year} {2007})},\ \Eprint {http://arxiv.org/abs/hep-ph/0606286}
  {arXiv:hep-ph/0606286 [hep-ph]} \BibitemShut {NoStop}%
\bibitem [{\citenamefont {Anselmino}\ \emph {et~al.}(2014)\citenamefont
  {Anselmino}, \citenamefont {Boglione}, \citenamefont {Gonzalez~Hernandez},
  \citenamefont {Melis},\ and\ \citenamefont {Prokudin}}]{Anselmino:2013lza}%
  \BibitemOpen
  \bibfield  {author} {\bibinfo {author} {\bibfnamefont {M.}~\bibnamefont
  {Anselmino}}, \bibinfo {author} {\bibfnamefont {M.}~\bibnamefont {Boglione}},
  \bibinfo {author} {\bibfnamefont {J.~O.}\ \bibnamefont {Gonzalez~Hernandez}},
  \bibinfo {author} {\bibfnamefont {S.}~\bibnamefont {Melis}}, \ and\ \bibinfo
  {author} {\bibfnamefont {A.}~\bibnamefont {Prokudin}},\ }\href {\doibase
  10.1007/JHEP04(2014)005} {\bibfield  {journal} {\bibinfo  {journal} {JHEP}\
  }\textbf {\bibinfo {volume} {04}},\ \bibinfo {pages} {005} (\bibinfo {year}
  {2014})},\ \Eprint {http://arxiv.org/abs/1312.6261} {arXiv:1312.6261
  [hep-ph]} \BibitemShut {NoStop}%
\bibitem [{\citenamefont {Signori}\ \emph {et~al.}(2013)\citenamefont
  {Signori}, \citenamefont {Bacchetta}, \citenamefont {Radici},\ and\
  \citenamefont {Schnell}}]{Signori:2013mda}%
  \BibitemOpen
  \bibfield  {author} {\bibinfo {author} {\bibfnamefont {A.}~\bibnamefont
  {Signori}}, \bibinfo {author} {\bibfnamefont {A.}~\bibnamefont {Bacchetta}},
  \bibinfo {author} {\bibfnamefont {M.}~\bibnamefont {Radici}}, \ and\ \bibinfo
  {author} {\bibfnamefont {G.}~\bibnamefont {Schnell}},\ }\href {\doibase
  10.1007/JHEP11(2013)194} {\bibfield  {journal} {\bibinfo  {journal} {JHEP}\
  }\textbf {\bibinfo {volume} {1311}},\ \bibinfo {pages} {194} (\bibinfo {year}
  {2013})},\ \Eprint {http://arxiv.org/abs/1309.3507} {arXiv:1309.3507
  [hep-ph]} \BibitemShut {NoStop}%
\bibitem [{\citenamefont {Sun}\ and\ \citenamefont {Yuan}(2013)}]{Sun:2013hua}%
  \BibitemOpen
  \bibfield  {author} {\bibinfo {author} {\bibfnamefont {P.}~\bibnamefont
  {Sun}}\ and\ \bibinfo {author} {\bibfnamefont {F.}~\bibnamefont {Yuan}},\
  }\href {\doibase 10.1103/PhysRevD.88.114012} {\bibfield  {journal} {\bibinfo
  {journal} {Phys. Rev.}\ }\textbf {\bibinfo {volume} {D88}},\ \bibinfo {pages}
  {114012} (\bibinfo {year} {2013})},\ \Eprint {http://arxiv.org/abs/1308.5003}
  {arXiv:1308.5003 [hep-ph]} \BibitemShut {NoStop}%
\bibitem [{\citenamefont {Sun}\ \emph {et~al.}(2014)\citenamefont {Sun},
  \citenamefont {Isaacson}, \citenamefont {Yuan},\ and\ \citenamefont
  {Yuan}}]{Su:2014wpa}%
  \BibitemOpen
  \bibfield  {author} {\bibinfo {author} {\bibfnamefont {P.}~\bibnamefont
  {Sun}}, \bibinfo {author} {\bibfnamefont {J.}~\bibnamefont {Isaacson}},
  \bibinfo {author} {\bibfnamefont {C.~P.}\ \bibnamefont {Yuan}}, \ and\
  \bibinfo {author} {\bibfnamefont {F.}~\bibnamefont {Yuan}},\ }\href@noop {}
  {\  (\bibinfo {year} {2014})},\ \Eprint {http://arxiv.org/abs/1406.3073}
  {arXiv:1406.3073 [hep-ph]} \BibitemShut {NoStop}%
\bibitem [{\citenamefont {Bacchetta}\ \emph {et~al.}(2017)\citenamefont
  {Bacchetta}, \citenamefont {Delcarro}, \citenamefont {Pisano}, \citenamefont
  {Radici},\ and\ \citenamefont {Signori}}]{Bacchetta:2017gcc}%
  \BibitemOpen
  \bibfield  {author} {\bibinfo {author} {\bibfnamefont {A.}~\bibnamefont
  {Bacchetta}}, \bibinfo {author} {\bibfnamefont {F.}~\bibnamefont {Delcarro}},
  \bibinfo {author} {\bibfnamefont {C.}~\bibnamefont {Pisano}}, \bibinfo
  {author} {\bibfnamefont {M.}~\bibnamefont {Radici}}, \ and\ \bibinfo {author}
  {\bibfnamefont {A.}~\bibnamefont {Signori}},\ }\href@noop {} {\  (\bibinfo
  {year} {2017})},\ \Eprint {http://arxiv.org/abs/1703.10157} {arXiv:1703.10157
  [hep-ph]} \BibitemShut {NoStop}%
\bibitem [{\citenamefont {Anselmino}\ \emph {et~al.}(2017)\citenamefont
  {Anselmino}, \citenamefont {Boglione}, \citenamefont {D'Alesio},
  \citenamefont {Murgia},\ and\ \citenamefont {Prokudin}}]{Anselmino:2016uie}%
  \BibitemOpen
  \bibfield  {author} {\bibinfo {author} {\bibfnamefont {M.}~\bibnamefont
  {Anselmino}}, \bibinfo {author} {\bibfnamefont {M.}~\bibnamefont {Boglione}},
  \bibinfo {author} {\bibfnamefont {U.}~\bibnamefont {D'Alesio}}, \bibinfo
  {author} {\bibfnamefont {F.}~\bibnamefont {Murgia}}, \ and\ \bibinfo {author}
  {\bibfnamefont {A.}~\bibnamefont {Prokudin}},\ }\href {\doibase
  10.1007/JHEP04(2017)046} {\bibfield  {journal} {\bibinfo  {journal} {JHEP}\
  }\textbf {\bibinfo {volume} {04}},\ \bibinfo {pages} {046} (\bibinfo {year}
  {2017})},\ \Eprint {http://arxiv.org/abs/1612.06413} {arXiv:1612.06413
  [hep-ph]} \BibitemShut {NoStop}%
\bibitem [{\citenamefont {Anselmino}\ \emph {et~al.}(2015)\citenamefont
  {Anselmino}, \citenamefont {Boglione}, \citenamefont {D'Alesio},
  \citenamefont {Gonzalez~Hernandez}, \citenamefont {Melis}, \citenamefont
  {Murgia},\ and\ \citenamefont {Prokudin}}]{Anselmino:2015sxa}%
  \BibitemOpen
  \bibfield  {author} {\bibinfo {author} {\bibfnamefont {M.}~\bibnamefont
  {Anselmino}}, \bibinfo {author} {\bibfnamefont {M.}~\bibnamefont {Boglione}},
  \bibinfo {author} {\bibfnamefont {U.}~\bibnamefont {D'Alesio}}, \bibinfo
  {author} {\bibfnamefont {J.~O.}\ \bibnamefont {Gonzalez~Hernandez}}, \bibinfo
  {author} {\bibfnamefont {S.}~\bibnamefont {Melis}}, \bibinfo {author}
  {\bibfnamefont {F.}~\bibnamefont {Murgia}}, \ and\ \bibinfo {author}
  {\bibfnamefont {A.}~\bibnamefont {Prokudin}},\ }\href {\doibase
  10.1103/PhysRevD.92.114023} {\bibfield  {journal} {\bibinfo  {journal} {Phys.
  Rev.}\ }\textbf {\bibinfo {volume} {D92}},\ \bibinfo {pages} {114023}
  (\bibinfo {year} {2015})},\ \Eprint {http://arxiv.org/abs/1510.05389}
  {arXiv:1510.05389 [hep-ph]} \BibitemShut {NoStop}%
\bibitem [{\citenamefont {Bacchetta}\ \emph {et~al.}(2015)\citenamefont
  {Bacchetta}, \citenamefont {Echevarria}, \citenamefont {Mulders},
  \citenamefont {Radici},\ and\ \citenamefont {Signori}}]{Bacchetta:2015ora}%
  \BibitemOpen
  \bibfield  {author} {\bibinfo {author} {\bibfnamefont {A.}~\bibnamefont
  {Bacchetta}}, \bibinfo {author} {\bibfnamefont {M.~G.}\ \bibnamefont
  {Echevarria}}, \bibinfo {author} {\bibfnamefont {P.~J.~G.}\ \bibnamefont
  {Mulders}}, \bibinfo {author} {\bibfnamefont {M.}~\bibnamefont {Radici}}, \
  and\ \bibinfo {author} {\bibfnamefont {A.}~\bibnamefont {Signori}},\ }\href
  {\doibase 10.1007/JHEP11(2015)076} {\bibfield  {journal} {\bibinfo  {journal}
  {JHEP}\ }\textbf {\bibinfo {volume} {11}},\ \bibinfo {pages} {076} (\bibinfo
  {year} {2015})},\ \Eprint {http://arxiv.org/abs/1508.00402} {arXiv:1508.00402
  [hep-ph]} \BibitemShut {NoStop}%
\bibitem [{\citenamefont {Scimemi}\ and\ \citenamefont
  {Vladimirov}(2018)}]{Scimemi:2017etj}%
  \BibitemOpen
  \bibfield  {author} {\bibinfo {author} {\bibfnamefont {I.}~\bibnamefont
  {Scimemi}}\ and\ \bibinfo {author} {\bibfnamefont {A.}~\bibnamefont
  {Vladimirov}},\ }\href {\doibase 10.1140/epjc/s10052-018-5557-y} {\bibfield
  {journal} {\bibinfo  {journal} {Eur. Phys. J.}\ }\textbf {\bibinfo {volume}
  {C78}},\ \bibinfo {pages} {89} (\bibinfo {year} {2018})},\ \Eprint
  {http://arxiv.org/abs/1706.01473} {arXiv:1706.01473 [hep-ph]} \BibitemShut
  {NoStop}%
\bibitem [{\citenamefont {Kang}\ \emph {et~al.}(2017)\citenamefont {Kang},
  \citenamefont {Prokudin}, \citenamefont {Ringer},\ and\ \citenamefont
  {Yuan}}]{Kang:2017btw}%
  \BibitemOpen
  \bibfield  {author} {\bibinfo {author} {\bibfnamefont {Z.-B.}\ \bibnamefont
  {Kang}}, \bibinfo {author} {\bibfnamefont {A.}~\bibnamefont {Prokudin}},
  \bibinfo {author} {\bibfnamefont {F.}~\bibnamefont {Ringer}}, \ and\ \bibinfo
  {author} {\bibfnamefont {F.}~\bibnamefont {Yuan}},\ }\href {\doibase
  10.1016/j.physletb.2017.10.031} {\bibfield  {journal} {\bibinfo  {journal}
  {Phys. Lett.}\ }\textbf {\bibinfo {volume} {B774}},\ \bibinfo {pages} {635}
  (\bibinfo {year} {2017})},\ \Eprint {http://arxiv.org/abs/1707.00913}
  {arXiv:1707.00913 [hep-ph]} \BibitemShut {NoStop}%
\bibitem [{\citenamefont {Kang}\ \emph
  {et~al.}(2015{\natexlab{a}})\citenamefont {Kang}, \citenamefont {Prokudin},
  \citenamefont {Sun},\ and\ \citenamefont {Yuan}}]{Kang:2014zza}%
  \BibitemOpen
  \bibfield  {author} {\bibinfo {author} {\bibfnamefont {Z.-B.}\ \bibnamefont
  {Kang}}, \bibinfo {author} {\bibfnamefont {A.}~\bibnamefont {Prokudin}},
  \bibinfo {author} {\bibfnamefont {P.}~\bibnamefont {Sun}}, \ and\ \bibinfo
  {author} {\bibfnamefont {F.}~\bibnamefont {Yuan}},\ }\href {\doibase
  10.1103/PhysRevD.91.071501} {\bibfield  {journal} {\bibinfo  {journal} {Phys.
  Rev.}\ }\textbf {\bibinfo {volume} {D91}},\ \bibinfo {pages} {071501}
  (\bibinfo {year} {2015}{\natexlab{a}})},\ \Eprint
  {http://arxiv.org/abs/1410.4877} {arXiv:1410.4877 [hep-ph]} \BibitemShut
  {NoStop}%
\bibitem [{\citenamefont {Landry}\ \emph {et~al.}(2003)\citenamefont {Landry},
  \citenamefont {Brock}, \citenamefont {Nadolsky},\ and\ \citenamefont
  {Yuan}}]{Landry:2002ix}%
  \BibitemOpen
  \bibfield  {author} {\bibinfo {author} {\bibfnamefont {F.}~\bibnamefont
  {Landry}}, \bibinfo {author} {\bibfnamefont {R.}~\bibnamefont {Brock}},
  \bibinfo {author} {\bibfnamefont {P.~M.}\ \bibnamefont {Nadolsky}}, \ and\
  \bibinfo {author} {\bibfnamefont {C.-P.}\ \bibnamefont {Yuan}},\ }\href
  {\doibase 10.1103/PhysRevD.67.073016} {\bibfield  {journal} {\bibinfo
  {journal} {Phys. Rev.}\ }\textbf {\bibinfo {volume} {D67}},\ \bibinfo {pages}
  {073016} (\bibinfo {year} {2003})},\ \Eprint
  {http://arxiv.org/abs/hep-ph/0212159} {arXiv:hep-ph/0212159 [hep-ph]}
  \BibitemShut {NoStop}%
\bibitem [{\citenamefont {Sun}\ \emph {et~al.}(2017)\citenamefont {Sun},
  \citenamefont {Isaacson}, \citenamefont {Yuan},\ and\ \citenamefont
  {Yuan}}]{Sun:2016kkh}%
  \BibitemOpen
  \bibfield  {author} {\bibinfo {author} {\bibfnamefont {P.}~\bibnamefont
  {Sun}}, \bibinfo {author} {\bibfnamefont {J.}~\bibnamefont {Isaacson}},
  \bibinfo {author} {\bibfnamefont {C.~P.}\ \bibnamefont {Yuan}}, \ and\
  \bibinfo {author} {\bibfnamefont {F.}~\bibnamefont {Yuan}},\ }\href {\doibase
  10.1016/j.physletb.2017.02.037} {\bibfield  {journal} {\bibinfo  {journal}
  {Phys. Lett.}\ }\textbf {\bibinfo {volume} {B769}},\ \bibinfo {pages} {57}
  (\bibinfo {year} {2017})},\ \Eprint {http://arxiv.org/abs/1602.08133}
  {arXiv:1602.08133 [hep-ph]} \BibitemShut {NoStop}%
\bibitem [{\citenamefont {Sun}\ \emph {et~al.}(2011)\citenamefont {Sun},
  \citenamefont {Xiao},\ and\ \citenamefont {Yuan}}]{Sun:2011iw}%
  \BibitemOpen
  \bibfield  {author} {\bibinfo {author} {\bibfnamefont {P.}~\bibnamefont
  {Sun}}, \bibinfo {author} {\bibfnamefont {B.-W.}\ \bibnamefont {Xiao}}, \
  and\ \bibinfo {author} {\bibfnamefont {F.}~\bibnamefont {Yuan}},\ }\href
  {\doibase 10.1103/PhysRevD.84.094005} {\bibfield  {journal} {\bibinfo
  {journal} {Phys. Rev.}\ }\textbf {\bibinfo {volume} {D84}},\ \bibinfo {pages}
  {094005} (\bibinfo {year} {2011})},\ \Eprint {http://arxiv.org/abs/1109.1354}
  {arXiv:1109.1354 [hep-ph]} \BibitemShut {NoStop}%
\bibitem [{\citenamefont {Echevarria}\ \emph {et~al.}(2015)\citenamefont
  {Echevarria}, \citenamefont {Kasemets}, \citenamefont {Mulders},\ and\
  \citenamefont {Pisano}}]{Echevarria:2015uaa}%
  \BibitemOpen
  \bibfield  {author} {\bibinfo {author} {\bibfnamefont {M.~G.}\ \bibnamefont
  {Echevarria}}, \bibinfo {author} {\bibfnamefont {T.}~\bibnamefont
  {Kasemets}}, \bibinfo {author} {\bibfnamefont {P.~J.}\ \bibnamefont
  {Mulders}}, \ and\ \bibinfo {author} {\bibfnamefont {C.}~\bibnamefont
  {Pisano}},\ }\href {\doibase 10.1007/JHEP07(2015)158,
  10.1007/JHEP05(2017)073} {\bibfield  {journal} {\bibinfo  {journal} {JHEP}\
  }\textbf {\bibinfo {volume} {07}},\ \bibinfo {pages} {158} (\bibinfo {year}
  {2015})},\ \bibinfo {note} {[Erratum: JHEP05,073(2017)]},\ \Eprint
  {http://arxiv.org/abs/1502.05354} {arXiv:1502.05354 [hep-ph]} \BibitemShut
  {NoStop}%
\bibitem [{\citenamefont {Guzzi}\ \emph {et~al.}(2014)\citenamefont {Guzzi},
  \citenamefont {Nadolsky},\ and\ \citenamefont {Wang}}]{Guzzi:2013aja}%
  \BibitemOpen
  \bibfield  {author} {\bibinfo {author} {\bibfnamefont {M.}~\bibnamefont
  {Guzzi}}, \bibinfo {author} {\bibfnamefont {P.~M.}\ \bibnamefont {Nadolsky}},
  \ and\ \bibinfo {author} {\bibfnamefont {B.}~\bibnamefont {Wang}},\ }\href
  {\doibase 10.1103/PhysRevD.90.014030} {\bibfield  {journal} {\bibinfo
  {journal} {Phys. Rev.}\ }\textbf {\bibinfo {volume} {D90}},\ \bibinfo {pages}
  {014030} (\bibinfo {year} {2014})},\ \Eprint {http://arxiv.org/abs/1309.1393}
  {arXiv:1309.1393 [hep-ph]} \BibitemShut {NoStop}%
\bibitem [{\citenamefont {Echevarria}\ \emph {et~al.}(2016)\citenamefont
  {Echevarria}, \citenamefont {Scimemi},\ and\ \citenamefont
  {Vladimirov}}]{Echevarria:2016scs}%
  \BibitemOpen
  \bibfield  {author} {\bibinfo {author} {\bibfnamefont {M.~G.}\ \bibnamefont
  {Echevarria}}, \bibinfo {author} {\bibfnamefont {I.}~\bibnamefont {Scimemi}},
  \ and\ \bibinfo {author} {\bibfnamefont {A.}~\bibnamefont {Vladimirov}},\
  }\href {\doibase 10.1007/JHEP09(2016)004} {\bibfield  {journal} {\bibinfo
  {journal} {JHEP}\ }\textbf {\bibinfo {volume} {09}},\ \bibinfo {pages} {004}
  (\bibinfo {year} {2016})},\ \Eprint {http://arxiv.org/abs/1604.07869}
  {arXiv:1604.07869 [hep-ph]} \BibitemShut {NoStop}%
\bibitem [{\citenamefont {Echevarria}\ \emph {et~al.}(2014)\citenamefont
  {Echevarria}, \citenamefont {Idilbi}, \citenamefont {Kang},\ and\
  \citenamefont {Vitev}}]{Echevarria:2014xaa}%
  \BibitemOpen
  \bibfield  {author} {\bibinfo {author} {\bibfnamefont {M.~G.}\ \bibnamefont
  {Echevarria}}, \bibinfo {author} {\bibfnamefont {A.}~\bibnamefont {Idilbi}},
  \bibinfo {author} {\bibfnamefont {Z.-B.}\ \bibnamefont {Kang}}, \ and\
  \bibinfo {author} {\bibfnamefont {I.}~\bibnamefont {Vitev}},\ }\href
  {\doibase 10.1103/PhysRevD.89.074013} {\bibfield  {journal} {\bibinfo
  {journal} {Phys. Rev.}\ }\textbf {\bibinfo {volume} {D89}},\ \bibinfo {pages}
  {074013} (\bibinfo {year} {2014})},\ \Eprint {http://arxiv.org/abs/1401.5078}
  {arXiv:1401.5078 [hep-ph]} \BibitemShut {NoStop}%
\bibitem [{\citenamefont {Boglione}\ \emph {et~al.}(2018)\citenamefont
  {Boglione}, \citenamefont {D'Alesio}, \citenamefont {Flore},\ and\
  \citenamefont {Gonzalez-Hernandez}}]{Boglione:2018dqd}%
  \BibitemOpen
  \bibfield  {author} {\bibinfo {author} {\bibfnamefont {M.}~\bibnamefont
  {Boglione}}, \bibinfo {author} {\bibfnamefont {U.}~\bibnamefont {D'Alesio}},
  \bibinfo {author} {\bibfnamefont {C.}~\bibnamefont {Flore}}, \ and\ \bibinfo
  {author} {\bibfnamefont {J.~O.}\ \bibnamefont {Gonzalez-Hernandez}},\ }\href
  {\doibase 10.1007/JHEP07(2018)148} {\bibfield  {journal} {\bibinfo  {journal}
  {JHEP}\ }\textbf {\bibinfo {volume} {07}},\ \bibinfo {pages} {148} (\bibinfo
  {year} {2018})},\ \Eprint {http://arxiv.org/abs/1806.10645} {arXiv:1806.10645
  [hep-ph]} \BibitemShut {NoStop}%
\bibitem [{\citenamefont {Boglione}\ \emph {et~al.}(2017)\citenamefont
  {Boglione}, \citenamefont {Gonzalez-Hernandez},\ and\ \citenamefont
  {Taghavi}}]{Boglione:2017jlh}%
  \BibitemOpen
  \bibfield  {author} {\bibinfo {author} {\bibfnamefont {M.}~\bibnamefont
  {Boglione}}, \bibinfo {author} {\bibfnamefont {J.~O.}\ \bibnamefont
  {Gonzalez-Hernandez}}, \ and\ \bibinfo {author} {\bibfnamefont
  {R.}~\bibnamefont {Taghavi}},\ }\href {\doibase
  10.1016/j.physletb.2017.06.034} {\bibfield  {journal} {\bibinfo  {journal}
  {Phys. Lett.}\ }\textbf {\bibinfo {volume} {B772}},\ \bibinfo {pages} {78}
  (\bibinfo {year} {2017})},\ \Eprint {http://arxiv.org/abs/1704.08882}
  {arXiv:1704.08882 [hep-ph]} \BibitemShut {NoStop}%
\bibitem [{\citenamefont {Kang}\ \emph
  {et~al.}(2015{\natexlab{b}})\citenamefont {Kang}, \citenamefont {Prokudin},
  \citenamefont {Sun},\ and\ \citenamefont {Yuan}}]{Kang:2015msa}%
  \BibitemOpen
  \bibfield  {author} {\bibinfo {author} {\bibfnamefont {Z.-B.}\ \bibnamefont
  {Kang}}, \bibinfo {author} {\bibfnamefont {A.}~\bibnamefont {Prokudin}},
  \bibinfo {author} {\bibfnamefont {P.}~\bibnamefont {Sun}}, \ and\ \bibinfo
  {author} {\bibfnamefont {F.}~\bibnamefont {Yuan}},\ }\href@noop {} {\
  (\bibinfo {year} {2015}{\natexlab{b}})},\ \Eprint
  {http://arxiv.org/abs/1505.05589} {arXiv:1505.05589 [hep-ph]} \BibitemShut
  {NoStop}%
\bibitem [{\citenamefont {Lin}\ \emph {et~al.}(2018)\citenamefont {Lin},
  \citenamefont {Melnitchouk}, \citenamefont {Prokudin}, \citenamefont {Sato},\
  and\ \citenamefont {Shows}}]{Lin:2017stx}%
  \BibitemOpen
  \bibfield  {author} {\bibinfo {author} {\bibfnamefont {H.-W.}\ \bibnamefont
  {Lin}}, \bibinfo {author} {\bibfnamefont {W.}~\bibnamefont {Melnitchouk}},
  \bibinfo {author} {\bibfnamefont {A.}~\bibnamefont {Prokudin}}, \bibinfo
  {author} {\bibfnamefont {N.}~\bibnamefont {Sato}}, \ and\ \bibinfo {author}
  {\bibfnamefont {H.}~\bibnamefont {Shows}},\ }\href {\doibase
  10.1103/PhysRevLett.120.152502} {\bibfield  {journal} {\bibinfo  {journal}
  {Phys. Rev. Lett.}\ }\textbf {\bibinfo {volume} {120}},\ \bibinfo {pages}
  {152502} (\bibinfo {year} {2018})},\ \Eprint
  {http://arxiv.org/abs/1710.09858} {arXiv:1710.09858 [hep-ph]} \BibitemShut
  {NoStop}%
\bibitem [{\citenamefont {Ye}\ \emph {et~al.}(2017)\citenamefont {Ye},
  \citenamefont {Sato}, \citenamefont {Allada}, \citenamefont {Liu},
  \citenamefont {Chen}, \citenamefont {Gao}, \citenamefont {Kang},
  \citenamefont {Prokudin}, \citenamefont {Sun},\ and\ \citenamefont
  {Yuan}}]{Ye:2016prn}%
  \BibitemOpen
  \bibfield  {author} {\bibinfo {author} {\bibfnamefont {Z.}~\bibnamefont
  {Ye}}, \bibinfo {author} {\bibfnamefont {N.}~\bibnamefont {Sato}}, \bibinfo
  {author} {\bibfnamefont {K.}~\bibnamefont {Allada}}, \bibinfo {author}
  {\bibfnamefont {T.}~\bibnamefont {Liu}}, \bibinfo {author} {\bibfnamefont
  {J.-P.}\ \bibnamefont {Chen}}, \bibinfo {author} {\bibfnamefont
  {H.}~\bibnamefont {Gao}}, \bibinfo {author} {\bibfnamefont {Z.-B.}\
  \bibnamefont {Kang}}, \bibinfo {author} {\bibfnamefont {A.}~\bibnamefont
  {Prokudin}}, \bibinfo {author} {\bibfnamefont {P.}~\bibnamefont {Sun}}, \
  and\ \bibinfo {author} {\bibfnamefont {F.}~\bibnamefont {Yuan}},\ }\href
  {\doibase 10.1016/j.physletb.2017.01.046} {\bibfield  {journal} {\bibinfo
  {journal} {Phys. Lett.}\ }\textbf {\bibinfo {volume} {B767}},\ \bibinfo
  {pages} {91} (\bibinfo {year} {2017})},\ \Eprint
  {http://arxiv.org/abs/1609.02449} {arXiv:1609.02449 [hep-ph]} \BibitemShut
  {NoStop}%
\bibitem [{\citenamefont {Bastami}\ \emph {et~al.}(2018)\citenamefont {Bastami}
  \emph {et~al.}}]{Bastami:2018xqd}%
  \BibitemOpen
  \bibfield  {author} {\bibinfo {author} {\bibfnamefont {S.}~\bibnamefont
  {Bastami}} \emph {et~al.},\ }\href@noop {} {\  (\bibinfo {year} {2018})},\
  \Eprint {http://arxiv.org/abs/1807.10606} {arXiv:1807.10606 [hep-ph]}
  \BibitemShut {NoStop}%
\bibitem [{\citenamefont {Bertone}\ \emph {et~al.}(2019)\citenamefont
  {Bertone}, \citenamefont {Scimemi},\ and\ \citenamefont
  {Vladimirov}}]{Bertone:2019nxa}%
  \BibitemOpen
  \bibfield  {author} {\bibinfo {author} {\bibfnamefont {V.}~\bibnamefont
  {Bertone}}, \bibinfo {author} {\bibfnamefont {I.}~\bibnamefont {Scimemi}}, \
  and\ \bibinfo {author} {\bibfnamefont {A.}~\bibnamefont {Vladimirov}},\
  }\href@noop {} {\  (\bibinfo {year} {2019})},\ \Eprint
  {http://arxiv.org/abs/1902.08474} {arXiv:1902.08474 [hep-ph]} \BibitemShut
  {NoStop}%
\bibitem [{\citenamefont {Airapetian}\ \emph {et~al.}(2013)\citenamefont
  {Airapetian} \emph {et~al.}}]{Airapetian:2012ki}%
  \BibitemOpen
  \bibfield  {author} {\bibinfo {author} {\bibfnamefont {A.}~\bibnamefont
  {Airapetian}} \emph {et~al.} (\bibinfo {collaboration} {HERMES}),\ }\href
  {\doibase 10.1103/PhysRevD.87.074029} {\bibfield  {journal} {\bibinfo
  {journal} {Phys. Rev.}\ }\textbf {\bibinfo {volume} {D87}},\ \bibinfo {pages}
  {074029} (\bibinfo {year} {2013})},\ \Eprint {http://arxiv.org/abs/1212.5407}
  {arXiv:1212.5407 [hep-ex]} \BibitemShut {NoStop}%
\bibitem [{\citenamefont {Adolph}\ \emph {et~al.}(2013)\citenamefont {Adolph}
  \emph {et~al.}}]{Adolph:2013stb}%
  \BibitemOpen
  \bibfield  {author} {\bibinfo {author} {\bibfnamefont {C.}~\bibnamefont
  {Adolph}} \emph {et~al.} (\bibinfo {collaboration} {COMPASS}),\ }\href
  {\doibase 10.1140/epjc/s10052-013-2531-6, 10.1140/epjc/s10052-014-3255-y}
  {\bibfield  {journal} {\bibinfo  {journal} {Eur. Phys. J.}\ }\textbf
  {\bibinfo {volume} {C73}},\ \bibinfo {pages} {2531} (\bibinfo {year}
  {2013})},\ \bibinfo {note} {[Erratum: Eur. Phys. J.C75,no.2,94(2015)]},\
  \Eprint {http://arxiv.org/abs/1305.7317} {arXiv:1305.7317 [hep-ex]}
  \BibitemShut {NoStop}%
\bibitem [{\citenamefont {Adolph}\ \emph {et~al.}(2017)\citenamefont {Adolph}
  \emph {et~al.}}]{Adolph:2016bga}%
  \BibitemOpen
  \bibfield  {author} {\bibinfo {author} {\bibfnamefont {C.}~\bibnamefont
  {Adolph}} \emph {et~al.} (\bibinfo {collaboration} {COMPASS}),\ }\href
  {\doibase 10.1016/j.physletb.2016.09.042} {\bibfield  {journal} {\bibinfo
  {journal} {Phys. Lett.}\ }\textbf {\bibinfo {volume} {B764}},\ \bibinfo
  {pages} {1} (\bibinfo {year} {2017})},\ \Eprint
  {http://arxiv.org/abs/1604.02695} {arXiv:1604.02695 [hep-ex]} \BibitemShut
  {NoStop}%
\bibitem [{\citenamefont {Seidl}\ \emph {et~al.}(2019)\citenamefont {Seidl}
  \emph {et~al.}}]{Seidl:2019jei}%
  \BibitemOpen
  \bibfield  {author} {\bibinfo {author} {\bibfnamefont {R.}~\bibnamefont
  {Seidl}} \emph {et~al.} (\bibinfo {collaboration} {Belle}),\ }\href@noop {}
  {\bibfield  {journal} {\bibinfo  {journal} {Submitted to: Phys. Rev. D}\ }
  (\bibinfo {year} {2019})},\ \Eprint {http://arxiv.org/abs/1902.01552}
  {arXiv:1902.01552 [hep-ex]} \BibitemShut {NoStop}%
\bibitem [{\citenamefont {Aschenauer}\ \emph {et~al.}(2019)\citenamefont
  {Aschenauer}, \citenamefont {Borsa}, \citenamefont {Sassot},\ and\
  \citenamefont {Van~Hulse}}]{Aschenauer:2019kzf}%
  \BibitemOpen
  \bibfield  {author} {\bibinfo {author} {\bibfnamefont {E.~C.}\ \bibnamefont
  {Aschenauer}}, \bibinfo {author} {\bibfnamefont {I.}~\bibnamefont {Borsa}},
  \bibinfo {author} {\bibfnamefont {R.}~\bibnamefont {Sassot}}, \ and\ \bibinfo
  {author} {\bibfnamefont {C.}~\bibnamefont {Van~Hulse}},\ }\href@noop {} {\
  (\bibinfo {year} {2019})},\ \Eprint {http://arxiv.org/abs/1902.10663}
  {arXiv:1902.10663 [hep-ph]} \BibitemShut {NoStop}%
\bibitem [{\citenamefont {Bradamante}(2018)}]{Bradamante:2018ick}%
  \BibitemOpen
  \bibfield  {author} {\bibinfo {author} {\bibfnamefont {F.}~\bibnamefont
  {Bradamante}} (\bibinfo {collaboration} {COMPASS}),\ }in\ \href@noop {}
  {\emph {\bibinfo {booktitle} {{23rd International Symposium on Spin Physics
  (SPIN 2018) Ferrara, Italy, September 10-14, 2018}}}}\ (\bibinfo {year}
  {2018})\ \Eprint {http://arxiv.org/abs/1812.07281} {arXiv:1812.07281
  [hep-ex]} \BibitemShut {NoStop}%
\bibitem [{\citenamefont {van Neerven}(1986)}]{vanNeerven:1985xr}%
  \BibitemOpen
  \bibfield  {author} {\bibinfo {author} {\bibfnamefont {W.~L.}\ \bibnamefont
  {van Neerven}},\ }\href {\doibase 10.1016/0550-3213(86)90165-3} {\bibfield
  {journal} {\bibinfo  {journal} {Nucl. Phys.}\ }\textbf {\bibinfo {volume}
  {B268}},\ \bibinfo {pages} {453} (\bibinfo {year} {1986})}\BibitemShut
  {NoStop}%
\bibitem [{\citenamefont {Berger}\ and\ \citenamefont
  {Forde}(2010)}]{Berger:2009zb}%
  \BibitemOpen
  \bibfield  {author} {\bibinfo {author} {\bibfnamefont {C.~F.}\ \bibnamefont
  {Berger}}\ and\ \bibinfo {author} {\bibfnamefont {D.}~\bibnamefont {Forde}},\
  }\href {\doibase 10.1146/annurev.nucl.012809.104538} {\bibfield  {journal}
  {\bibinfo  {journal} {Ann. Rev. Nucl. Part. Sci.}\ }\textbf {\bibinfo
  {volume} {60}},\ \bibinfo {pages} {181} (\bibinfo {year} {2010})},\ \Eprint
  {http://arxiv.org/abs/0912.3534} {arXiv:0912.3534 [hep-ph]} \BibitemShut
  {NoStop}%
\bibitem [{\citenamefont {Britto}(2011)}]{Britto:2010xq}%
  \BibitemOpen
  \bibfield  {author} {\bibinfo {author} {\bibfnamefont {R.}~\bibnamefont
  {Britto}},\ }\href {\doibase 10.1088/1751-8113/44/45/454006} {\bibfield
  {journal} {\bibinfo  {journal} {J. Phys.}\ }\textbf {\bibinfo {volume}
  {A44}},\ \bibinfo {pages} {454006} (\bibinfo {year} {2011})},\ \Eprint
  {http://arxiv.org/abs/1012.4493} {arXiv:1012.4493 [hep-th]} \BibitemShut
  {NoStop}%
\bibitem [{\citenamefont {Ellis}\ \emph {et~al.}(2012)\citenamefont {Ellis},
  \citenamefont {Kunszt}, \citenamefont {Melnikov},\ and\ \citenamefont
  {Zanderighi}}]{Ellis:2011cr}%
  \BibitemOpen
  \bibfield  {author} {\bibinfo {author} {\bibfnamefont {R.~K.}\ \bibnamefont
  {Ellis}}, \bibinfo {author} {\bibfnamefont {Z.}~\bibnamefont {Kunszt}},
  \bibinfo {author} {\bibfnamefont {K.}~\bibnamefont {Melnikov}}, \ and\
  \bibinfo {author} {\bibfnamefont {G.}~\bibnamefont {Zanderighi}},\ }\href
  {\doibase 10.1016/j.physrep.2012.01.008} {\bibfield  {journal} {\bibinfo
  {journal} {Phys. Rept.}\ }\textbf {\bibinfo {volume} {518}},\ \bibinfo
  {pages} {141} (\bibinfo {year} {2012})},\ \Eprint
  {http://arxiv.org/abs/1105.4319} {arXiv:1105.4319 [hep-ph]} \BibitemShut
  {NoStop}%
\bibitem [{\citenamefont {Passarino}\ and\ \citenamefont
  {Veltman}(1979)}]{Passarino:1978jh}%
  \BibitemOpen
  \bibfield  {author} {\bibinfo {author} {\bibfnamefont {G.}~\bibnamefont
  {Passarino}}\ and\ \bibinfo {author} {\bibfnamefont {M.~J.~G.}\ \bibnamefont
  {Veltman}},\ }\href {\doibase 10.1016/0550-3213(79)90234-7} {\bibfield
  {journal} {\bibinfo  {journal} {Nucl. Phys.}\ }\textbf {\bibinfo {volume}
  {B160}},\ \bibinfo {pages} {151} (\bibinfo {year} {1979})}\BibitemShut
  {NoStop}%
\bibitem [{\citenamefont {Ellis}\ and\ \citenamefont
  {Zanderighi}(2008)}]{Ellis:2007qk}%
  \BibitemOpen
  \bibfield  {author} {\bibinfo {author} {\bibfnamefont {R.~K.}\ \bibnamefont
  {Ellis}}\ and\ \bibinfo {author} {\bibfnamefont {G.}~\bibnamefont
  {Zanderighi}},\ }\href {\doibase 10.1088/1126-6708/2008/02/002} {\bibfield
  {journal} {\bibinfo  {journal} {JHEP}\ }\textbf {\bibinfo {volume} {02}},\
  \bibinfo {pages} {002} (\bibinfo {year} {2008})},\ \Eprint
  {http://arxiv.org/abs/0712.1851} {arXiv:0712.1851 [hep-ph]} \BibitemShut
  {NoStop}%
\bibitem [{\citenamefont {Beenakker}\ \emph {et~al.}(1989)\citenamefont
  {Beenakker}, \citenamefont {Kuijf}, \citenamefont {van Neerven},\ and\
  \citenamefont {Smith}}]{Beenakker:1988bq}%
  \BibitemOpen
  \bibfield  {author} {\bibinfo {author} {\bibfnamefont {W.}~\bibnamefont
  {Beenakker}}, \bibinfo {author} {\bibfnamefont {H.}~\bibnamefont {Kuijf}},
  \bibinfo {author} {\bibfnamefont {W.~L.}\ \bibnamefont {van Neerven}}, \ and\
  \bibinfo {author} {\bibfnamefont {J.}~\bibnamefont {Smith}},\ }\href
  {\doibase 10.1103/PhysRevD.40.54} {\bibfield  {journal} {\bibinfo  {journal}
  {Phys. Rev.}\ }\textbf {\bibinfo {volume} {D40}},\ \bibinfo {pages} {54}
  (\bibinfo {year} {1989})}\BibitemShut {NoStop}%
\end{thebibliography}%


\providecommand{\noopsort}[1]{}
%

\end{document}